\documentclass[aip,reprint,nofootinbib]{revtex4-2}

\pdfoutput=1

\usepackage{graphicx}
\usepackage{dcolumn}
\usepackage{bm}
\usepackage{amsmath}
\usepackage{amssymb}
\usepackage{booktabs}
\usepackage{multirow}
\usepackage{setspace}
\usepackage{enumitem}
\usepackage{xr}
\usepackage{physics}
\usepackage{qcircuit}
\usepackage{hyperref}
\usepackage{url}

\usepackage{xcolor}
\usepackage{siunitx}
\usepackage{afterpage}
\usepackage{mleftright}

\makeatletter
\AtBeginDocument{\let\LS@rot\@undefined}
\newcommand{\smallvast}{\bBigg@{2}}
\newcommand{\midvast}{\bBigg@{3}}
\newcommand{\vast}{\bBigg@{4}}
\newcommand{\Vast}{\bBigg@{6}}
\makeatother

\protected\def\verythinspace{
  \ifmmode
    \mskip0.5\thinmuskip
  \else
    \ifhmode
      \kern0.08334em
    \fi
  \fi
}

\hypersetup{
    pdfborder={0 0 0}
}

\begin{document}

\title{Tutorial: Gate-based superconducting quantum computing}

\author{Sangil Kwon}
\email{kwon2866@gmail.com}
\affiliation{\mbox{Department of Physics, Tokyo University of Science, Shinjuku, Tokyo 162-0825, Japan}}

\author{Akiyoshi Tomonaga}
\affiliation{\mbox{Department of Physics, Tokyo University of Science, Shinjuku, Tokyo 162-0825, Japan}}
\affiliation{\mbox{RIKEN, Center for Emergent Matter Science (CEMS), Wako, Saitama 351-0198, Japan}}

\author{Gopika Lakshmi Bhai}
\affiliation{\mbox{Department of Physics, Tokyo University of Science, Shinjuku, Tokyo 162-0825, Japan}}
\affiliation{\mbox{RIKEN, Center for Emergent Matter Science (CEMS), Wako, Saitama 351-0198, Japan}}

\author{Simon J. Devitt}
\affiliation{Centre for Quantum Software and Information (QSI), Faculty of Engineering and Information Technology, University of Technology Sydney, Sydney, NSW, 2007, Australia}

\author{Jaw-Shen Tsai}
\affiliation{\mbox{Department of Physics, Tokyo University of Science, Shinjuku, Tokyo 162-0825, Japan}}
\affiliation{\mbox{RIKEN, Center for Emergent Matter Science (CEMS), Wako, Saitama 351-0198, Japan}}

\date{\today}

\begin{abstract}
In this tutorial, we introduce basic conceptual elements to understand and build a gate-based superconducting quantum computing system.
\end{abstract}

\maketitle

\tableofcontents

\section{Introduction}
\label{sec:intro}

Quantum computing is considered as a next-generation information processing technology.
The basic element of a quantum computing system is a quantum bit, often called a qubit.
Over the last few decades, considerable progress has been made toward realizing quantum computing systems by physically implementing a qubit in various systems such as ion traps, quantum dots, nuclear spins, and cavity quantum electrodynamics.
The scalability of such a qubit is considered to be a prerequisite for a practical quantum computer of the future. 
In this regard, a solid-state qubit has been considered to be indispensable.
Superconducting quantum systems are one of the most promising candidates because, in these systems, qubits are intrinsically integrated in a solid-state device, and their wide range of choice for the qubit parameters is a considerable advantage, which in turn gives flexibility in designing such quantum circuits.

In this tutorial, we try to provide basic conceptual elements to understand and build a potentially scalable superconducting quantum computing system based on gate operations.
The logical flow is roughly from principle to practice.
After introducing the qubit and structure of a universal quantum computing system (Sec.~\ref{sec:univQC}), we explain a superconducting circuit that can be used as a qubit (Secs.~\ref{sec:SCqubit} and \ref{sec:relaxation}) and 
how to implement basic functions that are required for quantum computation (Secs.~\ref{sec:coupling} and \ref{sec:implementationQC}).
Then, we introduce a quantum error correction scheme, called the surface code, that is believed to be suitable for superconducting qubit systems (Sec.~\ref{sec:QEC}).
Lastly, we deal with practical topics, such as how to characterize and control a quantum system (Secs.~\ref{sec:characterize} and \ref{sec:control}).
The contents of this tutorial are briefly summarized in Table~\ref{tab:TOC}.

\begin{table}
\caption{Brief description of the contents.}
\label{tab:TOC}
\centering
\begin{ruledtabular}
\begin{tabular}{r l}
\noalign{\smallskip}
\textbf{\ref{sec:univQC}.} & 
\begin{minipage}[t]{0.42\textwidth}\raggedright
\textbf{Universal Quantum Computing System} \\
This section introduces a quantum bit, quantum gates, and a possible structure of a universal quantum computing system.
\end{minipage} \\ \noalign{\smallskip}

\textbf{\ref{sec:SCqubit}.} & 
\begin{minipage}[t]{0.42\textwidth}\raggedright
\textbf{Superconducting Qubit} \\
This section describes elementary circuits that can be used as qubits and their properties in various circuit parameter regimes.
\end{minipage} \\ \noalign{\smallskip}

\textbf{\ref{sec:relaxation}.} & 
\begin{minipage}[t]{0.42\textwidth}\raggedright
\textbf{Effect of Noise} \\
This section discusses the loss mechanisms of quantum information and several noise-resilient circuit designs.
\end{minipage} \\ \noalign{\smallskip}

\textbf{\ref{sec:coupling}.} & 
\begin{minipage}[t]{0.42\textwidth}\raggedright
\textbf{Coupling} \\
This section explains coupling schemes between a qubit and other quantum systems using classical analogies.
\end{minipage} \\ \noalign{\smallskip}

\textbf{\ref{sec:implementationQC}.} & 
\begin{minipage}[t]{0.42\textwidth}\raggedright
\textbf{Implementation of Quantum Computation} \\
This section explains how to implement basic functions that are required for quantum computation, such as readout, gate operation, and initialization.
\end{minipage} \\ \noalign{\smallskip}

\textbf{\ref{sec:QEC}.} & 
\begin{minipage}[t]{0.42\textwidth}\raggedright
\textbf{Quantum Error Correction} \\
This section explains how to construct an error-free logical qubit and how to perform logical gate operations in the context of the surface code.
\end{minipage} \\ \noalign{\smallskip}

\textbf{\ref{sec:characterize}.} & 
\begin{minipage}[t]{0.42\textwidth}\raggedright
\textbf{Characterizing a Quantum System} \\
This section describes standard procedures for quantum system characterization.
\end{minipage} \\ \noalign{\smallskip}

\textbf{\ref{sec:control}.} & 
\begin{minipage}[t]{0.42\textwidth}\raggedright
\textbf{Controlling a Quantum System} \\
This section explains several useful techniques for controlling a quantum system and their working principles.
\end{minipage} \\ \noalign{\smallskip}
\end{tabular}
\end{ruledtabular}
\end{table}

Since this is a tutorial, the topics covered here are very selective rather than comprehensive.
Hence, we cite references that are more accessible to readers.
Another reason for this is that many concepts and experimental techniques for superconducting circuits were originally developed in other branches of science---tracing all historical literature is not meaningful for readers.
For comprehensive reviews on this field, see Refs.~\onlinecite{YN, wendin, gu, mit1, mit2, yale}.

Regarding the difficulty of this tutorial, we assume that readers are somewhat familiar with quantum mechanics, especially the Dirac notation and the occupation number representation (second quantization), and elementary statistical mechanics, such as the Boltzmann distribution.
Since superconducting quantum computing systems are electrical circuits, knowledge on basic electrical engineering will be helpful, especially the S-parameters.
However, readers do not need to be masters of these topics.
Reading this tutorial does not require deep physical insights---it is more like learning a new language.\cite{quantumAdvanture}
Once you get used to it, you will enjoy it.

Before entering the main part, we would like to point out that the word ``scaling'' in quantum engineering is different from that in the semiconductor industry. In the semiconductor industry, scaling means reducing the size of the information processing device used, such as a transistor, and the energy cost per bit, so that we can integrate more and more devices into a chip.
In quantum engineering, ``scaling'' simply means adding more qubits because physical quantities involved in operations of a superconducting quantum computing platform, such as the charge of a Cooper pair and magnetic flux quantum, are already at the quantum limit, and a quantum information processing device is lossless.
Thus, the dramatic size reduction as demonstrated in Moore's law may not be expected for superconducting qubits.

A set of formulas for deriving equations in this tutorial are summarized in Table~\ref{tab:formula} at the end of the main text.

\section{Universal Quantum Computing System}
\label{sec:univQC}

\subsection{Essential Elements}

\subsubsection{Quantum Bit}
\label{sec:quantumBit}

\begin{figure}
\centering
\includegraphics{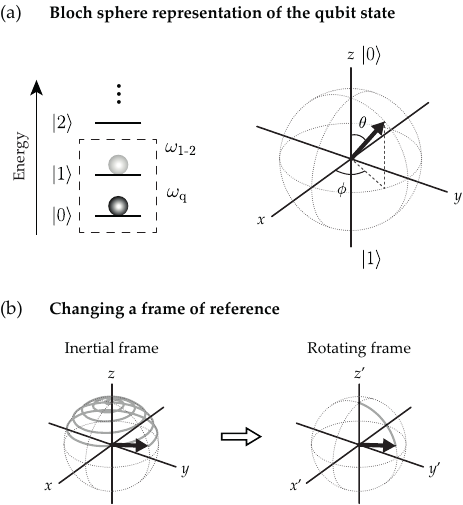}
\caption{(a) Bloch sphere representation of the qubit state.
In the energy level diagram, the two lowest states in the dashed boundary are used for computation.
This subspace is called the computational subspace.
To selectively control these two levels, $\omega_\textrm{q} \neq \omega_\textrm{1-2}$ is required, where $\omega_\textrm{q}$ is the transition frequency between $\ket{0}$ and $\ket{1}$, and $\omega_\textrm{1-2}$ is the transition frequency between $\ket{1}$ and $\ket{2}$.
(b) In the rotating frame, ``trivial evolution'' is eliminated such that we can concentrate on the dynamics we are interested in. The axes with the prime indicate that we are in the rotating frame.
As in the majority of the literature, all Bloch spheres in this tutorial are in the rotating frame.
}
\label{fig:BlochSphere}
\end{figure}

A qubit is a two-level system whose quantum mechanical state displays phase coherence between two basis states, $\ket{0}$ and $\ket{1}$.
The phase coherence between two quantum states can be defined as follows.
The quantum state of a qubit, in general, is a linear superposition of the two basis states $\ket{0}$ and $\ket{1}$, 
\begin{equation}\label{eq:superposition}
\ket{\psi} = \alpha\ket{0} + \beta\ket{1},
\end{equation}
where $\alpha$ and $\beta$ are complex numbers and $|\alpha|^2 + |\beta|^2 = 1$.
We can define the relative quantum phase $\varphi$ of the two basis states as
\begin{equation}
\varphi \equiv \arg(\alpha^*\beta).
\end{equation}
If $\varphi$ of a given state has a definite value, we say that a given state displays phase coherence (often just called coherence) between the states $\ket{0}$ and $\ket{1}$.

An arbitrary qubit state can also be expressed in the density matrix form:
\begin{equation}\label{eq:qubitDM}
\hat{\rho}\, = \ketbra{\psi}{\psi} =
\raise 1.6ex \hbox{$
\begin{array}{l c cc c}
& & \bra{0} & \bra{1} & \\ [3pt]
\ket{0}\phantom{,}\!\!\!\!\! & \multirow{2}{*}{\smallvast(} & |\alpha|^2 & \alpha\beta^* & \multirow{2}{*}{\smallvast)} \\[2pt]
\ket{1}\!\!\!\!\! & & \alpha^*\beta & |\beta|^2 &
\end{array}$}
\end{equation}
where kets and bras around the density matrix indicate the basis. 
Note that the diagonal elements represent the populations of the basis states and the off-diagonal elements represent the phase coherences between these states.
Hence, the populations and coherences depend on the chosen basis.

Note that a qubit and a spin-1/2 system are mathematically identical.
This allows us to represent the qubit state conveniently as an arrow, called the Bloch vector, in the Bloch sphere [Fig.~\ref{fig:BlochSphere}(a)].
Conventionally, the qubit quantization axis is set as the $z$-axis, and the north and south poles represent $\ket{0}$ and $\ket{1}$, respectively.
Hence, the longitudinal component of the Bloch vector corresponds to the polarization of the qubit, and the transverse component corresponds to the coherence between the two basis states.

In Cartesian coordinates, the Bloch vector $\vec{s}$ can be parameterized by the polar angle $\theta$ and the azimuthal angle $\phi$ as
\begin{align}
\vec{s} = (\sin\theta \cos\phi,\, \sin\theta \sin\phi,\, \cos\theta),
\end{align}
where $\theta$ and $\phi$ are the polar and azimuthal angles, respectively [Fig.~\ref{fig:BlochSphere}(a)].
Conversion from the Bloch vector to the density matrix can be done by using the Pauli matrices:
\begin{align}
\hat{\rho}\verythinspace
&= \frac{1}{2}(\hat{I}+s_x\hat{\sigma}_x+s_y\hat{\sigma}_y+s_z\hat{\sigma}_z) \nonumber\\
&= \begin{pmatrix}
\cos^2\!\verythinspace\left(\dfrac{\theta}{2}\right) &
\mathrm{e}^{-\mathrm{i}\phi} \cos(\dfrac{\theta}{2})\sin(\dfrac{\theta}{2}) \\[10pt]
\mathrm{e}^{\mathrm{i}\phi} \cos(\dfrac{\theta}{2})\sin(\dfrac{\theta}{2}) &
\sin^2\!\verythinspace\left(\dfrac{\theta}{2}\right)
\end{pmatrix}\! \label{eq:qubitDMsph}.
\end{align}
One possible mapping between Eqs.~\eqref{eq:qubitDM} and \eqref{eq:qubitDMsph} is $\alpha = \cos(\theta/2)$ and $\beta = \mathrm{e}^{\mathrm{i}\phi}\sin(\theta/2)$.

The rotation of the Bloch vector with the angle $\eta$ about the $k$-axis is done by the rotation operator $\hat{R}_k(\eta)$:
\begin{align}\label{eq:rotOper}
\hat{R}_k(\eta) 
\equiv
\exp(-\textrm{i}\eta\frac{\hat{\sigma}_k}{2})
= 
\cos(\frac{\eta}{2})\hat{I} - \textrm{i}\sin(\frac{\eta}{2})\hat{\sigma}_k
\end{align}
Here, we used the formula
\begin{align}
\exp(-\textrm{i}\eta\hat{A})
&= \left(1-\frac{\eta^2}{2!} + \cdots \right) \hat{I} 
- \textrm{i}\left(\eta - \frac{\eta^3}{3!} + \cdots \right) \hat{A} \nonumber\\
&= \cos(\eta)\hat{I} - \textrm{i}\sin(\eta) \hat{A}, \quad 
\textrm{if } \hat{A}^2 = \hat{I}. \label{eq:matrixExp}
\end{align}

When we use the Bloch sphere, we are free to choose a frame of reference.
In the majority of the literature, including this tutorial, the dynamics of the qubit state are described in the rotating frame [Fig.~\ref{fig:BlochSphere}(b)].
To determine the rotating frame frequency, we have to know the dynamics we want to focus on.
Then, we eliminate the trivial evolution by performing a unitary transformation, which changes our frame of reference.
Note that this is conceptually and mathematically identical to switching into the interaction picture.
Usually, the qubit frequency, the resonator frequency (see Sec.~\ref{sec:circuitAtom}), or the external drive frequency (Sec.~\ref{sec:SQG}) is chosen as the rotating frame frequency.

A qubit is often implemented by the two lowest states of a quantum system, such as (artificial or natural) atoms [Fig.~\ref{fig:BlochSphere}(a)].
This subspace is called the computational subspace.
In general, any Hilbert space whose dimension is truncated into two can be used as a qubit.
This generalized definition of a qubit is essential for constructing a logical qubit (Sec.~\ref{sec:QEC}).

In this tutorial, the notations denoting the qubit states, \{$\ket{0}$, $\ket{1}$, $\ket{2}$ (higher excitation level)\} and \{$\ket{\textrm{g}}$, $\ket{\textrm{e}}$, $\ket{\textrm{f\,}}$\}, are used interchangeably to avoid confusion with the photon or the charge number states.
In addition, $\omega_\textrm{q}$, which we call the qubit frequency, is the transition frequency between $\ket{0}$ and $\ket{1}$, and $\omega_{i\textrm{-}j}$ (with a hyphen in the subscript) is the transition frequency between $\ket{i}$ and $\ket{j}$;
$\omega_{ij}$ (without a hyphen in the subscript) indicates the energy level of the two-qubit state, $\ket{i}\otimes\ket{j}$ (or $\ket{ij}$ in the short form).

A generic two-qubit state can be written as
\begin{equation}
\ket{\psi} = \alpha\ket{00} + \beta\ket{01} + \gamma\ket{10} + \delta\ket{11},
\end{equation}
where $\alpha$, $\beta$, $\gamma$, and $\delta$ are complex numbers and $|\alpha|^2 + |\beta|^2 + |\gamma|^2 + |\delta|^2 = 1$.
In the density matrix form,
\begin{align}
\hat{\rho}\, = \!
\raise 1.6ex \hbox{$
\begin{array}{l c cccc c}
& & \bra{00} & \bra{01} & \bra{10} & \bra{11} & \\[3pt]
\ket{00}\phantom{,}\!\!\!\!\! & \multirow{4}{*}{\vast(} & |\alpha|^2 & \alpha\beta^* & \alpha\gamma^* & \alpha\delta^* & \multirow{4}{*}{\vast)} \\
\ket{01}\!\!\!\!\! & & \alpha^*\beta & |\beta|^2 & \beta\gamma^* & \beta\delta^* & \\
\ket{10}\!\!\!\!\! & & \alpha^*\gamma & \beta^*\gamma & |\gamma|^2 & \gamma\delta^* & \\
\ket{11}\!\!\!\!\! & & \alpha^*\delta & \beta^*\delta & \gamma^*\delta & |\delta|^2 & 
\end{array}$}.
\end{align}

\subsubsection{Quantum Gate}
\label{sec:quantumGate}

\begin{table}
\caption{Universal quantum gate set.
$\hat{R}_k(\eta)$ ($k=x,y,z$) is the rotation operator defined in Eq.~\eqref{eq:rotOper}.
$\ket{\psi}_\textrm{t(c)}$ indicates the quantum state of the target (control) qubit.
$\hat{X}$ is the $X$ gate in the operator form.
}
\label{tab:univGate}\centering
\begin{ruledtabular}
\begin{tabular}{c c c c}
\noalign{\smallskip}
Name	&	Function	& Symbol	&	Matrix \\
\noalign{\smallskip} \hline \noalign{\smallskip}
Pauli-$X$ ($X$) 	&	$\hat{R}_x(\pi)$	&	$\Qcircuit @C=1em @R=1em {&\gate{X}&\qw}$	&	$\begin{pmatrix} 0 & 1 \\ 1 & 0 \end{pmatrix}$	 \\
\noalign{\smallskip}
Pauli-$Y$ ($Y$)	&	$\hat{R}_y(\pi)$	&	$\Qcircuit @C=1em @R=1em {&\gate{Y}&\qw}$	&	$\begin{pmatrix} 0 & -\textrm{i} \\ \textrm{i} & 0 \end{pmatrix}$	 \\
\noalign{\smallskip}
Pauli-$Z$ ($Z$)	&	$\hat{R}_z(\pi)$	&	$\Qcircuit @C=1em @R=1em {&\gate{Z}&\qw}$	&	$\begin{pmatrix} 1 & 0 \\ 0 & -1 \end{pmatrix}$	 \\
\noalign{\smallskip}
Hadamard ($H$)	&	$\hat{R}_x(\pi) \hat{R}_y(\pi/2)$	&	$\Qcircuit @C=1em @R=1em {&\gate{H}&\qw}$	&	$\dfrac{1}{\sqrt{2}} \begin{pmatrix} 1 & 1 \\ 1 & -1 \end{pmatrix}$	 \\
\noalign{\smallskip}
Phase ($S$)	&	$\hat{R}_z(\pi/2)$	&	$\Qcircuit @C=1em @R=1em {&\gate{S}&\qw}$	&	$\begin{pmatrix} 1 & 0 \\ 0 & \textrm{i} \end{pmatrix}$	 \\
\noalign{\smallskip}
$\pi/8$ ($T$)	&	$\hat{R}_z(\pi/4)$	&	$\Qcircuit @C=1em @R=1em {&\gate{T}&\qw}$	&	$\begin{pmatrix} 1 & 0 \\ 0 & \textrm{e}^{\textrm{i}\pi/4} \end{pmatrix}$	 \\
\noalign{\smallskip}
$\begin{array}{c}\textrm{Controlled-NOT} \\ \textrm{(CNOT)} \end{array}$	&	$\begin{array}{c}\hat{X}\ket{\psi}_\textrm{t} \\ \textrm{if } \ket{\psi}_\textrm{c}=\ket{1} \end{array}$	&	$\def\arraystretch{0.6}\begin{array}{c} \Qcircuit @C=1em @R=1em {&\ctrl{1}&\qw\\&\targ&\qw} \\ \end{array}$	&	$\begin{pmatrix} 1 & 0 & 0 & 0 \\ 0 & 1 & 0 &  0 \\ 0 & 0 & 0 & 1 \\ 0 & 0 & 1 & 0 \end{pmatrix}$	 \\
\end{tabular}
\end{ruledtabular}
\end{table}

A quantum gate is a discrete control acting on qubits inducing the unitary evolution of the quantum states of the qubits.
Quantum computation is basically a series of quantum gate operations.

Consider a closed quantum system described by the time-independent Hamiltonian $\hat{\mathcal{H}}$.
The time evolution of such a system is described by a unitary operator $\hat{U}(t)$, which is called the time-evolution operator:\cite{shankar, sakurai}
\begin{equation}\label{eq:timeEvol}
\ket{\psi(t)} = \hat{U}(t)\ket{\psi(0)}.
\end{equation}
The Schr{\"o}dinger equation connects $\hat{U}(t)$ and $\hat{\mathcal{H}}$:
\begin{equation}\label{eq:uniDef}
\hat{U}(t) = \textrm{e}^{-\textrm{i}\hat{\mathcal{H}}t/\hbar}.
\end{equation}
Thus, a gate operation is implemented by engineering the system Hamiltonian such that the resulting unitary evolution of the qubits implements the target gate.

Any multiqubit gate operation can be decomposed into a set of single-qubit and controlled-NOT (CNOT) gates. Thus, the gate set \{single-qubit gates, CNOT\} is called a universal quantum gate set.
An arbitrary single-qubit gate can be well approximated by the discrete gate set \{$H$, $S$, $T$\} (Solovay--Kitaev theorem \cite{NC}).
Hence, we can rewrite a universal gate set as \{$H$, $S$, $T$, CNOT\}.
The definitions of these gates and other popular gates are summarized in Table~\ref{tab:univGate}.

Among these universal quantum gates, the quantum gates generated by the $H$, $S$, and CNOT gates form a group called the Clifford group.
This group is important in quantum computation, especially for quantum error correction (Sec.~\ref{sec:QEC}) and efficient gate qualification (Sec.~\ref{sec:RB}).
However, it is known that a quantum computer operated by only Clifford gates can be simulated efficiently on a probabilistic classical computer (Gottesman--Knill theorem\cite{NC}).
Thus, a non-Clifford gate, such as the $T$ gate, is required to show the advantage of quantum computation.

\subsection{Structure}
\label{sec:structure}

\begin{figure}
\centering
\includegraphics{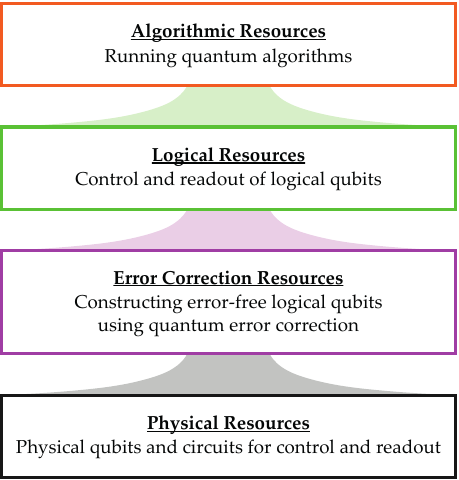}
\caption{Structure of a universal quantum computing system.}
\label{fig:structure}
\end{figure}

A gate-operation-based, universal, and scalable superconducting quantum computer will likely have the following structure (Fig.~\ref{fig:structure}):\cite{codyjones2012, gambetta2017}
\begin{description}
\item[Physical resources]
This layer is a collection of physical qubits and necessary circuits for the control and readout of the physical qubits.

\item[Error correction resources]
In this layer, errors acting on quantum information stored in a set of physical qubits are corrected. This operation produces a single error-free logical qubit.
For this, high fidelity controls, such as initialization, gate operation, readout, and feedback, for physical qubits are required.

\item[Logical resources]
Initialization, gate operation, and readout of logical qubits are performed in this layer.

\item[Algorithmic resources]
Quantum algorithms, such as Shor's factoring and Grover's search algorithms, are performed in this layer.

\end{description}

In this tutorial, the physical resources, the error correction resources, and part of the logical resources are briefly covered.
For quantum algorithms, see the standard textbooks on quantum computation, such as Refs.~\onlinecite{NC, BCRS}.


\section{Superconducting Qubit}
\label{sec:SCqubit}

\subsection{Design Criteria}
\label{sec:qubitCriteria}

A superconducting qubit is the two lowest energy eigenstates of an artificial atom made of a superconducting circuit.
To be a useful qubit, the circuit must be designed to satisfy the following conditions:
\begin{enumerate}
\itemsep-0.1em

\item \label{list:transFreq} \emph{Proper operating frequency range}:
A qubit must have a transition frequency that is significantly higher than the thermal energy of a typical solid-state system to observe quantum nature.
The only continuous refrigeration method for solid state devices below 0.3 K is to use a dilution refrigerator, whose base temperature is usually about 10 mK ($\sim$200 MHz).
This means that the transition frequency of a qubit must be at least a few gigahertz.
At the same time, the qubit transition frequency should be sufficiently lower than the superconducting energy gap of the host superconductor so as not to excite quasiparticles. For aluminum, which is the most popular material for superconducting qubit systems, the energy gap is about 100 GHz.

\item \label{list:anhar} \emph{Large anharmonicity}:
To be a well-defined two-level system, a qubit should have anharmonicity $\alpha \equiv \omega_\textrm{1-2}-\omega_\textrm{q}$ of at least $\sim$100 MHz to perform a reasonably fast gate operation. (See Sec.~\ref{sec:bandWidth} for the gate time and frequency selectivity.)
Recently, it has been found that having a third level in an accessible frequency range can be beneficial, such as for initialization or two-qubit gate operation. (See Secs.~\ref{sec:TQG} and \ref{sec:resetDumping}.)

\item \label{list:relaxTime} \emph{Long coherence time}:
The assigned quantum state should last for a long time compared with the time for gate operations.

\item \label{list:easyCoupling} \emph{Ease of coupling}:
For readout and (multi)qubit gate operation, a reasonably strong coupling between a qubit and another quantum system, such as a resonator or neighboring qubit, should be achieved easily.

\item \label{list:easyControl} \emph{Ease of control}:
The quantum state should be brought to a superposition easily and straightforwardly by an external mean.

\item \label{list:easyFab} \emph{Ease of fabrication}:
A qubit should be easy to fabricate with standard nanotechnology for good reproducibility.

\end{enumerate}

\subsection{Josephson Junction}
\label{sec:JJ}

\begin{figure}
\centering
\includegraphics{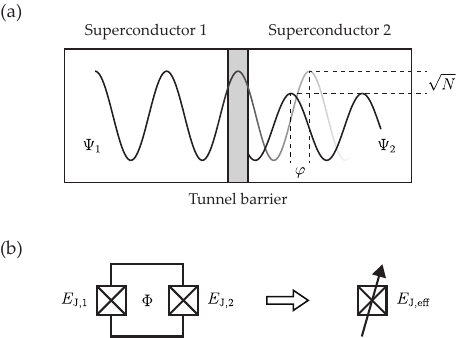}
\caption{(a) Schematic diagram of a Josephson junction where a pair of superconductors are weakly coupled via an oxide tunnel barrier.
The phase and the number difference of the macroscopic wavefunctions $\Psi_1$ and $\Psi_2$ fully determine the physics of the junction.
(b) A DC SQUID can be considered as a variable Josephson junction tuned by an external magnetic flux $\Phi$.
The symbol of a cross in a square represents a Josephson junction.}
\label{fig:junction}
\end{figure}

A superconductor is a macroscopic quantum mechanical system in the sense that it can be described by a single macroscopic wavefunction, i.e., the order parameter $\Psi$.
However, this property is not a sufficient condition for being a qubit;
we need a confinement potential to have discrete energy eigenstates such as electrons in the Coulomb potential forming an atom.
Moreover, to control the two lowest energy eigenstates selectively, the potential must be anharmonic to have distinct energy separation between eigenstates.

The solution for discrete energy eigenstates is to make an electrical circuit.
In a superconducting circuit, the quantized energy level emerges from the quantization of the charge and the magnetic flux stored in various electrical components just like the position and the momentum of electrons in a real atom.\footnote{Since the charge and the magnetic flux are collective coordinates that represent the cooperative motion of large numbers of electrons, the circuit quantization is essentially phenomenological.\cite{yurke1984}}

The solution for the anharmonicity is a Josephson junction where a pair of superconductors are weakly coupled [Fig.~\ref{fig:junction}(a)].
In a superconducting circuit, a Josephson junction acts as a nonlinear inductor, resulting in an anharmonic potential.
Since a superconductor is a macroscopic quantum mechanical system, only two quantities are required to describe the physics of a Josephson junction: the number imbalance of electrons $N$ and the relative phase $\varphi$ between the two superconductors.
Here, $N$ corresponds to the difference in $|\Psi|^2$ of the two superconductors.
The equations of motion regarding these two quantities, called the Josephson equations, are given by\cite{leggett}
\begin{equation}\label{eq:Jeq}
\frac{dN(t)}{dt} = \frac{2E_\textrm{J}}{\hbar} \sin(\varphi(t)) \quad \textrm{and} \quad
\frac{d\varphi(t)}{dt} = -\frac{2e}{\hbar}V(t),
\end{equation}
where
$E_\textrm{J}$ is the Josephson energy, which is a measure of the ability of Cooper pairs to tunnel through the junction;
$\hbar$ is the reduced Planck constant;
$e$ is the magnitude of the charge carried by a single electron; and 
$V$ is the voltage difference maintained across the junction.
The popular form of the left equation in Eq.~\eqref{eq:Jeq} is\cite{tinkham}
\begin{equation}\label{eq:Jcurrent}
I_\textrm{s}(t) = I_\textrm{c} \sin(\varphi(t)),
\end{equation}
where $I_\textrm{s}$ is a zero-voltage supercurrent flow through the junction and $I_\textrm{c} (=\! 2eE_\textrm{J}/\hbar)$ is the maximum current that can flow through the junction, i.e., the critical current of the junction.

Here, we point out that a DC Superconducting Quantum Interference Device (DC SQUID), which consists of two Josephson junctions and a superconducting loop [Fig.~\ref{fig:junction}(b)], can be considered as a variable Josephson junction whose effective Josephson energy $E_\textrm{J,eff}$ as a function of the external flux bias $\Phi$ is given by
\begin{equation}\label{eq:dcSQUID}
E_\textrm{J,eff}(\varphi_\textrm{ext}) = 
\sqrt{E_\textrm{J,1}^2 + E_\textrm{J,2}^2 
+ 2E_\textrm{J,1}E_\textrm{J,2} \cos(\varphi_\textrm{ext})},
\end{equation}
where $\varphi_\textrm{ext}$($\equiv\! 2\pi \Phi/\Phi_0$) is the phase offset due to the external flux bias.
This idea is useful for making tunable superconducting devices.

\subsection{Elementary Circuits}
\label{sec:qubitConcept}

\subsubsection{Generic Hamiltonian}

\begin{figure}
\centering
\includegraphics{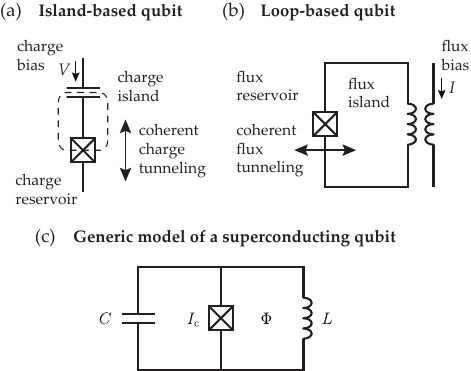}
\caption{(a) and (b) Elementary circuits of a superconducting qubit.
The dashed boundary line indicates the charge island.
In a certain parameter range, their operations as qubits can be understood as the coherent tunneling of charge or flux (see Secs.~\ref{sec:islandQubit} and \ref{sec:loopQubit}).
(c) Parallel circuit composed of an inductor with $L$, a capacitor with $C$, and a Josephson junction with the critical current $I_\textrm{c}$ as a generic model of a superconducting qubit.}
\label{fig:qubitCircuit}
\end{figure}

We can categorize elementary circuits of superconducting qubits into two groups, an island and a loop (Fig.~\ref{fig:qubitCircuit}).
In the early literature, these two kinds of qubits were called a charge qubit and a flux qubit, respectively, on the basis of the spread of the wavefunctions in the number (charge) and phase (flux) bases [typical wavefunctions of a charge qubit are shown in Fig.~\ref{fig:qubitIsland}(d)].\cite{tsai}
However, such a classification is valid only for a certain parameter range;
it does not work well for sophisticated qubits whose wavefunctions often show exotic distributions in both the number and the phase bases.
Therefore, we simply categorize circuits of superconducting qubits based on the geometry. Then, we will show how the qubit properties change as we tune the circuit parameters.
The knowledge acquired in this way can also be used for analyzing more complex qubits.

We introduce a Hamiltonian for a parallel circuit composed of 
a capacitor with the capacitance $C$ including the intrinsic capacitance of a junction, 
an inductor with the inductance $L$, and 
a Josephson junction, as a generic model of a superconducting qubit [Fig.~\ref{fig:qubitCircuit}(c)].
This particular circuit is easily quantized by treating $N$ and $\varphi$ as the operators $\hat{N}$ and $\hat{\varphi}$.
(For an introduction to superconducting circuit quantization, see Refs.~\onlinecite{vool, rasmussen2021}.)
Here, the number operator $\hat{N}$ is conjugate to the phase operator $\hat{\varphi}$: $\hat{N} = -\mathrm{i} \partial / \partial \varphi$.\footnote{They satisfy the relation $\mathrm{e}^{\mathrm{i}\hat{\varphi}} \hat{N} \mathrm{e}^{-\mathrm{i}\hat{\varphi}} = \hat{N} - 1$.
The popular form of this relation is $[\hat{\varphi}, \hat{N}] = \mathrm{i}$.
However, this form is not mathematically rigorous because the phase operator is not Hermitian.
It holds approximately only if $\hat{N}$ and $\hat{\varphi}$ are the \emph{relative} number and phase operators between two superconductors, and the number imbalance of electrons is much less than the number of electrons in each superconductor.\cite{leggett, MW}}
The resulting circuit Hamiltonian $\hat{\mathcal{H}}_\textrm{q}$ is given by\footnote{For a static flux bias, $\varphi_\mathrm{ext}$ can be assigned to either the inductor or the Josephson junction.
This freedom of choice is a consequence of gauge invariance.
However, for a time-dependent flux bias, the real circuit geometry must be considered.\cite{you2019, riwar2022, bryon2023}
Here, we assume that the time-dependent flux bias is primarily associated with the linear inductor.}
\begin{equation}\label{eq:qubitGeneral}
\hat{\mathcal{H}}_\textrm{q} =
4E_\textrm{C} (\hat{N} - N_\textrm{ext})^2
+ \frac{1}{2}E_\textrm{L} (\hat{\varphi} - \varphi_\textrm{ext})^2
- E_\textrm{J} \cos(\hat{\varphi}),
\end{equation}
where
$E_\textrm{C} (\equiv\! e^2 / 2C )$ is the capacitive energy, which is the energy cost to charge a capacitor with a single electron (the factor of 4 comes from the Cooper pairing), and
$E_\textrm{L} [\equiv\! (\Phi_0/2\pi)^2/L]$ is the inductive energy, which is the energy cost to ``charge'' an inductor with a single flux quantum $\Phi_0$.
The $E_\textrm{J}$ term, which represents the energy stored in the junction, was obtained by integrating the electrical work $\int I_\textrm{s} V dt$ with Eqs.~\eqref{eq:Jeq} and \eqref{eq:Jcurrent}.
Lastly, $N_\textrm{ext}$ is the charge offset due to the external voltage bias and $\varphi_\textrm{ext}$($\equiv\! 2\pi \Phi/\Phi_0$) is the phase offset due to the external flux bias $\Phi$.

Equation~\eqref{eq:qubitGeneral} suggests that characteristics of a superconducting qubit can be engineered by three circuit parameters, $E_\textrm{J}$, $E_\textrm{C}$, and $E_\textrm{L}$.
In Secs.~\ref{sec:islandQubit} and \ref{sec:loopQubit}, we explore how these circuit parameters determine the basic properties of a qubit.

\subsubsection{Island-Based Qubit}
\label{sec:islandQubit}

\begin{figure*}
\centering
\includegraphics{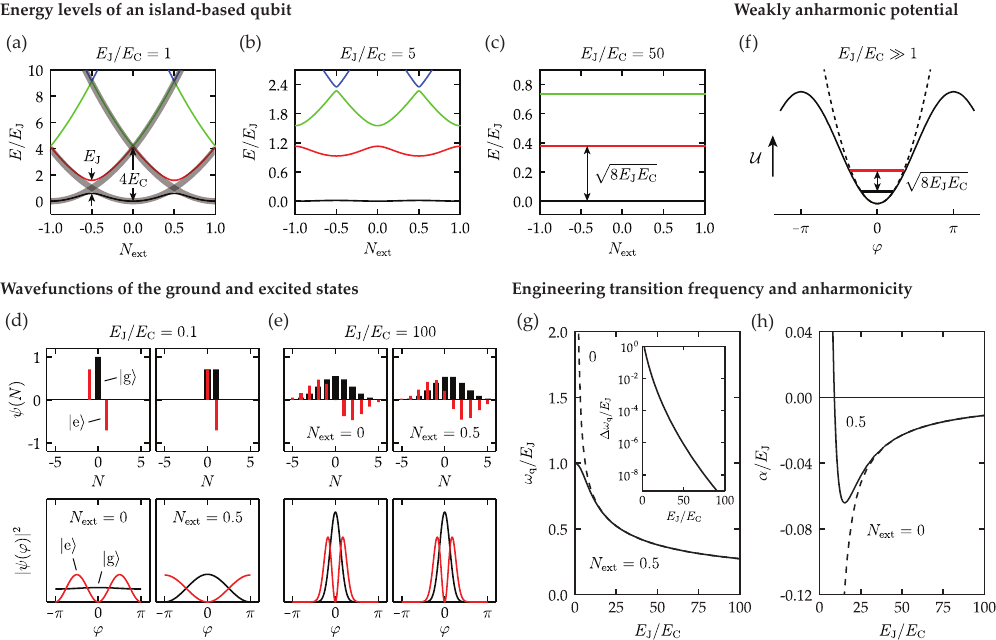}
\caption{(a)--(c) Energy levels of an island-based qubit with three different $E_\textrm{J}/E_\textrm{C}$ ratios.
(d) and (e) Wavefunctions of the ground state $\ket{\textrm{g}}$ and the excited state $\ket{\textrm{e}}$ in the number and phase bases.
(f) If the displacement in the phase basis is reasonably small, which is the case when $E_\textrm{J}/E_\textrm{C} \gg 1$, we can approximate the cosine potential $\mathcal{U}$ (solid line) as a weakly nonlinear harmonic potential (dashed line).
(g) Transition frequency between $\ket{\textrm{g}}$ and $\ket{\textrm{e}}$, $\omega_\textrm{q}$, as a function of $E_\textrm{J}/E_\textrm{C}$ ratio at $N_\textrm{ext} = 0$ (dashed line) and 0.5 (solid line).
The inset shows the $E_\textrm{J}/E_\textrm{C}$ dependence of $\Delta\omega_\textrm{q}$[$\equiv\,$$\omega_\textrm{q}(N_\textrm{ext}\! = \! 0) - \omega_\textrm{q}(0.5)$].
(h) Anharmonicity $\alpha$($\equiv\! \omega_\textrm{e-f}-\omega_\textrm{q}$, where $\omega_\textrm{e-f}$ is the transition frequency between $\ket{\textrm{e}}$ and the higher excitation level $\ket{\textrm{f\,}}$) as a function of $E_\textrm{J}/E_\textrm{C}$ ratio.
}
\label{fig:qubitIsland}
\end{figure*}

The Hamiltonian of an island-based qubit is Eq.~\eqref{eq:qubitGeneral} in the $E_\textrm{L} \rightarrow 0$ and $\varphi_\textrm{ext} \rightarrow 0$ limits:
\begin{equation}\label{eq:qubitIsland}
\hat{\mathcal{H}}_\textrm{q} =
4E_\textrm{C} (\hat{N} - N_\textrm{ext})^2
- E_\textrm{J} \cos(\hat{\varphi}).
\end{equation}
Therefore, the properties of an island-based qubit are mainly determined by the ratio $E_\textrm{J}/E_\textrm{C}$.

In the small $E_\textrm{J}/E_\textrm{C}$ limit, the $E_\textrm{C}$ term is dominant in Eq.~\eqref{eq:qubitGeneral}; as a result, the wavefunctions are localized in the number basis as shown in Fig.~\ref{fig:qubitIsland}(d), suggesting that the number basis will be more convenient to describe the physics in this regime.
In Fig.~\ref{fig:qubitIsland}(a), the gray lines indicate the $E_\textrm{C}$ term associated with $\ket{N=0}$ and $\ket{\pm 1}$.
At $N_\textrm{ext} = 0.5$, $\ket{N=0}$ and $\ket{1}$ are energetically degenerated.
Here, the $E_\textrm{J}$ term hybridizes these two states via coherent charge tunneling [Fig.~\ref{fig:qubitCircuit}(a)], resulting in an anticrossing whose size is approximately $E_\textrm{J}$.
At zero bias, a similar, but significantly smaller, hybridization occurs between $\ket{N=-1}$ and $\ket{1}$.
This results in the first excitation level at $\approx\,$4$E_\textrm{C}$.

As $E_\textrm{J}/E_\textrm{C}$ increases, the contribution from the anticrossing dominates [Fig.~\ref{fig:qubitIsland}(b)];
eventually, in the large $E_\textrm{J}/E_\textrm{C}$ limit, the energy levels become flat, i.e., insensitive to $N_\textrm{ext}$ [Fig.~\ref{fig:qubitIsland}(c)].
In this regime, the wavefunctions are localized in the phase basis as shown in Fig.~\ref{fig:qubitIsland}(e);
hence, it is reasonable to treat the $E_\textrm{J}$ term in Eq.~\eqref{eq:qubitGeneral} as the periodic potential and the $E_\textrm{C}$ term as the kinetic term.
In addition, since the kinetic term is much less than the potential term ($E_\textrm{J}/E_\textrm{C} \gg 1$), the displacement in the phase basis during the evolution of the qubit state is small.
Thus, as depicted in Fig.~\ref{fig:qubitIsland}(f), we can approximate the periodic potential (solid line) as a weakly nonlinear harmonic potential (dashed line).
Then, the qubit frequency can be obtained by approximating Eq.~\eqref{eq:qubitIsland} as a quantum harmonic oscillator ($\cos\hat{\varphi} \approx 1-\hat{\varphi}^2/2$),
\begin{align}\label{eq:qubitHO}
\hat{\mathcal{H}}_\textrm{q}
\approx
4E_\textrm{C} \hat{N}^2
+ \frac{1}{2} E_\textrm{J} \hat{\varphi}^2,
\end{align}
where $N_\textrm{ext}$ is ignored because the energy levels are insensitive to $N_\textrm{ext}$.
By comparing Eq.~\eqref{eq:qubitHO} with a standard spring-block harmonic oscillator with the spring constant $k$ and the mass $m$,
we obtain $\omega_\textrm{q} \approx \sqrt{8E_\textrm{J}E_\textrm{C}}/\hbar$ from $\omega_0 = \sqrt{k/m}$, where $\omega_0$ is the resonance frequency of the spring-block oscillator.
Since we are considering the regime where $\hat{\varphi}$ is localized, $C$ corresponds to $m$, and $L_\textrm{J}^{-1}$ corresponds to $k$, where $L_\textrm{J}$[$\equiv\! (\Phi_0/2\pi)^2/E_\textrm{J}$] is the effective inductance of the junction.
Moreover, in this large $E_\textrm{J}/E_\textrm{C}$ regime, the anharmonicity will decrease with an increase in $E_\textrm{J}/E_\textrm{C}$ because the Hamiltonian [Eq.~\eqref{eq:qubitIsland}] becomes closer to that of a quantum harmonic oscillator [Eq.~\eqref{eq:qubitHO}] in this direction.

More systematic plots regarding the two observations, (i) the flattening of the energy band and (ii) the suppression of the anharmonicity in the large $E_\textrm{J}/E_\textrm{C}$ limit, are given in Fig.~\ref{fig:qubitIsland}(g) and (h), respectively.
Note that the difference between the transition frequencies at $N_\textrm{ext} = 0$ and 0.5, denoted by $\Delta\omega_\textrm{q}$, decreases exponentially as shown in the inset of Fig.~\ref{fig:qubitIsland}(g).
This indicates that the energy levels are completely flat if $E_\textrm{J}/E_\textrm{C} \gtrsim 50$.


The anharmonicity at $N_\mathrm{ext} = 0$ and $0.5$ also collapses into a single curve because of the flattening of the energy band [Fig.~\ref{fig:qubitIsland}(h)].
The crucial observation is that, although $\alpha$ is also approaching zero, it decreases algebraically rather than exponentially.
This suggests that we can use the circuit in the large $E_\mathrm{J}/E_\mathrm{C}$ limit as a charge-insensitive qubit, known as a transmon\cite{koch2007} (see Sec.~\ref{sec:noiseResilient} for the implementation of a transmon).\footnote{Practically the same plot as Fig.~\ref{fig:qubitIsland}(a)--(c) can be found in K. K. Likharev's 1986 book.\cite{likharev}}

\subsubsection{Loop-Based Qubit}
\label{sec:loopQubit}

\begin{figure*}
\centering
\includegraphics{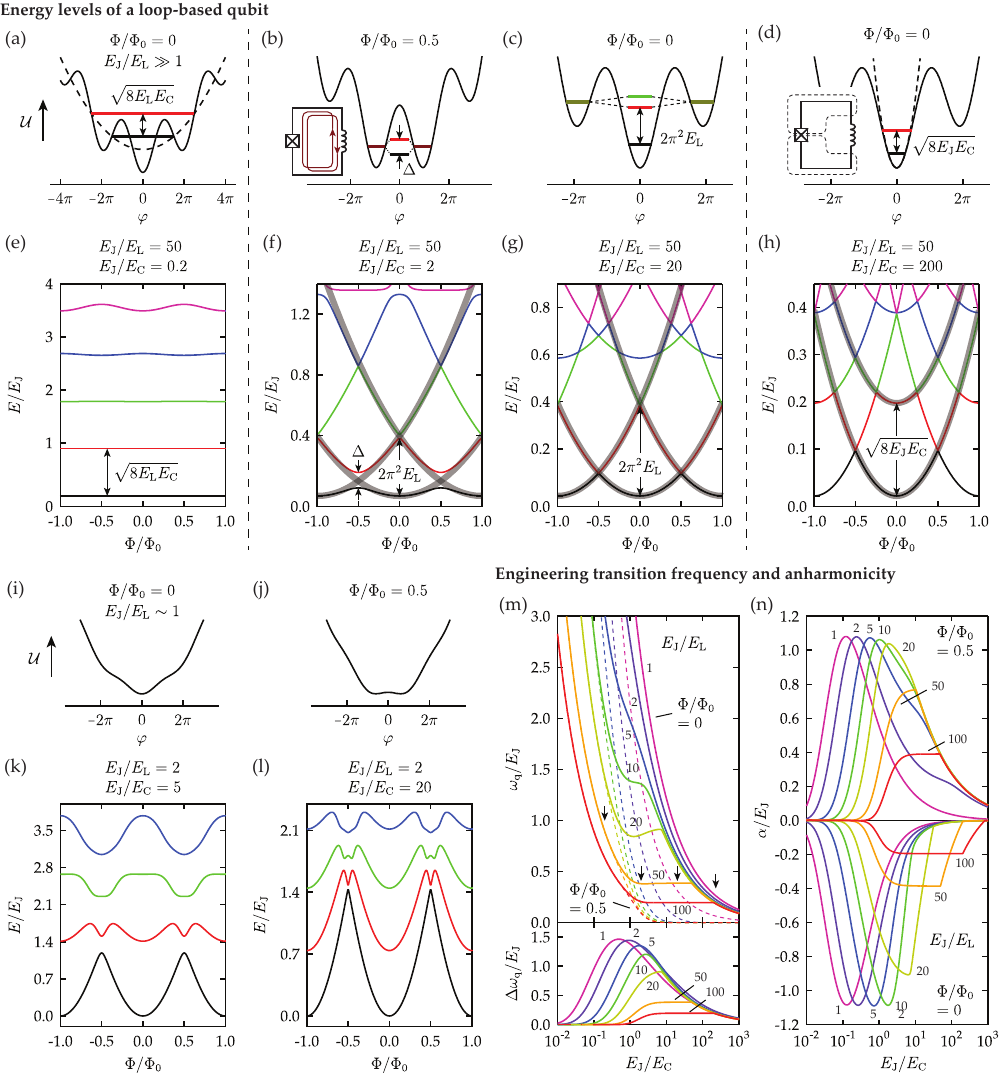}
\caption{(a)--(h) Potential $\mathcal{U}$ [(a)--(d)] and energy level diagrams [(e)--(h)] of a loop-based qubit when $E_\textrm{J}/E_\textrm{L} \gg 1$.
The potential-energy level correspondence is [(a) and (e)], [(b), (c) and (f), (g)], and [(d) and (h)].
$\Phi$ is the external flux bias and $\Phi_0$ is the flux quantum.
The inset in (b) shows the circulating currents in the circuit;
the inset in (d) shows a circuit describing a loop-based qubit as two charge islands (dashed lines) that are connected by an inductor.
Gray lines in (f)--(h) indicate how the ground and first excited levels appear from Eq.~\eqref{eq:qubitGeneral}.
(i)--(l) Similar diagrams for the potential and the energy levels when $E_\textrm{J}/E_\textrm{L} \sim 1$.
(m) Transition frequencies $\omega_\textrm{q}$ at $\Phi/\Phi_0 = 0$ (solid lines in the upper panel) and 0.5 (dashed lines), and their difference $\Delta\omega_\textrm{q}$ (lower panel) as a function of $E_\textrm{J}/E_\textrm{C}$.
Arrows in the upper panel indicate the $E_\textrm{J}/E_\textrm{L}$ and $E_\textrm{J}/E_\textrm{C}$ ratios employed in (e)--(h).
(n) Anharmonicity as a function of $E_\textrm{J}/E_\textrm{C}$.
At $\Phi/\Phi_0 = 0.5$, $\alpha$ is positive, while it is negative at zero bias.
In (m) and (n), the numbers near solid lines indicate the corresponding $E_\textrm{J}/E_\textrm{L}$ ratios.
}
\label{fig:qubitLoop}
\end{figure*}

A loop-based qubit is not as simple as an island-based qubit because we have to consider all terms in Eq.~\eqref{eq:qubitGeneral}.
We start with the effect of $E_\textrm{L}$.
Since $E_\textrm{L}$ is a function of $\varphi$, it is convenient to take the phase basis, and consequently, to treat $E_\textrm{J}$ and $E_\textrm{L}$ terms as the potential.
We first consider the regime in which $E_\textrm{J}/E_\textrm{L} \gg 1$.
In this regime, the periodic shape is prominent in the potential as shown in Fig.~\ref{fig:qubitLoop}(a)--(d).
When $E_\textrm{J}/E_\textrm{C} \ll 1$ [Fig.~\ref{fig:qubitLoop}(e)], the energy level diagram is almost independent of $\Phi$, and $\omega_\textrm{q} \approx \sqrt{8E_\textrm{L}E_\textrm{C}}/\hbar$.
The reason is that the oscillating potential is averaged out owing to the large kinetic energy [Fig.~\ref{fig:qubitLoop}(a)], and consequently, only the harmonic terms are effective in Eq.~\eqref{eq:qubitGeneral}.
In this regime, $\hat{N}$ is localized.
Thus, $\hat{N}$ corresponds to the position in the spring-block oscillator analogy,
$L$ corresponds to the mass,
and $C^{-1}$ corresponds to the spring constant.

For $E_\textrm{J}/E_\textrm{C} > 1$, the physics of a loop-based qubit can be understood in a similar way to that of an island-based qubit.
In Fig.~\ref{fig:qubitLoop}(f) and (g), the gray lines show the $E_\textrm{L}$ term in Eq.~\eqref{eq:qubitGeneral} associated with $\ket{\varphi = 0}$ and $\ket{\pm 2\pi}$, which means that the numbers of trapped fluxes in the loop are 0 and $\pm 1$, respectively.
Note that, at $\Phi/\Phi_0 = 0.5$ ($\varphi_\textrm{ext} = \pi$), the potential has a double-well shape, resulting in energy degeneracy between $\ket{\varphi \approx +\pi}$ and $\ket{\varphi \approx -\pi}$.
These degenerated states correspond to two superposed currents circulating in opposite directions [Fig.~\ref{fig:qubitLoop}(b) and its inset].
Similarly to the degeneracy point in an island-based qubit, the hybridization mediated by the kinetic energy ($E_\textrm{C}$ term) breaks the degeneracy, resulting in an anticrossing.
This process can be understood as coherent flux tunneling between the flux island (loop) and the flux reservoir [Fig.~\ref{fig:qubitCircuit}(b)].
On the basis of this explanation, it is easy to understand that $\omega_\textrm{q}$ at $\Phi/\Phi_0=0.5$ decreases monotonically as a function of $E_\textrm{J}/E_\textrm{C}$ [Fig.~\ref{fig:qubitLoop}(m), dashed lines].

At zero flux bias, the first excitation level is formed through the hybridization of states $\ket{\varphi \approx \pm 2\pi}$, as shown in Fig.~\ref{fig:qubitLoop}(c).
Since this hybridization requires the tunneling of two potential barriers, the energy gap is significantly smaller than that at $\Phi/\Phi_0 = 0.5$.
Hence, $\omega_\textrm{q}$ at zero bias is approximately $2\pi^2 E_\textrm{L}$[$=E_\textrm{L}(\pm 2\pi)^2/2$] and weakly depends on $E_\textrm{J}/E_\textrm{C}$.
This explains why $\omega_\textrm{q}$ at zero bias shows a plateau in Fig.~\ref{fig:qubitLoop}(m).

As $E_\textrm{C}$ decreases further [Fig.~\ref{fig:qubitLoop}(h)], the ground and excited states at zero bias become bound states within a well of the periodic potential. In this case, we can approximate the potential as a weakly nonlinear harmonic potential [Fig.~\ref{fig:qubitLoop}(d)] as we did in Sec.~\ref{sec:islandQubit}.
Hence, $\omega_\textrm{q} \approx \sqrt{8E_\textrm{J}E_\textrm{C}}/\hbar$.
This explanation suggests that the physics of a loop-based qubit in this regime is actually close to that of two island-based qubits connected by an inductor, i.e., inductively shunted junction [the inset of Fig.~\ref{fig:qubitLoop}(d)].
The reason is that $L$ is very large in the regime $E_\textrm{J}/E_\textrm{L} \gg 1$ such that the reactance at $\omega_\textrm{q}$ is significant, whereas the circuit is electrically shorted at the low-frequency limit.

In the regime $E_\textrm{J}/E_\textrm{L} \sim 1$, the harmonic contribution to the potential is substantial; thus, it is difficult to separate the contributions from the periodic and harmonic potentials to the energy levels.
One consequence is that the minimum of the potential becomes almost flat at $\Phi/\Phi_0 = 0.5$ as shown in Fig.~\ref{fig:qubitLoop}(j).
The other consequence is that, in Fig.~\ref{fig:qubitLoop}(k) and (l), the first excitation level near zero bias already has a parabolic shape rather than a cross shape because the first excitation level at zero bias is mostly governed by the physics shown in Fig.~\ref{fig:qubitLoop}(d), rather than the hybridization shown in Fig.~\ref{fig:qubitLoop}(c).
This explains why $\omega_\textrm{q}$ at $\Phi/\Phi_0 = 0$ in this regime decreases monotonically with increasing $E_\textrm{J}/E_\textrm{C}$ without any plateau in Fig.~\ref{fig:qubitLoop}(m).

The experimentally accessible range of $E_\textrm{J}/E_\textrm{C}$ is typically from $\sim\,$0.1 to $\sim\,$100.
In this range, $\omega_\textrm{q}$ of a loop-based qubit with $E_\textrm{J}/E_\textrm{L} \gg 1$ at $\Phi/\Phi_0 = 0.5$ is often too low to satisfy condition~\ref{list:transFreq} in Sec.~\ref{sec:qubitCriteria},
while $\omega_\textrm{q}$ of a qubit with $E_\textrm{J}/E_\textrm{L} \sim 1$ at zero bias is too high.
Regarding the anharmonicity, a loop-based qubit with $E_\textrm{J}/E_\textrm{L} \gg 1$ is more advantageous than that with $E_\textrm{J}/E_\textrm{L} \sim 1$ as shown in Fig.~\ref{fig:qubitLoop}(n).
If $E_\textrm{L}$ increases even further such that $E_\textrm{J}/E_\textrm{L} \ll 1$, the potential becomes almost harmonic, and as a result, the circuit does not show enough anharmonicity to be a qubit.


\section{Effect of Noise}
\label{sec:relaxation}

\subsection{Relaxation}

\subsubsection{Concept}
\label{sec:relaxConcept}

\begin{figure*}
\centering
\includegraphics{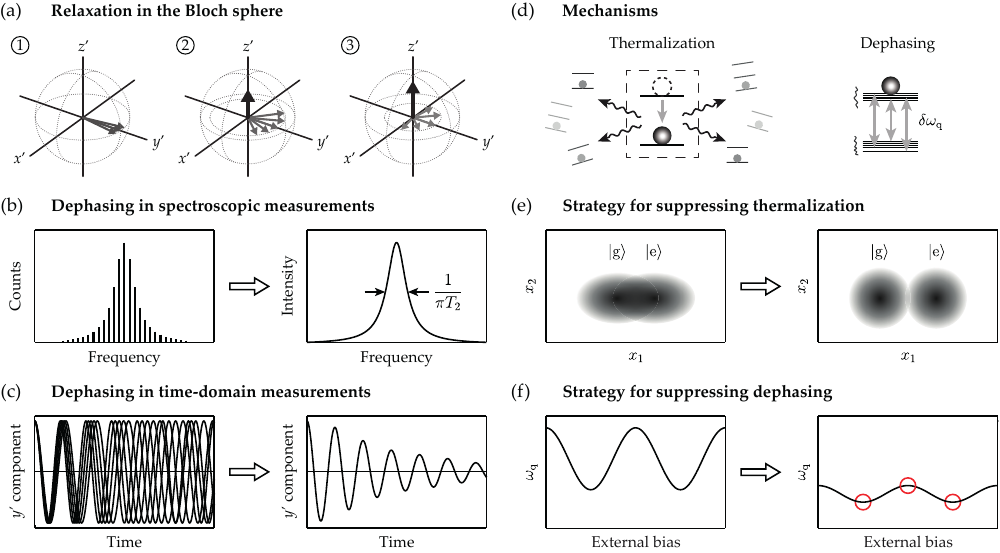}
\caption{(a) Relaxation of qubit states in a set of identical measurements represented in the Bloch sphere.
Each arrow represents the qubit state for each measurement.
Primes($'$) in the axes indicate the rotating frame with the average qubit transition frequency.
The numbers in circles indicate the time instant during a single measurement.
The thick arrow growing along the $z'$-axis represents the longitudinal relaxation, while spreading and shortening arrows in the $x'y'$-plane represent the transverse relaxation.
(b) In spectroscopic measurements, the qubit transition frequency varies with time because of noises from the surrounding environment.
In general, a large deviation from the center is unlikely to occur as shown in the histogram (left figure).
Such a fluctuation broadens the qubit spectrum (right figure).
(The measurement procedure is described in Sec.~\ref{sec:twoTone}.)
This phenomenon, called dephasing, corresponds to the spreading of arrows in (a).
(c) The temporal fluctuation in qubit transition frequency induces the loss of phase coherence between the signals obtained in each measurement (left figure).
The averaged signal is a decaying signal with the time constant $T_2$ (right figure). (The measurement procedure is described in Sec.~\ref{sec:Ramsey}.)
Note that the Fourier transform connects the decay in time-domain measurement and the spread in spectroscopic measurement; hence, the width of the qubit spectrum is about $1/\pi T_2$ in Hz as shown in (b).
In (b) and (c), $T_1$ is assumed to be much longer than $T_2$.
(d) Relaxation mechanisms. Thermalization is due to incoherent energy exchange between the qubit and the environment. Dephasing is due to fluctuations in the transition frequency of the qubit, $\delta\omega_\textrm{q}$.
(e) Thermalization can be suppressed by reducing the overlap between the ground-state and excited-state wavefunctions in the circuit variable space, such as $\hat{N}$ and $\hat{\varphi}$ in Eq.~\eqref{eq:qubitGeneral}. In this figure, the circuit variables are denoted by $x_1$ and $x_2$ for generality. 
(f) Dephasing can be suppressed by designing the qubit to be less sensitive to the external bias and operating the qubit at its sweet spot. The figure shows the schematic external bias dependence of the qubit transition frequency ($\omega_\textrm{q}$). Red circles indicate sweet spots.}
\label{fig:relaxation}
\end{figure*}

The states of a qubit cannot last forever;
after some time, they relax back to the ground state because of the interaction between the qubit and the surrounding environment.
This is the reason for having condition~\ref{list:relaxTime} in Sec.~\ref{sec:qubitCriteria}.
We can define two experimentally measurable time scales that characterize the relaxation of a quantum state [Fig.~\ref{fig:relaxation}(a)]:
one is the longitudinal relaxation time ($T_1$) and the other is the transverse relaxation time, referred to as the decoherence time ($T_2$).
As the name implies, $T_1$ is the time constant for recovering the longitudinal component of the Bloch vector to its thermal equilibrium value.
Thus, the physical process responsible for $T_1$ is thermalization of the qubit.\footnote{A considerable number of papers use the term ``depolarizing'' to describe the physical process responsible for $T_1$. We do not use this term to avoid confusion with the depolarizing channel, which contracts the Bloch vector \emph{independently} of its direction.\cite{NC, BCRS}}
$T_2$ is the time constant for the decay of the transverse component of the Bloch vector to zero.

Note that there are two contributions to $T_2$ in Fig.~\ref{fig:relaxation}(a):
one is the shortening of arrows and the other is the spreading of arrows.
The shortening of arrows is due to the growth of the longitudinal component, whereas the spreading is due to the loss of the phase coherence of the qubit, called dephasing.
As shown in Fig.~\ref{fig:relaxation}(b) and (c), dephasing is caused by the temporal fluctuation in qubit transition frequency.
Hence, both thermalization and dephasing contribute to $T_2$, while $T_1$ is entirely determined by thermalization.
This explanation can be written as\cite{ithier2005}
\begin{equation}\label{eq:relaxRate}
\frac{1}{T_1} = \Gamma_\parallel, \quad \frac{1}{T_2} = \frac{\Gamma_\parallel}{2} + \Gamma_\varphi,
\end{equation}
where $\Gamma_\parallel$ is the decay rate of the excited state population and $\Gamma_\varphi$ is the dephasing rate.
The reason for the factor of 2 in $\Gamma_\parallel /2$ will be given in Sec.~\ref{sec:EOM}.
The measurement procedures for $T_1$ and $T_2$ are described in Sec.~\ref{sec:timeDomain}.

The interaction with the surrounding environment is usually treated as various noise processes.
The effects of noises are explained further in Secs.~\ref{sec:thermalization} and \ref{sec:dephasing}.

\subsubsection{Thermalization}
\label{sec:thermalization}

Thermalization of a qubit occurs via incoherent energy exchange between the qubit and the environment [Fig.~\ref{fig:relaxation}(d)].
The effect of such an interaction is usually modeled as fluctuations in the qubit Hamiltonian.
In the qubit Hamiltonian, there are physical quantities that mediate the interaction between the qubit and the environment, such as external charge and flux biases.
If we denote such a physical quantity as $\lambda$, the susceptibility of the qubit Hamiltonian to the fluctuation in $\lambda$, denoted by $\hat{X}_\lambda$, is given by the derivative of the qubit Hamiltonian $\hat{\mathcal{H}}_\mathrm{q}$ with respect to $\lambda$.
For example, the noise caused by fluctuating charges nearby, called charge noise, is coupled to the qubit through the external charge bias;
hence, $\lambda = N_\mathrm{ext}$.
For the noise caused by fluctuating spins, called flux noise, $\lambda = \varphi_\mathrm{ext}$.
Similarly, the effect of the fluctuation in critical current can be estimated by $\lambda = E_\mathrm{J}$ (or $I_\mathrm{c}$).
Then, for $\hat{\mathcal{H}}_\mathrm{q}$ in Eq.~\eqref{eq:qubitGeneral}, $\hat{X}_\lambda$ for these quantities are given by
\begin{align}
&\hat{X}_N = \frac{\partial \hat{\mathcal{H}}_\mathrm{q}}{\partial N_\mathrm{ext}} 
= -8E_\mathrm{C}\hat{N}, \label{eq:effDipole1}\\
&\hat{X}_\Phi = \frac{\partial \hat{\mathcal{H}}_\mathrm{q}}{\partial \varphi_\mathrm{ext}} 
= -E_\mathrm{L}\hat{\varphi}, \label{eq:effDipole2}\\
&\hat{X}_{E_\mathrm{J}} = \frac{\partial \hat{\mathcal{H}}_\mathrm{q}}{\partial E_\mathrm{J}} \bar{E}_\mathrm{J}
= -\bar{E}_\mathrm{J} \cos(\hat{\varphi}). \label{eq:effDipole3}
\end{align}
Here, we insert the mean value of the junction energy $\bar{E}_\mathrm{J}$ in Eq.~\eqref{eq:effDipole3} to set the dimension of $\hat{X}_{E_\mathrm{J}}$ identical to the other two equations for fair comparison.

To thermalize a qubit, the environment must be able to dissipate an electromagnetic energy near $\omega_\textrm{q}$.
Thus, the thermalization process is governed by the noise whose frequency is near $\omega_\textrm{q}$.
Using Fermi's golden rule, $\Gamma_\parallel$ can be written as\cite{ithier2005}
\begin{equation}\label{eq:T1}
\Gamma_\parallel
= \frac{1}{\hbar^2} \sum_\lambda
\big|\!\mel{1}{\hat{X}_\lambda}{0}\!\big|^2 \,
\big| S_\lambda(\omega_\textrm{q})+S_\lambda(-\omega_\textrm{q}) \big|,
\end{equation}
where $S_\lambda$ is the noise power spectral density associated with the fluctuation in $\lambda$.
On the basis of Eq.~\eqref{eq:T1}, the thermalization process is often interpreted as the dipole transition associated with the effective dipole moment $\hat{X}_\lambda$.

In Eq.~\eqref{eq:T1}, $S_\lambda(\omega_\textrm{q})$ and $S_\lambda(-\omega_\textrm{q})$ represent emission (from $\ket{1}$ to $\ket{0}$) and absorption (from $\ket{0}$ to $\ket{1}$), respectively.
When the qubit frequency is much greater than the temperature of the environment $T$, i.e., $\hbar\omega_\textrm{q} \gg k_\textrm{B}T$, we can safely ignore the contribution from the absorption process.
Then, we have
\begin{equation}\label{eq:noisePSD}
S_\lambda(\omega_\textrm{q})+S_\lambda(-\omega_\textrm{q})
\approx 
A_\lambda^2 \left(\frac{2\pi \times 1\textrm{ [Hz]}}{\omega_\textrm{q}}\right)^{\mu},
\end{equation}
where $A_\lambda$ is the noise magnitude, and
$\mu \approx 1$ for $1/f$ noise and $\mu \approx -1$ for Ohmic noise.
It has been reported that Ohmic noise is chiefly responsible for $\Gamma_\parallel$ (Refs.~\onlinecite{astafiev2004, yan2015}) and $1/f$ noise is responsible for $\Gamma_\varphi$ (Refs.~\onlinecite{paladino2014, yoshihara2006, bylander2011}).

Equations~\eqref{eq:effDipole1}--\eqref{eq:T1} suggest that thermalization due to various noise processes acting on a qubit is determined by the circuit parameters and the off-diagonal matrix elements of $\hat{N}$ and $\hat{\varphi}$, i.e., the overlap between wavefunctions in the circuit variable space.

Currently, there are three approaches to suppressing thermalization:
\begin{enumerate}
\itemsep-0.1em

\item \label{list:material} \emph{Clean environment}:
This approach eliminates the noise source by removing any unnecessary quantum systems, such as defects, which could possibly couple to the qubit.
Naturally, this approach requires much knowledge and engineering regarding materials, such as host superconductors, substrates, and oxide layers.\cite{OW, barends2013, dunsworth2017}
For example, it is known that a qubit on a silicon substrate usually shows a shorter $T_1$ than that on a sapphire substrate, partly because of a lossy amorphous silicon oxide layer.\cite{chu2016}
For a comprehensive review for this approach, see Ref.~\onlinecite{TLSreview}.

\item \label{list:partRatio} \emph{Reducing participation ratio}:
This approach reduces losses in dielectric media, such as oxides or organics at the surface of a qubit, by minimizing the participation ratio that is defined as the fraction of the electric field energy stored within the volume of each dielectric medium.\cite{wang2015, minev2020}
Since an electric field in a planar device is highly concentrated near the edges, a qubit made of large superconducting pads with a simple design shows good performance in general.\cite{chu2016, wang2015}

\item \label{list:overlap} \emph{Reducing wavefunction overlap}:
We can engineer the potential by choosing the geometry and parameters of the circuit to minimize the effective dipole moment, i.e., the wavefunction overlap in the circuit variable space, as shown in Fig.~\ref{fig:relaxation}(e).
This is the strategy that the so-called protected qubit takes.\cite{PQreview, brooks2013, bell2014, dempster2014, groszkowski2018, gyenis2019}
However, reducing the effective dipole moment inevitably makes the qubit difficult to control [compare Eqs.~\eqref{eq:T1} and \eqref{eq:gDef}].

\end{enumerate}

\subsubsection{Dephasing}
\label{sec:dephasing}

Dephasing is due to the temporal fluctuation in the transition frequency $\delta \omega_\textrm{q}$ [Fig.~\ref{fig:relaxation}(d)], which can be expressed as\cite{ithier2005}
\begin{equation}\label{eq:T2}
\delta \omega_\textrm{q} \propto \big| \! \mel{1}{\hat{X}_\lambda}{1} - \mel{0}{\hat{X}_\lambda}{0} \! \big|.
\end{equation}
Equation~\eqref{eq:T2} suggests that the dephasing is determined by the diagonal matrix elements of $\hat{N}$ and $\hat{\varphi}$ [Eqs.~\eqref{eq:effDipole1}--\eqref{eq:effDipole3}].
In the Bloch sphere [Fig.~\ref{fig:relaxation}(a)], if $\omega_\textrm{q}$ is the same as the frequency of the rotating frame, the transverse component will lie along the $y'$-axis.
However, owing to the fluctuation in $\omega_\textrm{q}$, the transverse component rotates around the $z'$-axis with amount of rotation differing from measurement to another measurement, resulting in the spreading of arrows [Fig.~\ref{fig:relaxation}(a)].
As a result, the qubit loses the phase coherence and the averaged transverse component in the Bloch vector decays in time as shown in Fig.~\ref{fig:relaxation}(c).
To yield such a decay, the time scale of the fluctuation must be much slower than the qubit transition (thus, adiabatic) and should be a similar order of magnitude to the measurement time scale.
Therefore, the dephasing rate $\Gamma_\varphi$ is mainly determined by low-frequency noise.

To estimate $\Gamma_\varphi$, we have to perform an integration with respect to the frequency of the noise.
For this, we set the low-frequency $\omega_\textrm{low}$ and high-frequency $\omega_\textrm{high}$ cutoffs.
If our time scale of interest $\tau$ satisfies $\omega_\textrm{low}\tau \ll 1$ and $\omega_\textrm{high}\tau \gg 1$, $\Gamma_\varphi$ is roughly given by\cite{ithier2005, koch2007}
\begin{align}\label{eq:dephasingRate}
\Gamma_\varphi \sim 
A_\lambda \abs{\frac{\partial \omega_\textrm{q}}{\partial \lambda}},
\end{align}
where $A_\lambda$ is the noise magnitude defined in Eq.~\eqref{eq:noisePSD}.

On the basis of what we have learned thus far, we explain two approaches to suppressing dephasing.
\begin{enumerate}
\itemsep-0.1em

\item \label{list:strategyGeo} \emph{Geometry}:
We can select a circuit geometry that is insensitive to a certain type of noise.
A fixed-frequency island-based qubit is insensitive to flux noise simply because there is no loop that can contain a flux [Fig.~\ref{fig:qubitCircuit}(a)].
For a loop-based qubit, the sensitivity to flux noise depends on the circuit parameters.
If the qubit is in a circuit parameter range in which the qubit states are the circulating current states shown in Fig.~\ref{fig:qubitCircuit}(b) and the inset of Fig.~\ref{fig:qubitLoop}(b), the qubit is insensitive to charge noise.
The reason is that a continuously flowing DC supercurrent does not allow any charge offset within the current path, i.e., the circuit is electrically shorted in the low-frequency limit.
However, such a state is sensitive to flux noise.
If the circuit parameters are chosen such that the qubit states are similar to the island-like qubit shown in the inset of Fig.~\ref{fig:qubitLoop}(d), then the qubit states are sensitive to charge noise but less sensitive to flux noise.

\item \label{list:strategyBias} \emph{Bias dependence}:
Since $\Gamma_\varphi$ is proportional to $\partial_\lambda \omega_\textrm{q}$ [Eq.~\eqref{eq:dephasingRate}], we can choose the circuit parameters that give minimal bias dependence as shown in Fig.~\ref{fig:relaxation}(f).
In this regard, operating a qubit at a bias at which $\partial_\lambda \omega_\textrm{q} = 0$, called a sweet spot [red circles in Fig.~\ref{fig:relaxation}(f)], is necessary because the qubit is first-order insensitive to noise at this particular bias [$\Gamma_\varphi = 0$ in Eq.~\eqref{eq:T2}].

\end{enumerate}

Note that the energy of a qubit is conserved during dephasing, in contrast to thermalization.
This allows us to recover the phase coherence by applying pulses that can revert the direction of the time evolution.
Such a technique is called refocusing and will be discussed in Sec.~\ref{sec:refocusing}.

In Sec.~\ref{sec:noiseResilient}, we briefly explore several noise-resilient qubit designs and discuss how to improve the robustness of the qubit by tuning the circuit parameters.

\subsection{Noise-Resilient Designs}
\label{sec:noiseResilient}

\begin{figure*}
\centering
\includegraphics{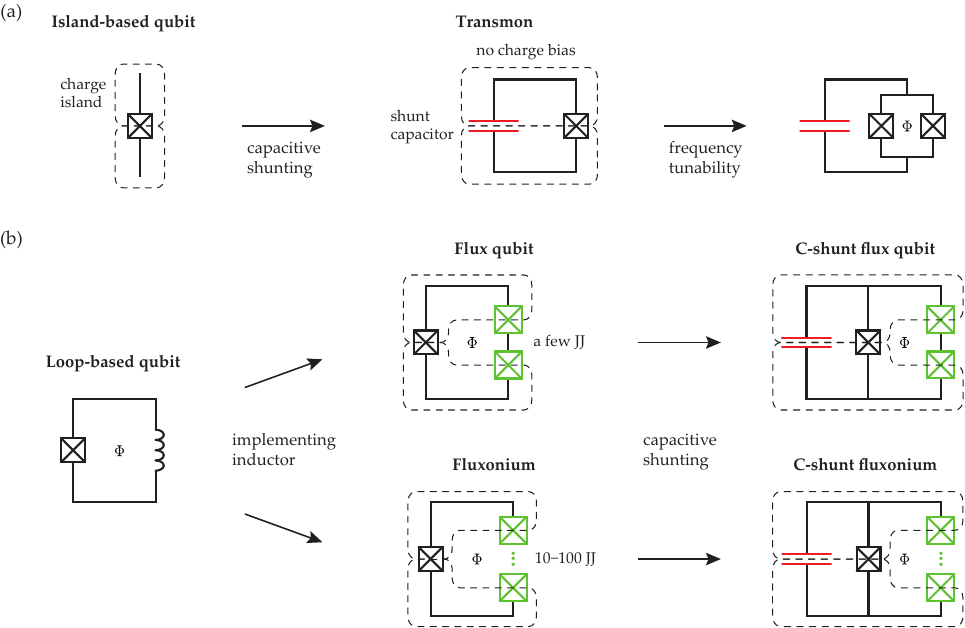}
\caption{Conceptual evolution of noise-resilient qubit designs from an island-based qubit (a) and a loop-based qubit (b). Dashed boundaries indicate islands. JJ stands for Josephson junction.}
\label{fig:qubitNoiseless}
\end{figure*}

\subsubsection{Island-Based Qubit}
\label{sec:transmon}

The most successful noise-resilient design of an island-based qubit is a transmon.
As mentioned in Sec.~\ref{sec:relaxation}, the dephasing rate of an island-based qubit in Fig.~\ref{fig:qubitNoiseless}(a) is insensitive to flux noise because of the absence of a loop.
To suppress the effect of charge noise, the transmon design pushes strategy~\ref{list:strategyBias} in Sec.~\ref{sec:dephasing} to the limit:
eliminating the $N_\textrm{ext}$ dependence by choosing $E_\textrm{J}/E_\textrm{C} = 50$--100 (Fig.~\ref{fig:qubitIsland}).

This limit can be achieved by adopting a shunt capacitor [red capacitor in Fig.~\ref{fig:qubitNoiseless}(a)].
The shunt capacitor takes the majority of the effect of the charge noise, thus minimizes this effect on the junction.
The physics of this idea is the same as adding a heavy mass to reduce the sensitivity to mechanical noise.

As mentioned in Sec.~\ref{sec:islandQubit}, the trade-off is the reduced anharmonicity:
in the large $E_\textrm{J}/E_\textrm{C}$ limit, the qubit wavefunctions are localized in the phase space; hence, a transmon is a weakly nonlinear harmonic oscillator.
From this reasoning, we can easily imagine that, if we treat the qubit wavefunction as a rolling glass bead in a potential well, the bead sees more anharmonicity as the kinetic energy ($E_\textrm{C}$) increases.
Indeed, the anharmonicity of a transmon is roughly given by $-E_\textrm{C}$ (see Sec.~\ref{sec:couplingConcept}).
$E_\textrm{C}$ is usually chosen 100--500 MHz to satisfy condition~\ref{list:anhar} in Sec.~\ref{sec:qubitCriteria}.
Then $E_\textrm{J}$ must be 10--30 GHz to satisfy condition~\ref{list:transFreq} in Sec.~\ref{sec:qubitCriteria}.
The resulting circuit parameters are summarized in Table~\ref{tab:circuitPara}.

A DC SQUID is employed to tune the qubit frequency as explained in Sec.~\ref{sec:JJ} [rightmost figure in Fig.~\ref{fig:qubitNoiseless}(a)].
However, in this case, the transmon is exposed to the flux noise.
Therefore, we need to design a DC SQUID with minimal flux dependence based on Eq.~\eqref{eq:dcSQUID}.\cite{hutchings2017}
In addition, we have to operate the tunable transmon at the flux bias sweet spot.

\subsubsection{Loop-Based Qubit}
\label{sec:fluxQubit}

The main difficulty in implementing a loop-based qubit is designing an inductor with sufficiently large inductance because the inductance of a superconducting loop made of aluminum or niobium is usually very small such that $E_\textrm{L} > E_\textrm{J}$.
Consequently, the resulting anharmonicity is too small to satisfy condition~\ref{list:anhar} in Sec.~\ref{sec:qubitCriteria} as explained in Sec.~\ref{sec:loopQubit}.

A popular strategy is to add multiple Josephson junctions, where the Josephson energy for each junction is $\beta E_\textrm{J}$, as an effective inductor.
Here, we still want to keep the current flowing in the loop dominated by the main junction [black junction in Fig.~\ref{fig:qubitNoiseless}(b)].\footnote{In the literature, the main junction is often called the ``$\alpha$ junction'', where $\alpha = \beta^{-1}$, for historical reasons.}
Since the flux tunneling rate through a Josephson junction is roughly proportional to $\exp\!\big(\!\!-\!\!\sqrt{E_\textrm{J}/E_\textrm{C}}\big)$ (Ref.~\onlinecite{rastelli2013}), $\beta$ must be larger than 1.

The resulting potential term $\hat{\mathcal{U}}$ is given by
\begin{equation}\label{eq:Cshunt}
\hat{\mathcal{U}} \approx 
- E_\textrm{J} \cos(\hat{\varphi})
- \beta E_\textrm{J} \sum_{i=1}^{N_\textrm{J}} \cos(\hat{\varphi}_i),
\end{equation}
where $N_\textrm{J}$ is the number of additional Josephson junctions
and $\hat{\varphi}_i$ is the phase difference across additional junction $i$.
Note that the loop inductance does not appear in Eq.~\eqref{eq:Cshunt}.
The reason is that the phase variable associated with the loop inductance is nearly zero because of the large $E_\textrm{L}$; 
thus, its contribution to the qubit dynamics is small compared with that from the additional junctions.
On the other hand, in the phase dimensions associated with the additional junctions, the potential has periodic modulations that can support coherent flux tunneling.\cite{orlando1999, robertson2006}

First, we consider the case when $N_\textrm{J}$ is 2 or 3 and $\beta \approx 2$ [upper figures in Fig.~\ref{fig:qubitNoiseless}(b)].
The circuit with these parameters, which roughly corresponds to a loop-based qubit with $E_\textrm{J}/E_\textrm{L} \sim 1$, is called a flux qubit.
Although the resulting energy level structure from Eq.~\eqref{eq:Cshunt} is not the same as that from Eq.~\eqref{eq:qubitGeneral}, the overall dependence of the energy levels on the circuit parameters is qualitatively similar to that in Fig.~\ref{fig:qubitLoop}(k)--(n).

For a noise-resilient qubit, we need to select $E_\textrm{J}$ to minimize the $\varphi_\textrm{ext}$ dependence as mentioned in Sec.~\ref{sec:relaxation}.
At the same time, we also need to satisfy condition~\ref{list:transFreq} in Sec.~\ref{sec:qubitCriteria}.
It was found that $E_\textrm{J} \sim 10$--100 GHz and $\beta \approx 2$ balance these two.\cite{yan2015}
However, there is a trade-off:
the qubit becomes sensitive to charge noise because the circulating currents are close to zero even at $\Phi/\Phi_0 = 0.5$.
To circumvent this, a shunt capacitance is added to the main junction as we did for the transmon;
thus, we have $E_\textrm{C} = 0.1\textrm{--}1$ GHz.
The final circuit shown in Fig.~\ref{fig:qubitNoiseless}(b) is called a capacitively shunted (C-shunt) flux qubit [upper rightmost figure in Fig.~\ref{fig:qubitNoiseless}(b)].\cite{you2007}

One might ask whether a flux-tunable transmon and a C-shunt flux qubit are the same type of qubit.
Considering the circuit structure, the two are certainly very similar, but the operating flux bias at which criterion~\ref{list:transFreq} in Sec.~\ref{sec:qubitCriteria} is satisfied differs:
a transmon is usually operated at zero flux bias, whereas a C-shunt flux qubit is operated at $\Phi/\Phi_0 = 0.5$.
Consequently, the details of their potential landscapes differ at their respective operating flux biases.
Thus, they are typically considered to be distinct qubits.

With a sufficiently large $N_\textrm{J}$ ($\sim$10--100), Eq.~\eqref{eq:Cshunt} can be treated as a linear inductor with $E_\textrm{L} \approx \beta E_\textrm{J}/N_\textrm{J}$ [lower figures in Fig.~\ref{fig:qubitNoiseless}(b)], if the self-resonance frequency of the junction array is sufficiently higher than that of each junction, $\sqrt{8E_\textrm{J} E_\textrm{C}}/h$ (Ref.~\onlinecite{manucharyan2009}).
This condition can be satisfied by limiting $N_\textrm{J}$ to $N_\textrm{J} \lesssim \sqrt{C_\textrm{J}/C_\textrm{g}}$, where $C_\textrm{J}$ is the capacitance across each junction and $C_\textrm{g}$ is the capacitance between the junction array and the ground.
By tuning $\beta$ and $N_\textrm{J}$, we can satisfy $E_\textrm{L} \ll E_\textrm{J}$.
A superconducting qubit in this regime is called a fluxonium or an RF SQUID qubit.
In this case, it is easy for $\omega_\textrm{q}$ at zero bias to satisfy condition~\ref{list:transFreq} in Sec.~\ref{sec:qubitCriteria};
at $\Phi/\Phi_0 = 0.5$, $\omega_\textrm{q}$ might be too low.
This drawback can be resolved by employing active qubit initialization protocols (see Sec.~\ref{sec:reset}).
According to Fig.~\ref{fig:qubitLoop}(n), the anisotropy is significantly larger than that of a flux qubit.
Capacitive shunting has also been applied to a fluxonium [lower rightmost figure in Fig.~\ref{fig:qubitNoiseless}(b)], resulting in improved $T_2$ (Ref.~\onlinecite{nguyen2019}).

Lastly, we would like to point out that a Josephson junction array as a linear inductor itself is an interesting system.
The reason is that it is difficult, although not impossible,\cite{peruzzo2020} to make a geometric inductor whose reactance exceeds the superconducting resistance quantum $R_\textrm{Q} = h/(2e)^{2} \approx 6.5$ k$\Omega$ because of stray capacitance and radiation to vacuum, whose impedance is about 377 $\Omega$.
Such a linear inductor whose impedance is similar to or larger than $R_\textrm{Q}$ is often called a superinductor.
Thus, implementations of superinductors have usually been based on kinetic inductance,\cite{manucharyan2009, astafiev2012, grunhaupt2019}
i.e., an inductive contribution to the impedance that arises from kinetic energy of the charge carrier, instead of geometric inductance.
(For further discussion about kinetic inductance, see Sec.~\ref{sec:paraamp}.)
Very recently, a qubit made of Josephson junction arrays with extremely high inductance ($E_\textrm{L} < 100$ MHz) succeeded in implementing the regime shown in Fig.~\ref{fig:qubitLoop}(a) and (e).\cite{blochnium}

\begin{table}
\caption{Circuit parameters of several noise-resilient qubit designs.
For the flux qubit and fluxonium, which have multiple junctions, $E_\textrm{J}$ is for the smallest junction [the black junctions in Fig.~\ref{fig:qubitNoiseless}(b)].
In addition, the flux qubit and fluxonium considered in this table are capacitively shunted ones.
}
\label{tab:circuitPara}\centering
\begin{ruledtabular}
\begin{tabular}{l c c c c c c}
\noalign{\smallskip}
Type	&	$E_\textrm{J}$ [GHz]	& $E_\textrm{J}/E_\textrm{C}$	& $E_\textrm{J}/E_\textrm{L}$	&	$\beta$	&	$N_\textrm{J}$	&	Ref. \\
\noalign{\smallskip} \hline \noalign{\smallskip}
Transmon		&	10--30	&	50--100	&				&				&					& \onlinecite{paik2011},\onlinecite{rigetti2012} \\
Flux qubit	&	10--100	&	10--100	&	$\sim$1	&	$\approx$2	&	2--3		& \onlinecite{yan2015} \\
Fluxonium	&	1--10		&	1--10		&	3--10	&	2--5	&	10--100	& \onlinecite{nguyen2019} \\
\end{tabular}
\end{ruledtabular}
\end{table}


\section{Coupling}
\label{sec:coupling}

Thus far, we have explained how to make a qubit out of superconductors.
To perform actual computation, a qubit must be coupled to other systems so that the qubit state can be controlled or read.
The most commonly used physics for these operations is the cavity quantum electrodynamics.\cite{haroche}
It provides an integrated control/readout scheme via the interaction between an atom and a cavity.
The same physics can be applied to a superconducting circuit as the interaction between a qubit and a resonator.
This circuit version of the cavity quantum electrodynamics is called the circuit Quantum ElectroDynamics (cQED).\cite{yale, LHlecture, blais2004}
In addition, for multiqubit gate operation, qubit-qubit coupling is required.
In this section, we discuss how to couple a qubit to other systems.

\subsection{Two Coupled Classical Oscillators}
\label{sec:classicalCoupling}

\begin{figure*}
\centering
\includegraphics{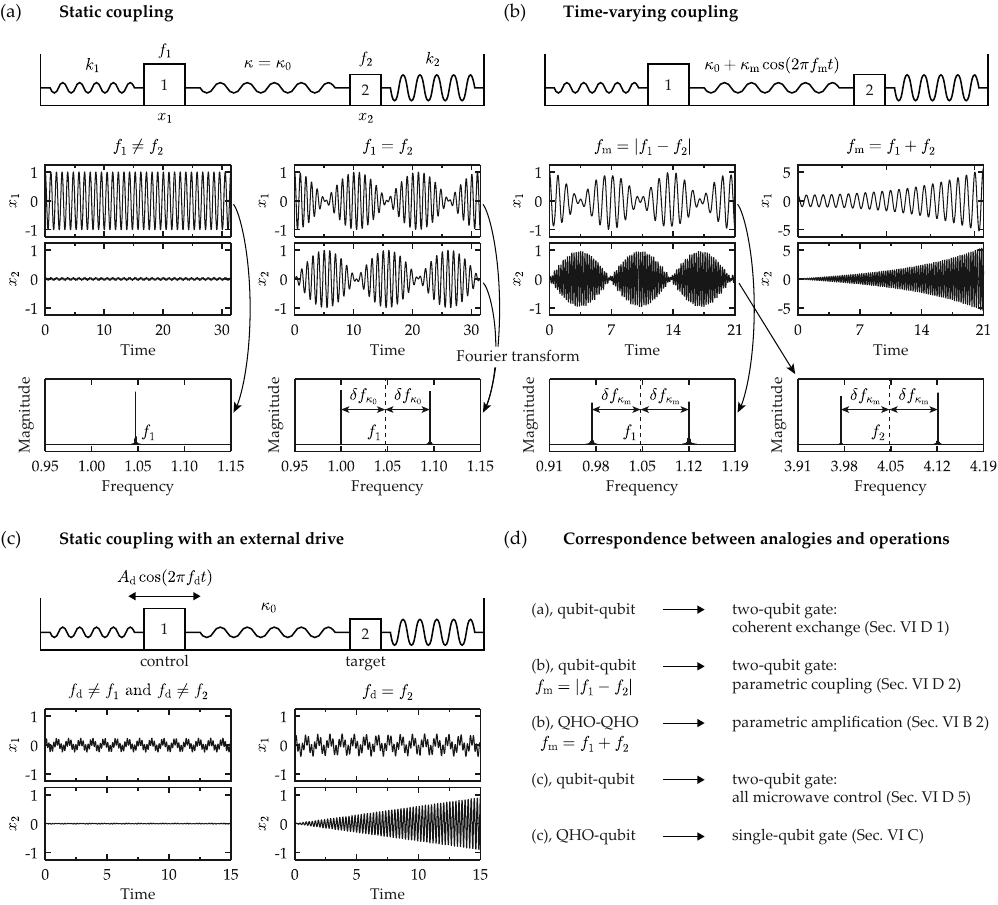}
\caption{Two coupled classical harmonic oscillators without damping.
Each oscillator is composed of a spring and a block.
$k_i$ is the spring constant of oscillator $i$;
$f_i$ is the resonance frequency; and
$x_i$ is the position.
Two oscillators are coupled via a coupling spring whose spring constant is $\kappa$.
The graphs show the solutions of Eq.~\eqref{eq:coupledHO_EOM} for various conditions.
(a) Evolution of the system when the coupling is static: $\kappa = \kappa_0$.
The two oscillators do not interact with each other if $f_1 \neq f_2$;
however, if $f_1 = f_2$, the oscillators exchange their energy at a rate of $2\delta f_{\kappa_0}$, whose quantity is determined by $\kappa_0$.
As shown by the Fourier-transformed solution, the energy exchange can be interpreted as a splitting of the resonance frequency with $2\delta f_{\kappa_0}$.
(b) Evolution of the system with a time-varying coupling constant $\kappa_0 + \kappa_\textrm{m} \cos(2\pi f_\textrm{m}t)$, where $f_\textrm{m}$ is the modulation frequency.
When $f_\textrm{m} = |f_1-f_2|$, the two oscillators exchange their energy even if $f_1 \neq f_2$.
The Fourier transform shows that the resonance frequency of each oscillator is split by $2\delta f_{\kappa_\textrm{m}}$ whose quantity is determined by $\kappa_\textrm{m}$.
(c) Evolution of the system with the static coupling and an external drive acting on oscillator 1.
Here, oscillator 1 is the control oscillator and oscillator 2 is the target oscillator.
The amplitude and frequency of the drive are denoted by $A_\textrm{d}$ and $f_\textrm{d}$, respectively.
The parameter sets are given as follows:
\{$m_1\!=\!10/(2\pi)^2$, $m_2\!=\!2.5/(2\pi)^2$,
$k_1\!=\!10$, $k_2\!=\!40$, $\kappa_0\!=\!1$\}, if not specified.
In (a), 
$k_2\!=\!10$ for $f_1\!=\!f_2$.
In (b),
\{$\kappa_\textrm{m}\!=\!3$, $f_\textrm{m}\!=\!|f_1-f_2|$\} and 
\{$\kappa_\textrm{m}\!=\!0.75$, $f_\textrm{m}\!=\! f_1+f_2$\}.
In (c), 
$A_\textrm{d}\!=\!30$, and
$f_\textrm{d}\!=\!f_2+1$ or $f_2$.
The initial conditions are as follows:
for (a) and (b), \{$x_1(t\!=\!0)\!=\!1$, $\dot{x}_1(0)\!=\!0$, $x_2(0)\!=\!0$, $\dot{x}_2(0)\!=\!0$\}; 
and for (c), \{0, 0, 0, 0\}.
All numbers are unitless.
(d) Correspondence between classical analogies in this figure and required operations for quantum computation covered in this tutorial.
The left column indicates the analogies in this figure and actual quantum oscillators;
the right column indicates the target operation.
For example, the last row means that ``we can understand the physics of the single-qubit gate operation by replacing oscillators 1 and 2 in (c) with a Quantum Harmonic Oscillator (QHO) and a qubit, respectively.''
}
\label{fig:coupledHO}
\end{figure*}

Before exploring a quantum system, considering a similar classical system is often helpful to understand the physics in the quantum regime.
As we will see soon, the physics behind various couplings associated with qubits can be captured using two simple classical harmonic oscillators.
Figure~\ref{fig:coupledHO} shows a schematic diagram of our model system:
it is composed of two classical simple harmonic oscillators, each made of a spring and a block.
The two oscillators interact via the coupling spring, whose spring constant can be either static [Fig.~\ref{fig:coupledHO}(a) and (c)] or time-varying [Fig.~\ref{fig:coupledHO}(b)].
In addition, oscillator 1 may be driven by an external force [Fig.~\ref{fig:coupledHO}(c)].
The equations of motion of this system are given by
\begin{equation}\label{eq:coupledHO_EOM}
\begin{split}
m_1 \ddot{x}_1 &= -(k_1 + \kappa)x_1 + \kappa x_2 + A_\textrm{d}\cos(2\pi f_\textrm{d}t), \\
m_2 \ddot{x}_2 &= -(k_2 + \kappa)x_2 + \kappa x_1,
\end{split}
\end{equation}
where $m_i$ is the mass of oscillator $i$, where $i=1,2$;
$k_i$ is the spring constant of oscillator $i$;
$x_i$ is the position of the center of block $i$;
$\kappa$ is the spring constant of the coupling spring, which can be decomposed into two parts, namely, the static $\kappa_0$ and the time-varying $\kappa_\textrm{m}\cos(2\pi f_\textrm{m}t)$; and
$A_\textrm{d}$ and $f_\textrm{d}$ are the amplitude and the frequency of the drive, respectively.

Note that, although the coupling spring is always present, its effect on the dynamics strongly depends on the system parameters.
When $\kappa$ is the static parameter $\kappa_0$, the two oscillators exchange their energy only when they are on-resonance [Fig.~\ref{fig:coupledHO}(a)].
Even if the oscillators are off-resonance, we can force them to exchange their energy by modulating $\kappa$ with the frequency difference between the two oscillators, $|f_1-f_2|$, where $2\pi f_i = \sqrt{(k_i+\kappa_0)/m_i}$ [Fig.~\ref{fig:coupledHO}(b)].
These two phenomena can be seen in both classical and quantum systems regardless of whether statistics is fermionic (qubit) or bosonic (resonator).

Next, we inject energy into the system by two methods.
One is to modulate the coupling constant with $f_\textrm{m} = f_1+f_2$.
In this case, as one can see in Fig.~\ref{fig:coupledHO}(b), both $x_1$ and $x_2$ increase exponentially with time.
This is parametric amplification, which is important for realizing noiseless amplification.
The concept and applications of parametric amplification will be discussed further in Sec.~\ref{sec:paraamp}.
The other is to drive oscillator 1 with the frequency $f_\textrm{d}$.
When $f_\textrm{d} = f_2$, $x_2$ increases linearly with time.

When we apply the physics learned from these energy injection processes, we need to consider the quantum statistics.
If two coupled quantum systems are bosonic, we can simply interpret the displacement of the blocks as the population.
However, if one or both of the systems are fermionic, we will see an oscillation in the population, instead of the linear increase that we saw in Fig.~\ref{fig:coupledHO}(c).
Such an oscillation is called the Rabi oscillation, which will be discussed further in Sec.~\ref{sec:SQG}.

\subsection{From Circuit to Atom}
\label{sec:circuitAtom}

\subsubsection{Qubit, Resonator, and Somewhere between Them}
\label{sec:couplingConcept}

Now that we are well equipped with the necessary physics, it is time to move back to quantum.
In cQED, the circuit Hamiltonians for a qubit and a resonator are mapped to spin-1/2 fermionic and bosonic Hamiltonians, respectively:
\begin{equation}\label{eq:bareMapping}
\hat{\mathcal{H}}_\textrm{q} \rightarrow -\hbar\omega_\textrm{q}\frac{\hat{\sigma}_z}{2}, \quad
\hat{\mathcal{H}}_\textrm{r} \rightarrow \hbar\omega_\textrm{r}\hat{a}^\dagger \hat{a},
\end{equation}
where $\hat{\sigma}_z$ is the Pauli $z$ operator and 
$\hat{a}^{\dagger}$ ($\hat{a}$) is the bosonic creation (annihilation) operator.
Note the $-$ sign in $\hat{\mathcal{H}}_\textrm{q}$.
In this tutorial, we represent the ground state of the qubit as $\ket{0} \doteq (1, 0)^\top$ and the excited state as $\ket{1} \doteq (0, 1)^\top$.
Therefore, the $-$ sign is necessary for $\ket{0}$ to have a physically lower energy.
This representation has the advantage of being mathematically consistent with the result obtained by restricting the Hilbert space dimension of the harmonic oscillator to two.

Note that this mapping assumes an ideal two-level system and a single-mode resonator.
As we will see in Secs.~\ref{sec:TQG} and \ref{sec:resetDumping}, however, higher excitation levels of a qubit have to be considered in many situations, especially for a transmon whose anharmonicity is weak.
On the other hand, the resonator may show a small nonlinearity that has to be considered for high fidelity control.
Hence, we sketch how to model a transmon in the second quantization formalism as an example of a weakly anharmonic/nonlinear system.
By expanding Eq.~\eqref{eq:qubitIsland}, we have
\begin{align}\label{eq:qubitTransmon}
\hat{\mathcal{H}}_\textrm{q}
\approx
4E_\textrm{C} \hat{N}^2
+ \frac{1}{2} E_\textrm{J} \hat{\varphi}^2
- \frac{1}{24} E_\textrm{J} \hat{\varphi}^4.
\end{align}
In Eq.~\eqref{eq:qubitTransmon}, the first two terms, i.e., harmonic terms, can be diagonalized by defining
\begin{align}
\hat{N} = -\textrm{i} N_0 (\hat{b}-\hat{b}^\dagger), \quad
\hat{\varphi} = \varphi_0 (\hat{b}+\hat{b}^\dagger), 
\end{align}
where $N_0^2 = \sqrt{E_\textrm{J}/32E_\textrm{C}}$ and $\varphi_0^2 = \sqrt{2E_\textrm{C}/E_\textrm{J}}$ are the zero-point fluctuations, and $\hat{b}^{\dagger}$ ($\hat{b}$) is the bosonic creation (annihilation) operator for a transmon.
Note that the zero-point fluctuations are determined by the $E_\textrm{J}/E_\textrm{C}$ ratio.
After normal ordering, we obtain
\begin{align}\label{eq:qubitTransmonQ}
\hat{\mathcal{H}}_\textrm{q} \approx 
\left( \sqrt{8E_\textrm{J}E_\textrm{C}} - E_\textrm{C} \right) \hat{b}^\dagger \hat{b}
- \frac{E_\textrm{C}}{2} \hat{b}^\dagger\hat{b}^\dagger\hat{b}\hat{b}.
\end{align}
Here, the terms whose mean excitation number is nonzero, such as $\hat{b}\hat{b}$ and $\hat{b}^\dagger\hat{b}^\dagger$, are ignored because the dynamics induced by these terms will be averaged out at the time scale we are interested in.
(This is the rotating wave approximation, which will be introduced formally in Sec.~\ref{sec:qrCoupling}).
Thus, a transmon can be mapped into a harmonic oscillator with the Kerr-type nonlinearity:
\begin{align}\label{eq:weakAnharQ}
\hat{\mathcal{H}}_\textrm{q} \rightarrow 
\hbar\omega_\textrm{q} \hat{b}^\dagger \hat{b}
+ \frac{\hbar K}{2} \hat{b}^\dagger\hat{b}^\dagger\hat{b}\hat{b},
\end{align}
where $K$ is the Kerr coefficient.
Equations~\eqref{eq:qubitTransmonQ} and \eqref{eq:weakAnharQ} suggests that $\omega_\textrm{q}$ is about $\sqrt{8E_\textrm{J}E_\textrm{C}}/\hbar$ ($\because E_\textrm{J}/E_\textrm{C} \gg 1$) and the anharmonicity, i.e., $\hbar K$, is $-E_\textrm{C}$.
These results are consistent with Secs.~\ref{sec:islandQubit} and \ref{sec:transmon}.

Although Eq.~\eqref{eq:weakAnharQ} models several important properties of a weakly anharmonic/nonlinear system successfully, we still need a unified description of superconducting circuits in a wide range of nonlinearity for complicated circuits.
Moreover, off-resonant resonator modes have been known to contribute substantially to the relaxation times of a qubit via the Purcell effect (see Sec.~\ref{sec:readout}).\cite{houck2008}
To remedy these issues, semiclassical superconducting circuit quantization methods have been proposed and showed a good agreement with experimental data.
Interested readers should see Refs.~\onlinecite{minev2020, nigg2012, bourassa2012, lieb2012, solgun2014}.

For the rest of this section, we model a superconducting qubit as an ideal two-level system because this provides a qualitatively satisfactory picture to understand the physics of various couplings associated with a qubit at the level of this tutorial.

\subsubsection{Qubit-Resonator Coupling}
\label{sec:qrCoupling}

\begin{figure}
\centering
\includegraphics{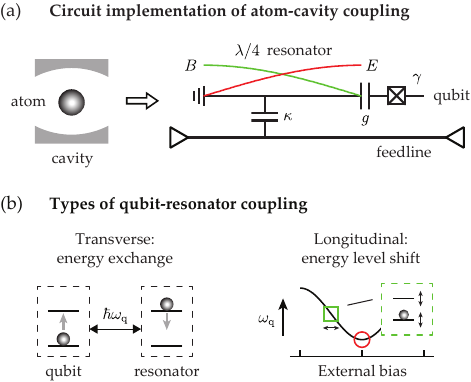}
\caption{(a) In a superconducting circuit, the atom-cavity interaction can be implemented by the qubit-resonator coupling. Here, the resonator can be either a planar resonator or a 3D cavity; in either case, it is usually modeled as an $LC$ circuit.
The circuit shows the capacitive coupling between a quarter-wavelength ($\lambda/4$) resonator and an island-based qubit.
Here, $g$ represents the strength of the transverse coupling between the resonator and the qubit;
$\kappa$ represents the loss rate of photons from the resonator; and
$\gamma$ represents the transverse relaxation rate of the qubit.
(b) Mechanisms for the transverse and longitudinal coupling between a qubit and a resonator.
For the longitudinal coupling, the change in bias shifts the qubit transition frequency $\omega_\textrm{q}$ at the bias shown by the green square, resulting in a strong longitudinal coupling;
at the bias shown by the red circle, the longitudinal coupling is zero.
}
\label{fig:couplingQR}
\end{figure}

Consider a single qubit capacitively coupled to a single-mode resonator without an external drive;\footnote{Usually capacitive coupling is easier to design because, for inductive coupling, we need to consider not only geometric inductance but also kinetic inductance, which is harder to simulate or estimate than the capacitance.}
one example is shown in Fig.~\ref{fig:couplingQR}(a).
In this case, the physical process of the coupling is the zero-point voltage fluctuation of the resonator acting on the net charge $2e\hat{N}$ ($2e$ is the charge of a Cooper pair) via the coupling capacitor between the qubit and the resonator [the capacitor labeled $g$ in Fig.~\ref{fig:couplingQR}(a)].
Then, $2e\hat{N}$ can be considered as the effective dipole moment of this artificial atom [see Eq.~\eqref{eq:effDipole1}].
The qubit-resonator coupling Hamiltonian $\hat{\mathcal{H}}_\textrm{qr}$ can be written as
\begin{equation}
\hat{\mathcal{H}}_\textrm{qr} = 
2e\hat{N} \beta V_\textrm{r,0} (\hat{a}+\hat{a}^\dagger).
\end{equation}
Here, $V_\textrm{r,0}$($=\!\sqrt{\hbar\omega_\textrm{r}/2C_\textrm{r}}$, where $C_\textrm{r}$ is the capacitance of the resonator) is the root-mean-square voltage of the zero-point fluctuation in the resonator;
$\beta$ is the ratio between the coupling and stored energies, which is the same as the ratio of the coupling capacitance to the total capacitance of the qubit; and
$(\hat{a}+\hat{a}^\dagger)$ represents the process of populating/depopulating the resonator.
Defining the coupling constant
\begin{equation}\label{eq:gDef}
\hbar g_{ij} = 2e \beta V_\textrm{r,0} \mel{i}{\hat{N}}{j},
\end{equation}
where $\ket{i}$ and $\ket{j}$ ($i,j\!\in\!\{0,1\}$) are the eigenstates of the bare qubit, yields
\begin{align}
\hat{\mathcal{H}}_\textrm{qr} 
&= \hbar \sum_{i,j} g_{ij} \ketbra{i}{j} (\hat{a}+\hat{a}^\dagger) \nonumber\\
&= \hbar (g_x \hat{\sigma}_x + g_z \hat{\sigma}_z) (\hat{a}+\hat{a}^\dagger), \label{eq:qrCoupling}
\end{align}
where $g_x$ and $g_z$ are defined by
\begin{equation}\label{eq:couplingMap}
g_x \equiv g_{01} (= g_{10}), \quad
g_z \equiv \frac{g_{00} - g_{11}}{2}.
\end{equation}
The $g_x$ term is called the transverse coupling because the axis for the Pauli operator is perpendicular to the qubit quantization axis;
the $g_z$ term is called the longitudinal coupling.
Here, a term associated with $y$ is omitted because the choice of $x$ or $y$ is just a matter of convention.

The transverse coupling mediates the energy exchange between the qubit and the resonator [Fig.~\ref{fig:couplingQR}(b)].
Thus, the transverse coupling is effective when the coupled system has a mode whose frequency is close to $\omega_\textrm{q}$ as we saw in Fig.~\ref{fig:coupledHO}(a).
The longitudinal coupling changes the qubit frequency.
It is effective when $\omega_\textrm{q}$ varies considerably with the external bias [Fig.~\ref{fig:couplingQR}(b)].
Note that the physics of the relaxation processes in Sec.~\ref{sec:relaxation} can be understood within this framework; 
the transverse and longitudinal couplings are actually the mechanisms for thermalization and dephasing, respectively.

Equations~\eqref{eq:gDef} and \eqref{eq:couplingMap} suggest that, if $\omega_\textrm{q}$ does not depend on the physical parameter that is coupled to the effective dipole moment, the voltage in this case, there is no longitudinal coupling.
Note that, because of this, the dominant coupling associated with a qubit at its sweet spot is the transverse coupling.
One consequence is that the only possible coupling associated with a capacitively coupled transmon is the transverse coupling because $\omega_\textrm{q}$ of a transmon is insensitive to the external voltage fluctuation, i.e., a transmon is always at its charge bias sweet spot.
To implement the longitudinal coupling, a transmon needs a flux-tunable element, such as a DC SQUID, and should be coupled to the target system inductively.\cite{didier2015}

Although Eq.~\eqref{eq:qrCoupling} captures all the physics regarding the capacitive coupling, solving Eq.~\eqref{eq:qrCoupling} using Eq.~\eqref{eq:bareMapping} is straightforward.
We ignore the $g_z$ term because the $g_x$ term is the dominant term at the sweet spot as mentioned above.
Then, we have
\begin{align}\label{eq:1q_H}
\hat{\mathcal{H}}_\mathrm{1q}
&= \hat{\mathcal{H}}_\mathrm{q} + \hat{\mathcal{H}}_\mathrm{r} 
+ \hat{\mathcal{H}}_\mathrm{qr} \nonumber\\
&=
- \hbar\omega_\mathrm{q}\frac{\hat{\sigma}_z}{2}
+ \hbar\omega_\mathrm{r} \hat{a}^\dagger \hat{a}
+ \hbar g \hat{\sigma}_x (\hat{a}+\hat{a}^\dagger).
\end{align}
Here, we omit the subscript $x$ in $g$ for simplicity.\footnote{Although $g$ can generally take a complex value, it can be treated as a real number without loss of generality by appropriately redefining the phases of the atomic basis states or the photon operator.
Here, we assume that $g$ is a positive real number.}

Now, we move to the rotating frame to focus on the dynamics induced by the coupling term.
The Hamiltonian defining this frame is
\begin{equation}\label{eq:bareQR}
\hat{\mathcal{H}}_0 = 
- \hbar\omega_\mathrm{q} \frac{\hat{\sigma}_z}{2}
+ \hbar\omega_\mathrm{r} \hat{a}^\dagger \hat{a}.
\end{equation}
The final single-qubit Hamiltonian in the rotating frame $\hat{\mathcal{H}}_\mathrm{1q}^\mathrm{rot}$ can be obtained by a unitary transformation with the unitary operator $\hat{U}_\mathrm{rot} \equiv \mathrm{e}^{\mathrm{i}\hat{\mathcal{H}}_0 t/\hbar}$:
\begin{align}
\hat{\mathcal{H}}_\mathrm{1q}^\mathrm{rot} &=
\hat{U}_\mathrm{rot}
(\hat{\mathcal{H}}_\mathrm{1q} - \hat{\mathcal{H}}_0)
\hat{U}_\mathrm{rot}^\dagger
\nonumber\\
&= \hbar g 
\big[
\hat{\sigma}_+ \hat{a}\,
\mathrm{e}^{\mathrm{i}(\omega_\mathrm{q}-\omega_\mathrm{r})t}
+ \hat{\sigma}_- \hat{a}^\dagger
\mathrm{e}^{-\mathrm{i}(\omega_\mathrm{q}-\omega_\mathrm{r})t} \nonumber\\
&\quad\quad\ \ 
+ \hat{\sigma}_+ \hat{a}^\dagger
\mathrm{e}^{\mathrm{i}(\omega_\mathrm{q}+\omega_\mathrm{r})t}
+ \hat{\sigma}_- \hat{a}\,
\mathrm{e}^{-\mathrm{i}(\omega_\mathrm{q}+\omega_\mathrm{r})t} 
\big],
\label{eq:1qRot}
\end{align}
where $\hat{\sigma}_+ \equiv \ketbra{1}{0} = (\hat{\sigma}_x - \mathrm{i}\hat{\sigma}_y)/2$
and $\hat{\sigma}_- \equiv \ketbra{0}{1} = (\hat{\sigma}_x + \mathrm{i}\hat{\sigma}_y)/2$.\footnote{Note that the flipped sign in front of $\hat{\sigma}_y$ compared to standard spin algebra is not a typo.
This is a direct consequence of consistently applying the qubit basis notation defined below Eq.~\eqref{eq:bareMapping}.
Furthermore, this approach coincides with the matrix representation of the ladder operators obtained when the Hilbert space of a harmonic oscillator is truncated to two dimensions.
Therefore, the operators $\hat{\sigma}_\pm$ used in this tutorial and the spin angular momentum operators $\hat{S}_\pm$ found in standard quantum mechanics textbooks have distinct matrix representations.}
Here, we used the relations
\begin{gather}
\hat{U}_\mathrm{rot} \hat{\sigma}_x \hat{U}_\mathrm{rot}^\dagger
= \hat{\sigma}_- \mathrm{e}^{-\mathrm{i}\omega_\mathrm{q} t}
+ \hat{\sigma}_+ \mathrm{e}^{\mathrm{i}\omega_\mathrm{q} t}, \\
\hat{U}_\mathrm{rot} (\hat{a}+\hat{a}^\dagger) \hat{U}_\mathrm{rot}^\dagger
= \hat{a}\mathrm{e}^{-\mathrm{i}\omega_\mathrm{r} t}
+ \hat{a}^\dagger \mathrm{e}^{\mathrm{i}\omega_\mathrm{r} t}.
\end{gather}
In Eq.~\eqref{eq:1qRot}, if $g$ is a constant, all terms average out over the time scale of interest unless $\omega_\mathrm{q}$ and $\omega_\mathrm{r}$ are reasonably close.
In this context, the transverse coupling is considered effective only if $\omega_\mathrm{q} \approx \omega_\mathrm{r}$.
Moreover, we can safely ignore the rapidly oscillating terms, $\hat{\sigma}_+ \hat{a}^\dagger$ and $\hat{\sigma}_- \hat{a}$, in most situations.
Such an approximation is known as the rotating wave approximation (RWA).

After applying the RWA, $\hat{\mathcal{H}}_\mathrm{qr}$ becomes 
(now we move back to the inertial frame)
\begin{equation}\label{eq:qr}
\hat{\mathcal{H}}_\mathrm{qr} \approx
\hbar g (\hat{\sigma}_+ \hat{a} + \hat{\sigma}_- \hat{a}^\dagger).
\end{equation}
The physical meaning of $g$ is the exchange of energy between a quantized electromagnetic field and a qubit at a rate of $g/2\pi$.
Such an energy exchange with a well-defined period and phase is called coherent exchange;
this will be useful for two-qubit gate operation (Sec.~\ref{sec:TQG}).
Equation~\eqref{eq:qr}, together with Eq.~\eqref{eq:bareMapping}, is called the Jaynes--Cummings Hamiltonian $\hat{\mathcal{H}}_\mathrm{JC}$ (Refs.~\onlinecite{haroche, agarwal}):
\begin{align}\label{eq:JCH}
\hat{\mathcal{H}}_\mathrm{JC}
=
- \hbar\omega_\mathrm{q}\frac{\hat{\sigma}_z}{2}
+ \hbar\omega_\mathrm{r} \hat{a}^\dagger \hat{a}
+ \hbar g(\hat{\sigma}_+ \hat{a} + \hat{\sigma}_- \hat{a}^\dagger).
\end{align}
Equation~\eqref{eq:JCH} will be the central equation in Sec.~\ref{sec:readout}.

\subsubsection{Qubit-Qubit Coupling}
\label{sec:qqCoupling}

\begin{figure}
\centering
\includegraphics{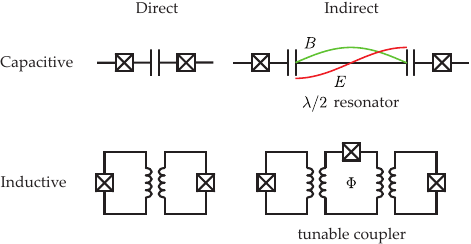}
\caption{Implementations of qubit-qubit coupling.
Two qubits can be coupled either directly or indirectly via a coupling resonator.
Here, a half-wavelength ($\lambda/2$) resonator and a tunable coupler are shown as examples.
Solid lines labeled $B$ and $E$ represent magnetic and electric field profiles in the resonators, respectively.
For fixed-frequency island-based qubits, the capacitive coupling is the only available coupling scheme.
However, for loop-based qubits, not only inductive coupling but also capacitive coupling is possible because a loop-based qubit can also be understood as two superconducting islands as shown in the inset of Fig.~\ref{fig:qubitLoop}(d).
Note that the circuit for the tunable coupler is the same as that of the neighboring qubits. The qubit-qubit coupling constant is tuned by the external flux bias $\Phi$.}
\label{fig:couplingQQ}
\end{figure}

The physics of qubit-qubit coupling is similar to that of qubit-resonator coupling.
The Hamiltonian describing the qubit-qubit coupling can be written as
\begin{align}\label{eq:qqCoupling}
\hat{\mathcal{H}}_\textrm{qq} &= 
\hbar \big(
J_{xx} \hat{\sigma}_x^{(1)} \hat{\sigma}_x^{(2)}
+ J_{zx} \hat{\sigma}_z^{(1)} \hat{\sigma}_x^{(2)} \nonumber\\
&\quad\quad\,
+ J_{xz} \hat{\sigma}_x^{(1)} \hat{\sigma}_z^{(2)}
+ J_{zz} \hat{\sigma}_z^{(1)} \hat{\sigma}_z^{(2)}
\big),
\end{align}
where $\sigma_i^{(j)}$ ($i=x,z$) represents the Pauli operators for qubit $j$ and $J_{kl}$ is the qubit-qubit coupling constant.
Note that we have four terms in the qubit-qubit coupling because both systems are fermions (see the last paragraph of Sec.~\ref{sec:couplingConcept}).
For better visibility, we call each term with its subscripts, for example, the $J_{zz}$ term as the $ZZ$ term or the $ZZ$ interaction.

In Eq.~\eqref{eq:qqCoupling}, the $XX$ interaction corresponds to the transverse interaction.
Regarding the longitudinal interaction, there is ambiguity in its definition.
If we follow the convention in the qubit-resonator interaction consistently, only the $XZ$ and $ZX$ interactions must be called the longitudinal interactions.
However, a considerable number of papers designate all non-transverse interactions, which includes the $ZZ$ interaction, as the longitudinal interactions.
In this tutorial, we use the term ``longitudinal interaction'' for the qubit-resonator interaction only.
For the qubit-qubit interaction, we call the type of interaction explicitly, such as the $XZ$ interaction, for clarity.

We consider transmons coupled directly and capacitively (Fig.~\ref{fig:couplingQQ}).
In this case, the transverse ($XX$) interaction is the dominant interaction as discussed in Sec.~\ref{sec:qrCoupling}.
Hence, we consider the $XX$ term only and omit the subscript $xx$ for simplicity.
Similarly to the capacitive qubit-resonator coupling, $J$ is determined by the coupling capacitance $C_\textrm{12}$ and the voltage fluctuations of the ground states,
\begin{equation}\label{eq:g}
\hbar J = C_{12} V_\textrm{q,0}^{(1)} V_\textrm{q,0}^{(2)}
\approx \frac{\hbar}{2} \frac{C_{12}}{\sqrt{C_\textrm{q1} C_\textrm{q2}}} \sqrt{\omega_\textrm{q1} \omega_\textrm{q2}},
\end{equation}
where
$V_\textrm{q,0}^{(i)}\,$($i\!=\!1,2$) is the root-mean-square voltage of the ground state of qubit $i$;
$\omega_{\textrm{q}i}$ and $C_{\textrm{q}i}$ are the transition frequency and total capacitance of qubit $i$, respectively.
Since a transmon is a weakly nonlinear harmonic oscillator (Sec.~\ref{sec:transmon}), $V_\textrm{q,0}^{(i)} \approx \sqrt{\hbar\omega_{\textrm{q}i}/2C_{\textrm{q}i}}$ if $C_{12} \ll C_{\textrm{q}i}$.
Note that $J$ depends on the transition frequency.
Hence, for a coupling associated with a frequency-tunable qubit, $J$ is also tunable.

The resulting two-qubit Hamiltonian is
\begin{align}\label{eq:2q_H}
\hat{\mathcal{H}}_\mathrm{2q} =
- \hbar\omega_\mathrm{q1}\frac{\hat{\sigma}_z^{(1)}}{2}
- \hbar\omega_\mathrm{q2}\frac{\hat{\sigma}_z^{(2)}}{2}
+ \hbar J \hat{\sigma}_x^{(1)}\hat{\sigma}_x^{(2)}.
\end{align}
We move to the rotating frame defined by
\begin{equation}
\hat{\mathcal{H}}_0
=
- \hbar\omega_\mathrm{q1} \frac{\hat{\sigma}_z^{(1)}}{2}
- \hbar\omega_\mathrm{q2} \frac{\hat{\sigma}_z^{(2)}}{2}.
\end{equation}
Then, we have an equation similar to Eq.~\eqref{eq:1qRot}:
\begin{align}
\hat{\mathcal{H}}_\mathrm{2q}^\mathrm{rot} &=
\hat{U}_\mathrm{rot}
(\hat{\mathcal{H}}_\mathrm{2q} - \hat{\mathcal{H}}_0)
\hat{U}_\mathrm{rot}^\dagger
\nonumber\\
&= \hbar J
\big[
\hat{\sigma}_+^{(1)} \hat{\sigma}_-^{(2)}
\mathrm{e}^{\mathrm{i}(\omega_\mathrm{q1}-\omega_\mathrm{q2})t}
+ \hat{\sigma}_-^{(1)} \hat{\sigma}_+^{(2)}
\mathrm{e}^{-\mathrm{i}(\omega_\mathrm{q1}-\omega_\mathrm{q2})t} \nonumber\\
&\quad\quad\ \ 
+ \hat{\sigma}_+^{(1)} \hat{\sigma}_+^{(2)}
\mathrm{e}^{\mathrm{i}(\omega_\mathrm{q1}+\omega_\mathrm{q2})t}
+ \hat{\sigma}_-^{(1)} \hat{\sigma}_-^{(2)}
\mathrm{e}^{-\mathrm{i}(\omega_\mathrm{q1}+\omega_\mathrm{q2})t} 
\big].
\label{eq:2qRot}
\end{align}

If $J$ is static and $|\omega_\textrm{q1}-\omega_\textrm{q2}| \gg J$, it is clear that the coupling term will be averaged out, and consequently cannot be used for two-qubit gate operation unless one of the following actions is taken:
(i) tuning $\omega_\textrm{q1}$ or $\omega_\textrm{q2}$ so that $\omega_\textrm{q1} \approx \omega_\textrm{q2}$;
(ii) modulating $J$ with the frequency $|\omega_\textrm{q1} \pm \omega_\textrm{q2}|$ to cancel out oscillating factors; or
(iii) adding an additional drive term.
These strategies are based on the lessons learned in Sec.~\ref{sec:classicalCoupling} and will be the basis of two-qubit gates in Sec.~\ref{sec:TQG}.

It is often necessary to couple two qubits separated by a macroscopic distance.
In this case, a resonator or even a qubit is employed as a coupler---such a scheme is called indirect coupling (Fig.~\ref{fig:couplingQQ}).
Here, we need to be careful not to excite the coupler itself;
otherwise, the information will leak to the Hilbert space of the coupler.
Hence, the resonance frequency of the coupler must be significantly far from the transition frequency of the qubits such that $\big|\omega_\textrm{r}-\omega_{\textrm{q}i}\big| \gg g^{(i)}$, where $g^{(i)}$ is the transverse coupling constant associated with the resonator and qubit $i$.
The coupler mediates the exchange of virtual photons between the two qubits.
Such a system can also be modeled as Eq.~\eqref{eq:2q_H}.\cite{majer2007, lu2012}

\subsection{Strong (Transverse) Coupling}
\label{sec:strongCoupling}

\begin{figure}
\centering
\includegraphics{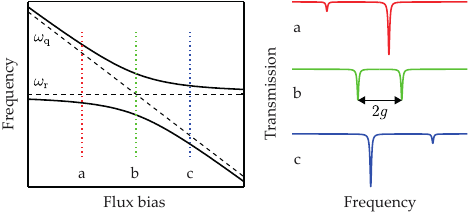}
\caption{Anticrossing due to strong qubit-resonator coupling for the circuit shown in Fig.~\ref{fig:couplingQR}(a).
The splitting when $\omega_\textrm{q} = \omega_\textrm{r}$ is $2g$.}
\label{fig:antiCrossing}
\end{figure}

In this subsection, we consider how to quantify the strength of the transverse coupling because the current standard qubit control and readout methods are based on the transverse coupling.\footnote{There are many studies on the potential use of the longitudinal qubit-resonator coupling for quantum computation. Interested readers should see Refs.~\onlinecite{didier2015, billangeon2015, richer2016, wang2019}}
For efficient qubit control and readout, we need a reasonably strong qubit-resonator coupling; otherwise, the signal will be too small and the control will be too slow.
Similarly, we also need a strong qubit-qubit interaction for efficient two-qubit gate operation. (See Secs.~\ref{sec:readout} and \ref{sec:TQG} for further explanation.)
Then, what are the criteria that must be satisfied to be called a strong coupling?

The strength of the qubit-resonator coupling is usually characterized by three quantities: $g$, $\kappa$, and $\gamma$ [Fig.~\ref{fig:couplingQR}(a)].
Here,
$g/2\pi$ is the transverse coupling strength in Hz,
$\kappa/2\pi$ is the loss rate of photons from the resonator, i.e., the spectral linewidth of the resonator, in Hz
($\kappa=\omega_\mathrm{r}/Q$, where $Q$ is the quality factor of the resonator), and
$\gamma/2\pi$($=\!1/2\pi T_2$) is the transverse relaxation rate, i.e., the spectral linewidth, of the qubit in Hz.
When the system satisfies $g > \kappa/2,\ \gamma/2$, the coupling is considered to be in the strong coupling regime.
The physical meaning is clear: to ensure a strong qubit-resonator interaction, the photon must stay in the resonator and the qubit needs to keep its coherence while the two systems exchange their energy.

The experimental signature of a strong qubit-resonator or qubit-qubit coupling is an anticrossing called the vacuum Rabi splitting (Fig.~\ref{fig:antiCrossing}).
Such a situation is well described by the Jaynes--Cummings Hamiltonian [Eq.~\eqref{eq:JCH}].
In the Jaynes--Cummings Hamiltonian, when the qubit and the resonator are far off-resonance, the signal we observe is the transition between $\ket{\mathrm{g} 0}$, which is the ground state, and $\ket{\mathrm{e} 0}$ (bias points a and c in Fig.~\ref{fig:antiCrossing}), where $\ket{i j}$ denotes the quantum state where the $i$th state of the bare qubit and the $j$th state of the bare resonator are occupied.
At resonance, the signal represents the transition between $\ket{\mathrm{g} 0}$ and $(\ket{\mathrm{g} 1}+\ket{\mathrm{e} 0})/\sqrt{2}$ because of the hybridization between the qubit state and the resonator state (bias point b in Fig.~\ref{fig:antiCrossing}).
In the time domain, the populations of the two systems oscillate out of phase.
This oscillation is called the vacuum Rabi oscillation.

The physics of the vacuum Rabi splitting/oscillation can be understood using our classical oscillators in Sec.~\ref{sec:classicalCoupling}.
When the coupling is static and there is no external drive,
Eq.~\eqref{eq:coupledHO_EOM} can be written in the Hamiltonian form
\begin{align}
\mathcal{H}
=
\underbrace{\frac{p_1^2}{2m_1} + \frac{k_1+\kappa_0}{2} x_1^2}_{\mathcal{H}_\textrm{osc1}}
\,\underbrace{+\,\frac{p_2^2}{2m_2} + \frac{k_2+\kappa_0}{2} x_2^2}_{\mathcal{H}_\textrm{osc2}}
\,\underbrace{-\ \kappa_0\, x_1 x_2}_{\mathcal{H}_\textrm{c}}.
\label{eq:coupledHO_H}
\end{align}
The energy exchange and frequency splitting shown in Fig.~\ref{fig:coupledHO}(a) is caused by the coupling term $\mathcal{H}_\textrm{c}$.
Thus, this coupling term corresponds to the transverse coupling.\footnote{There is no longitudinal coupling in our classical oscillators because this system is harmonic.
The state of a harmonic system, i.e., boson, cannot be represented in the Bloch sphere because there is no well-defined geometrical quantization axis.
However, bosons can couple to each other and exchange their energy;
we just call this coupling transverse to be consistent with that for fermionic systems.}
Although our classical oscillators show essentially the same physics, the vacuum Rabi splitting/oscillation is a highly quantum phenomenon because it is a consequence of coupling between the qubit and the \emph{vacuum} mode of the resonator.

The strong transverse qubit-qubit coupling also yields a similar anticrossing.
However, the transition probability, i.e., the strength of the signal, near the anticrossing is more complex than that of the qubit-resonator coupling.
The reason is that there are two types of symmetry, triplet and singlet, associated with the quantum states of two entangled qubits, and transitions between different symmetries are forbidden.\cite{majer2007, filipp2011}

Note that, compared with other quantum systems, superconducting planar circuit is particularly convenient system for realizing a strong coupling because the low-dimensional nature of this system results in a strongly concentrated electromagnetic field profile and consequently produces a large $V_\textrm{r,0}$ in Eq.~\eqref{eq:gDef}.


\section{Implementation of Quantum Computation}
\label{sec:implementationQC}

\subsection{Equation of Motion}
\label{sec:EOM}

To implement functions required for quantum computation, we need to know how our qubits evolve during various operations.
For a closed quantum system, the evolution of a density matrix $\hat{\rho}$ is fully described by the von Neumann equation:
\begin{equation}\label{eq:neumann}
\frac{d \hat{\rho}}{dt} =
\frac{1}{\textrm{i}\hbar}[\hat{\mathcal{H}},\hat{\rho}],
\end{equation}
where $\hat{\mathcal{H}}$ is the Hamiltonian of the system that the density matrix represents.
Note that Eq.~\eqref{eq:neumann} is in the Schr{\"o}dinger picture; in the Heisenberg picture, the density matrix is not time-dependent.
If $\hat{\mathcal{H}}$ is time-independent, the solution of Eq.~\eqref{eq:neumann} is given by
\begin{equation}\label{eq:neumannSol}
\hat{\rho}(t) 
= \textrm{e}^{-\textrm{i}\hat{\mathcal{H}}t/\hbar} \hat{\rho}(0) 
\textrm{e}^{\textrm{i}\hat{\mathcal{H}}t/\hbar}
= \hat{U}(t) \hat{\rho}(0) \hat{U}^\dagger(t),
\end{equation}
which is the density matrix version of Eq.~\eqref{eq:timeEvol}.

However, a qubit is actually an open quantum system---it is always interacting with the environment, a readout resonator, and other control lines, resulting in the relaxation of a quantum state as discussed in Sec.~\ref{sec:relaxation}.
Thus, this relaxation phenomenon have to be included in the equation of motion for precise control.
To simplify the situation, we introduce three assumptions.
The first one is the Born approximation, which assumes that the interaction between the qubit and the environment is reasonably weak such that the environment is practically unaffected by the system.
The second one is the Markovian approximation, which assumes that the noise process acting on the system is memoryless.
In other words, the internal dynamics of the environment hides any entanglement with the system as quickly as it arises.
The last one is that the initial states of the system and environment are not entangled, i.e., $\hat{\rho}(t=0) = \hat{\rho}_\textrm{sys} \otimes \hat{\rho}_\textrm{env}$.
With these approximations, the dynamics of the qubit can be well described by the Lindblad master equation, which is given by\cite{haroche, schumacher, manzano}
\begin{equation}\label{eq:lindblad}
\frac{d \hat{\rho}}{dt} =
\frac{1}{\textrm{i}\hbar}[\hat{\mathcal{H}},\hat{\rho}]
+\sum_k \left( 
\hat{\mathcal{L}}_k \hat{\rho} \hat{\mathcal{L}}_k^\dagger 
- \frac{1}{2} \left\{ \hat{\mathcal{L}}_k^\dagger\hat{\mathcal{L}}_k, \hat{\rho} \right\} \right).
\end{equation}

To describe the dynamics of the qubit properly, we need to choose the Lindblad operator $\hat{\mathcal{L}}_k$ based on the model we have.
For example, the effects of an environment on a single qubit can be modeled by
\begin{equation}\label{eq:Lqubit}
\hat{\mathcal{L}}_1 = \sqrt{\Gamma_\parallel}\hat{\sigma}_-, \quad
\hat{\mathcal{L}}_2 = \sqrt{\frac{\Gamma_\varphi}{2}}\hat{\sigma}_z.
\end{equation}
Here, $\hat{\mathcal{L}}_1$ describes the thermalization process, i.e., the transition from $\ket{1}$ to $\ket{0}$ ($\hat{\sigma}_-$), and
$\hat{\mathcal{L}}_2$ describes the dephasing process.
Other effects can also be considered by introducing additional Lindblad operators.
The unit of $\hat{\mathcal{L}}_k$ is [s$^{-1/2}$].

Let us solve the equation with $\hat{\mathcal{L}}_1$ only for simplicity.
Using the identity $\hat{\mathcal{L}}_1^\dagger \hat{\mathcal{L}}_1 = \Gamma_\parallel \ketbra{1}{1}$ in Eq.~\eqref{eq:lindblad}, we obtain
\begin{align}
\frac{d}{dt}
\begin{pmatrix}
\rho_{00} & \rho_{01} \\
\rho_{10} & \rho_{11} \\
\end{pmatrix}
=
\Gamma_\parallel
\begin{pmatrix}
\rho_{11} & -\rho_{01}/2 \\
-\rho_{10}/2 & -\rho_{11} \\
\end{pmatrix}.
\end{align}
Solving this is straightforward:
\begin{align}
\begin{pmatrix}
\rho_{00}(t) & \rho_{01}(t) \\
\rho_{10}(t) & \rho_{11}(t) \\
\end{pmatrix}
=
\begin{pmatrix}
1-\rho_{11}(0)\textrm{e}^{-\Gamma_\parallel t} & \rho_{01}(0)\textrm{e}^{-\Gamma_\parallel t/2} \\
\rho_{10}(0)\textrm{e}^{-\Gamma_\parallel t/2} & \rho_{11}(0)\textrm{e}^{-\Gamma_\parallel t} \\
\end{pmatrix}.
\end{align}
Note that the diagonal elements decay with the time constant $\Gamma_\parallel$, whereas the off-diagonal elements decay with $\Gamma_\parallel/2$.
This explains Eq.~\eqref{eq:relaxRate}.

The Lindblad master equation can also be applied to a resonator.
In this case,
\begin{equation}\label{eq:Lres}
\hat{\mathcal{L}} = \sqrt{\kappa}\hat{a}.
\end{equation}

Although the Lindblad master equation is an appropriate tool to describe the dynamics of a quantum system induced by uncontrolled interactions with the environment, we need another formalism that describes the interaction between the system and a ``controlled'' environment, such as traveling electromagnetic fields through transmission lines, to model an actual experiment.
Input-output theory is a theory for this.
Here, ``input'' refers to the field that drives the system and ``output'' refers to the field that propagates away from the system.
Interested readers should see Refs.~\onlinecite{GZ, yurke2004, clerk2010}.

\subsection{Readout}

\subsubsection{Dispersive Readout}
\label{sec:readout}

\begin{figure}
\centering
\includegraphics{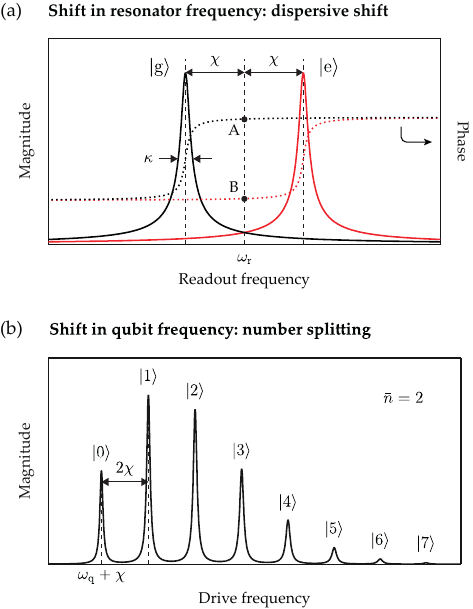}
\caption{(a) Qubit-state-dependent shift in resonator frequency.
Here, we assume $\Delta_\mathrm{qr} (\equiv \omega_\mathrm{q} - \omega_\mathrm{r}) > 0$, and thus $\chi > 0$.
This frequency shift, called the dispersive shift, allows us to detect the qubit state by monitoring the S-parameters of the circuit.
For the circuit shown in Fig.~\ref{fig:couplingQR}(a), the qubit state can be inferred by measuring the transmission of the circuit at $\omega_\mathrm{r}$.
If the measured phase is A, then the qubit state is $\ket{\mathrm{g}}$;
if the phase is B, the qubit state is $\ket{\mathrm{e}}$.
(b) Resonator-state-dependent shift in qubit frequency.
In the strong dispersive regime with $\chi > \kappa, \gamma$, the qubit frequency can be split by the photon-state-dependent frequency shift.
The resonator state is assumed to be a coherent state whose average photon number is $\bar{n}$.
This figure was obtained by solving the Lindblad equation [Eq.~\eqref{eq:lindblad}] with Eqs.~\eqref{eq:Lqubit}, \eqref{eq:Lres}, and \eqref{eq:dispersiveJC}. 
For the solution, QuTiP was used.\cite{qutip1, qutip2}
}
\label{fig:readout}
\end{figure}

Readout of a qubit state means to transfer the information of the qubit state to a change in a physical quantity of a classical device.
At the time of writing, the standard method of detecting the superconducting qubit state is dispersive readout, i.e., detecting the qubit state by observing the shift in the resonance frequency of a readout resonator interacting with the qubit [Fig.~\ref{fig:readout}(a)].

Advantages of dispersive readout are that
(i) it does not rely on the dominant degree of freedom of a qubit, such as charge or flux, and
(ii) its nondestructive nature.
Before dispersive readout, a single-electron transistor was employed for island-based qubit readout and a DC SQUID was used for loop-based qubit readout because of their excellent sensitivity to charge and flux, respectively.
The problem was that if the eigenstates of the qubit show significant spread or superposition in the number or phase basis [Fig.~\ref{fig:qubitIsland}(d) and (e)], which happens in all noise-resilient qubits mentioned in Sec.~\ref{sec:noiseResilient}, these quantity-specific detection methods are not effective and often suffer from a strong backaction that disturbs the subsequent evolution of the measured observable.
As a result, the qubit state becomes uncertain after the readout.
This prevents any feedback scheme based on the measurement outcome.

In the dispersive readout scheme, a qubit state is detected and controlled by a resonator via a strong qubit-resonator interaction.
However, near resonance ($\omega_\mathrm{q} \approx \omega_\mathrm{r}$), we cannot selectively detect or control the qubit state because, in this regime, the strong qubit-resonator interaction hybridizes the qubit and resonator states (see Sec.~\ref{sec:strongCoupling}).
Hence, we detune $\omega_\mathrm{q}$ such that the qubit-resonator detuning $\Delta_\mathrm{qr} (\equiv \omega_\mathrm{q} - \omega_\mathrm{r})$ is much greater than $g$ and $\kappa$.
This limit is called the dispersive limit.
In this off-resonant regime, a qubit transition induced by photon exchange with the resonator is negligible.
However, the qubit shows small but easily measurable frequency shifts that depend on the resonator state;
at the same time, the resonator also shows a small frequency shift that depends on the qubit state.
The qubit state is detected by measuring this frequency shift of the resonator.

To see the physics in the dispersive limit more clearly, we consider the Jaynes--Cummings Hamiltonian $\hat{\mathcal{H}}_\mathrm{JC}$ [Eq.~\eqref{eq:JCH}].
Here, we treat the qubit-resonator interaction $\hat{\mathcal{H}}_\mathrm{qr}$ [Eq.~\eqref{eq:qr}] as a perturbation to the uncoupled qubit-resonator Hamiltonian $\hat{\mathcal{H}}_0$ [Eq.~\eqref{eq:bareQR}].
Then, we take a unitary transformation that diagonalizes $\hat{\mathcal{H}}_\mathrm{JC}$ perturbatively to first order in $\hat{\mathcal{H}}_\mathrm{qr}$.
Such a transformation is called the Schrieffer--Wolff transformation.\cite{phillips, cohen1998}
(The same results can be obtained using standard perturbation theory.\cite{haroche, agarwal})
A unitary operator $\hat{U}_\mathrm{disp} = \mathrm{e}^{\hat{S}}$ for the Schrieffer--Wolff transformation is defined such that $\hat{S}^\dagger = -\hat{S}$ and $[\hat{S}, \hat{\mathcal{H}}_0] = -\hat{\mathcal{H}}_\mathrm{qr}$.
Then, we have (using the formulas in Table~\ref{tab:formula})
\begin{align}
\hat{\mathcal{H}}_\mathrm{JC}^\mathrm{disp} 
&= \hat{U}_\mathrm{disp} \hat{\mathcal{H}}_\mathrm{JC} \hat{U}_\mathrm{disp}^\dagger \nonumber\\
&= \hat{\mathcal{H}}_0 + \hat{\mathcal{H}}_\mathrm{qr}
+ [\hat{S}, \hat{\mathcal{H}}_0] + [\hat{S}, \hat{\mathcal{H}}_\mathrm{qr}] \nonumber\\
&\quad 
+ \frac{1}{2}[\hat{S},[\hat{S}, \hat{\mathcal{H}}_0]] + \frac{1}{2}[\hat{S},[\hat{S}, \hat{\mathcal{H}}_\mathrm{qr}]] + \cdots \nonumber\\
&\approx
\hat{\mathcal{H}}_0 + \frac{1}{2} [\hat{S}, \hat{\mathcal{H}}_\mathrm{qr}].
\label{eq:SWtransform}
\end{align}
In our case, $\hat{U}_\mathrm{disp}$ is given by
\begin{equation}\label{eq:dispersiveTrans}
\hat{U}_\mathrm{disp} =
\exp \!\verythinspace
\bigg[\underbrace{\frac{g}{\Delta_\mathrm{qr}} (\hat{\sigma}_+ \hat{a} - \hat{\sigma}_- \hat{a}^\dagger)}_{=\hat{S}} \bigg].
\end{equation}
From Eq.~\eqref{eq:SWtransform}, we obtain
\begin{align}
\hat{\mathcal{H}}_\mathrm{JC}^\mathrm{disp}
\approx
- \hbar(\omega_\mathrm{q} + \chi) \frac{\hat{\sigma}_z}{2}
+ \hbar(\omega_\mathrm{r} - \chi\hat{\sigma}_z)\hat{a}^\dagger \hat{a}, \label{eq:dispersiveJC}
\end{align} 
where $\chi \equiv g^2/\Delta_\mathrm{qr}$.
The second term in Eq.~\eqref{eq:dispersiveJC} shows that the resonator frequency shifts by $\mp \chi$, depending on the qubit state [Fig.~\ref{fig:readout}(a)].
Therefore, dispersive readout detects the longitudinal component of the Bloch vector.
This is the novel feature of the dispersive readout---creating a qubit-resonator interaction with $\hat{\sigma}_z$, which enables us to detect the qubit state, from a purely transverse interaction by taking the dispersive limit.

Note that the dispersive term [$-\hbar\chi \hat{\sigma}_z \hat{a}^\dagger \hat{a}$ in Eq.~\eqref{eq:dispersiveJC}] commutes with the bare qubit and bare resonator terms;
in other words, measuring the qubit state does not disturb the subsequent evolution of the qubit and resonator, meaning that the dispersive readout scheme is nondestructive.
Such a measurement scheme is called Quantum NonDemolition (QND) measurement.\cite{braginsky, braginsky1996, haroche}
Here, we emphasize that the term ``nondemolition'' does not mean the absence of wavefunction collapse.
If the measurement scheme is QND-type, a measured qubit remains in the eigenstate that we record as a measurement outcome, and subsequent measurements reproduce the outcome of the first measurement.

For quantum error correction, the state of an ancilla qubit (see Sec.~\ref{sec:QEC}) must be determined in a single shot---without averaging the output signals of repeated identical measurements.
Thus, maximizing the Signal-to-Noise Ratio (SNR) is crucial.
While we can enhance the SNR by increasing the probe power, i.e., the average number of photons $\bar{n}$ for the detection of the resonator state, $\bar{n}$ must be significantly less than the critical photon number $n_\mathrm{crit}$, which is given by $\Delta_\mathrm{qr}^2/4g^2$ (Ref.~\onlinecite{blais2004}).
If not, Eq.~\eqref{eq:dispersiveJC} would no longer be valid.
Then, the readout process is no longer QND, and the photons induce an unwanted qubit transition as a backaction.\cite{johnson2012}
The resulting change in qubit population during the readout process reduces the readout fidelity.
In experiments, it was found that $\bar{n}$ must be $\lesssim\,$10 for high-fidelity readout.\cite{johnson2012, walter2017}

The next question to consider is the following:
with a given $\bar{n}$, what is the optimal readout condition that ensures fast readout with high fidelity?
From Fig.~\ref{fig:readout}(a), we can see that $\kappa$ should not be too large compared to $\chi$;
otherwise, the frequency shift would be difficult to observe.
The opposite limit---a small $\kappa$---is also undesirable because it makes the readout process inefficient; photons would stay too long in the resonator, resulting in a small signal.
Careful theoretical and experimental studies\cite{walter2017, gambetta2008} found that the condition for the best SNR is $2\chi = \kappa$.
In experiments, the quality factor ($=\!\omega_\mathrm{r}/\kappa$) of the readout resonator is usually designed to be on the order of 100--1000 (Ref.~\onlinecite{heinsoo2018}).
If we lower the quality factor further, the readout process will be faster;
however, achieving a comparable $\chi$ might not satisfy the dispersive limit because a large $\chi$ is achieved by either enhancing $g$ or reducing $\Delta_\mathrm{qr}$.
Moreover, if $g$ is on the order of $0.1\omega_\mathrm{q}$ or larger, the qubit-resonator coupling enters the so-called ultrastrong coupling regime, in which the RWA is no longer applicable.\cite{USC1, USC2}

Another important phenomenon to consider is the Purcell effect, which refers to the enhancement of the spontaneous emission rate of a qubit, i.e., the reduction in $T_1$, due to its coupling to a resonator.
The presence of the resonator imposes boundary conditions on the electromagnetic field modes with which the qubit interacts, altering the density of states accessible to the emitted photon.
Consequently, the Purcell effect is maximized when $\omega_\mathrm{q} = \omega_\mathrm{r}$ because the resonator strongly enhances the local density of states at its resonant frequency, thereby increasing the decay probability of the qubit.\cite{agarwal}
Therefore, if the detuning $\Delta_\mathrm{qr}$ is not large enough, $T_1$ will be severely limited.\cite{houck2008}
However, simply increasing $\Delta_\mathrm{qr}$ is not always desirable, as the resulting dispersive shift $\chi$ might become too small to ensure high readout fidelity.
To maximize readout fidelity while maintaining fast readout capability, the Purcell filter was developed.
By modifying the electromagnetic environment, the Purcell filter suppresses the density of states at the qubit transition frequency; hence, it protects the qubit from the unwanted acceleration of energy relaxation.
For details, see Refs.~\onlinecite{reed2010, sete2015}.

In real experiments, the measured dispersive shift can be significantly different from the ideal value of $g^2/\Delta_\mathrm{qr}$ due to the contributions from higher excitation levels interacting with the resonator.
To fully account for these contributions, we compute $g_{ij}$ using Eq.~\eqref{eq:gDef} and define $\chi_{ij} = |g_{ij}|^2/(\omega_\mathrm{r} - \omega_{i\text{-}j})$, where $\omega_{i\text{-}j}$ is the energy \emph{released} during the transition from $\ket{i}$ to $\ket{j}$ ($i \neq j$).
(Note that $\omega_{i\text{-}j}$ is negative when $i < j$.)
The total dispersive shift $\chi_\mathrm{total}$ observed in experiments is then given by $\sum_{j=0}^M [(\chi_{j1}-\chi_{1j})-(\chi_{j0}-\chi_{0j})]/2$, where $M$ is the cutoff energy level.\cite{yamamoto2014, zhu2013}
Fortunately, by treating $\chi_\mathrm{total}$ as an empirical parameter $\chi$, the readout process can still be understood within the framework established in this section.

We can also rewrite Eq.~\eqref{eq:dispersiveJC} in the following form:
\begin{equation}\label{eq:dispersiveJC2}
\hat{\mathcal{H}}_\mathrm{JC}^\mathrm{disp} \approx
- \hbar\verythinspace
[\verythinspace \omega_\mathrm{q} + \chi(1+2\hat{a}^\dagger \hat{a})]\verythinspace \frac{\hat{\sigma}_z}{2}
+ \hbar\omega_\mathrm{r}\hat{a}^\dagger \hat{a}.
\end{equation}
In this form, the $\chi$ term can be interpreted as the Lamb shift (represented by the constant 1 in the parentheses) and the AC Stark shift (represented by $2\hat{a}^\dagger \hat{a}$) in the qubit frequency.
This implies that each photon number state of the resonator yields a distinct shift in the qubit frequency, splitting the qubit spectrum into discrete peaks—a phenomenon known as photon number splitting [Fig.~\ref{fig:readout}(b)].
Note that the AC Stark shift induces a loss of qubit phase coherence; because the qubit interacts with a superposition of different photon number states, each component acquires a different phase during the time evolution.
This process is known as measurement-induced dephasing.
Extending this concept, any fluctuation in the resonator photon number, such as photon shot noise, causes dephasing in the qubit.
For detailed quantitative analyses of this kind of dephasing, see Refs.~\onlinecite{yale, sears2012}.

On the basis of the physics we have explored thus far, the dispersive readout process can be described as follows:\cite{gambetta2008}
(i) Photons, each with an energy of approximately $\hbar\omega_\mathrm{r}$, enter the resonator.
(ii) The qubit state information is encoded onto the photons, for example, as the phase of the transmission, via the qubit-resonator interaction.
The measurement-induced dephasing caused by this same interaction makes the qubit state lose its phase coherence and collapse into $\ket{\mathrm{g}}$ or $\ket{\mathrm{e}}$.
(iii) The photons escape from the resonator and are then detected.

\subsubsection{Josephson Parametric Amplifier}
\label{sec:paraamp}

\begin{figure*}
\centering
\includegraphics{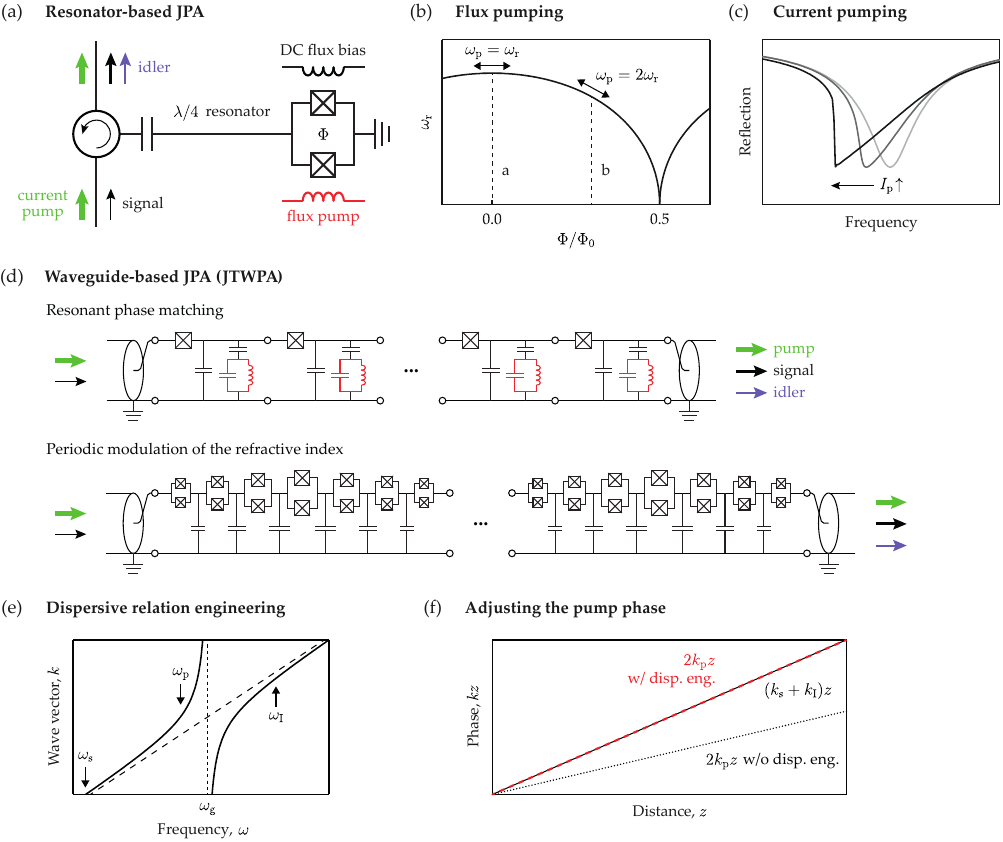}
\caption{(a) A Josephson parametric amplifier (JPA) based on a quarter-wavelength resonator terminated by a DC SQUID, where the tunability is provided by a DC flux bias $\Phi$.
During the amplification process, another frequency tone, called the idler, is generated to satisfy energy conservation.
This type of JPA can be operated by either a flux pump or a current pump.
(b) DC flux bias $\Phi$ dependence of $\omega_\mathrm{r}$.
At zero DC bias (bias a), the pump frequency $\omega_\mathrm{p}$ must equal $\omega_\mathrm{r}$ for parametric amplification, 
whereas $\omega_\mathrm{p}$ can be $2\omega_\mathrm{r}$ at an appreciable DC bias (bias b).
%
(c) Pump current $I_\mathrm{p}$ dependence of the resonator line shape.
As $I_\mathrm{p}$ increases, the resonance frequency decreases due to the inductive contribution from $I_\mathrm{p}$.
(d) Schematic circuit diagram of a Josephson traveling-wave parametric amplifier (JTWPA).
The phases of the interacting tones can be matched either by resonant structures (in red) or by periodic modulation of the refractive index, i.e., the junction size.
The inductance of this amplifier is modulated by a current pump.
(e) Dispersion relation engineered to have a gap at the frequency $\omega_\mathrm{g}$ for phase matching.
The dashed diagonal line shows the bare dispersion relation without the gap.
(f) The phase of the pump tone (red dashed line) can be adjusted via dispersion relation engineering to match the sum of the phases of the signal and idler tones (black solid line), where $z$ represents the propagation distance of the interacting tones along the amplifier.
The black dotted line represents the phase of the pump tone without dispersion relation engineering.
}
\label{fig:paraamp}
\end{figure*}

Although dispersive readout can give a reasonably good SNR, achieving single-shot readout is still difficult because the SNR is degraded by thermal noise as the signal travels from the chip to the room-temperature instruments.
This is an inevitable consequence of the limited number of microwave photons, each having an energy orders of magnitude smaller than the room-temperature thermal energy.
Hence, the pre-amplification of the signal before detection is indispensable.
One might think that using multiple amplifiers will solve the problem.
However, this is not a good option because, if a signal passes through a chain of amplifiers, the SNR is primarily determined by the noise figure of the first amplifier in the chain (Friis formula).
Therefore, placing a low-noise amplifier immediately after the qubit is crucial.

Commercially available High-Electron-Mobility Transistor (HEMT) amplifiers are widely used and installed in the output microwave wiring (at the 4K plate of the dilution refrigerator), typically providing a gain of 30--40 dB.
Although a HEMT amplifier has a high gain and a broad operating bandwidth, it adds an average of 10--20 noise photons to the signal, which in turn worsens the SNR. 
To overcome this, practically noiseless parametric amplifiers were developed using Josephson junctions.

A parametric amplifier essentially transfers energy from a strong pump to a weak signal by mixing the signal and pump frequencies via reactance modulation.
Hence, the reactance is a time-varying parameter, from which the amplifier derives its name.
This type of amplifier has low noise because it modulates the reactance instead of resistance.
In superconducting circuits, a variable inductor can be implemented using Josephson junctions, hence the name Josephson Parametric Amplifier (JPA).
(A pedagogical introduction to the JPA can be found in Ref.~\onlinecite{aumentado2020}.)

The physics of parametric amplification was introduced in Fig.~\ref{fig:coupledHO}(b):
$x_1$ and $x_2$ correspond to the signal (non-zero initial value) and the idler (zero initial value), respectively.
Here, the idler is a tone generated during the amplification process as a consequence of energy conservation:
$\omega_\mathrm{m} = \omega_\mathrm{s} + \omega_\mathrm{I}$,
where $\omega_\mathrm{m}$ is the modulation frequency,
$\omega_\mathrm{s}$ is the signal frequency, and
$\omega_\mathrm{I}$ is the idler frequency.

In general, JPAs can be categorized on the basis of two factors.
One is the method used to maximize the interaction time for energy transfer from the pump to the signal.
This can be achieved by using either a resonator (multiple bounces in a cavity) or a long waveguide.
The other is the method of inductance modulation.
Depending on the modulation method or operating conditions, $\omega_\mathrm{m}$ can be either the same as or twice the pump frequency $\omega_\mathrm{p}$.

For further explanation, we present a resonator-based JPA in Fig.~\ref{fig:paraamp}(a) as an example.
Note that the $\lambda/4$ resonator is terminated by a DC SQUID with a flux bias.
As mentioned in Sec.~\ref{sec:JJ}, a DC SQUID is a variable junction whose effective inductance $L_\mathrm{J,eff}$ is given by
\begin{align}\label{eq:triIden}
L_\mathrm{J,eff}(\varphi_\mathrm{ext}) 
&= 
\left(\frac{\Phi_0}{2\pi} \right)^2 
\frac{1}{E_\mathrm{J,eff}(\varphi_\mathrm{ext})} \nonumber\\
&=
\left(\frac{\Phi_0}{2\pi} \right)^2 
\frac{1}{2 E_\mathrm{J}}
\frac{1}{\left|\cos (\varphi_\mathrm{ext}/2)\right|},
\end{align}
where $\varphi_\textrm{ext} \equiv 2\pi \Phi/\Phi_0$.
Here, Eq.~\eqref{eq:dcSQUID} is used with the assumption that $E_\mathrm{J1} = E_\mathrm{J2} = E_\mathrm{J}$.
By varying the flux bias, we can modulate the inductance [red coil in Fig.~\ref{fig:paraamp}(a)].
This type of modulation is called flux pumping.\cite{yamamoto2008}
Here, we decompose $\varphi_\mathrm{ext}$ into the DC component $\varphi_\mathrm{ext}^\mathrm{dc}$ and the pump component $\varphi_\mathrm{ext}^\mathrm{p}$.
We can operate the amplifier in two regimes depending on $\varphi_\mathrm{ext}^\mathrm{dc}$:
\begin{enumerate}
\itemsep-0.1em

\item If $\varphi_\mathrm{ext}^\mathrm{dc} = 0$ [bias a in Fig.~\ref{fig:paraamp}(b)], $L_\mathrm{J,eff}$ varies quadratically with $\varphi_\mathrm{ext}^\mathrm{p}$ because $1/\cos (x/2) \approx 1 + x^2/8$ for small $x$.
This results in $\omega_\mathrm{m}=2\omega_\mathrm{p}$, where $\omega_\mathrm{p}=\omega_\mathrm{r}$.
Such a process is called a four-wave mixing process ($\omega_\mathrm{s}$, $\omega_\mathrm{I}$, and two $\omega_\mathrm{p}$).
However, frequency modulation at this bias is inefficient because the first derivative of the resonance frequency with respect to the flux bias is zero;
thus, current pumping is typically applied, which is discussed below.

\item For a suitable value of $\varphi_\mathrm{ext}^\mathrm{dc}$, we can have an appreciable contribution from the linear term in Eq.~\eqref{eq:triIden}.
Bias b in Fig.~\ref{fig:paraamp}(b) is an example of this.
In this case, $\omega_\mathrm{m} = \omega_\mathrm{p}$.
The parametric amplification of the signals is then achieved by applying a pump tone with $\omega_\mathrm{p} = 2\omega_\mathrm{r}$.
This process is known as a three-wave mixing process ($\omega_\mathrm{s}$, $\omega_\mathrm{I}$, and one $\omega_\mathrm{p}$). 
The advantages of this operation are that 
(i) we can easily separate the pump tone and the signal in the frequency domain, and 
(ii) we can tune $\omega_\mathrm{r}$ by adjusting $\varphi_\mathrm{ext}^\mathrm{dc}$.

\end{enumerate}

There is another method for inductance modulation called current pumping [green arrows in Fig.~\ref{fig:paraamp}(a)].\cite{vijay2009}
In this method, the inductance of a JPA is modulated by applying a large current, i.e., a pump current, flowing through the Josephson junctions.
Qualitatively, the number of charge carriers, i.e., Cooper pairs, is locally and partially reduced in a Josephson junction owing to its weak link nature.
Because of this, the charge carriers must move faster near the junctions to maintain the same current in and out of the junctions.
The resulting large kinetic energy of the charge carriers contributes to the inductance of the circuit, in addition to the geometric inductance.
This additional inductance is called the kinetic inductance.
Since the kinetic energy is proportional to the square of the velocity of the charge carriers, the kinetic inductance is roughly proportional to the square of the pump current.
Thus, current-pumped amplification is typically a four-wave mixing process regardless of the DC flux bias.\cite{casbel2007}
Such an inductive contribution from current pumping shifts the resonance frequency of the resonator, as shown in Fig.~\ref{fig:paraamp}(c).
This characteristic line shape can be modeled as a Duffing oscillator.\cite{duffing, roy2018}

A resonator-based JPA is relatively easy to make but suffers from a gain-bandwidth trade-off.
The reason is that, to achieve a higher gain, the signal must undergo multiple reflections within the resonator;
this requires a higher quality factor and inevitably reduces the bandwidth.
A waveguide-based JPA is called a Josephson Traveling-Wave Parametric Amplifier (JTWPA).
A JTWPA is free from this gain-bandwidth trade-off.
This allows us to achieve high-bandwidth (several gigahertz) and high-gain ($> 20~\mathrm{dB}$) amplification.\cite{macklin2015}
Since the waveguide has to be nonlinear to mix the signal and pump tones, it is implemented using a long Josephson junction array with current pumping.

The primary challenge in realizing a JTWPA is phase matching, i.e., momentum conservation:
the phases of all interacting tones must be matched.
Considering a JTWPA operated by current pumping (a four-photon process), the wave vectors of the pump $k_\mathrm{p}$, the signal $k_\mathrm{s}$, and the idler $k_\mathrm{I}$ must satisfy $2k_\mathrm{p}=k_\mathrm{s}+k_\mathrm{I}$.
The phase mismatch results from changes in the refractive index caused by the interaction between the strong pump tone and the nonlinear medium.
This leads to $\Delta k=k_\mathrm{s}+k_\mathrm{I}-2k_\mathrm{p} > 0$, as represented by the discrepancy between the dotted and solid black lines in Fig.~\ref{fig:paraamp}(f).
(Resonator-based JPAs are free from this problem due to their geometric confinement.)
To match the phases, we create a local distortion in the dispersion relation.
Then, we can select a pump frequency $\omega_\mathrm{p}$ at which $k_\mathrm{p}$ is larger than its value without dispersion relation engineering [Fig.~\ref{fig:paraamp}(e)], thereby compensating for the mismatch and yielding $\Delta k=0$, as shown by the overlap between the red dashed line and the black solid line in Fig.~\ref{fig:paraamp}(f).
For actual implementation, we can either insert a resonant structure near each Josephson junction\cite{macklin2015, obrien2015, white2015} or periodically modulate the refractive index, similar to what is done in photonic crystals [Fig.~\ref{fig:paraamp}(d)].\cite{planat2020, howe2026}
Here, modulating the refractive index can be achieved by varying the sizes of the junctions in the DC SQUIDs.

These JPAs can perform as quantum-limited amplifiers, which add only the minimum noise allowed by the laws of quantum mechanics.
For a phase-preserving linear amplifier, whose gain is the same regardless of the phase of the input signal, this minimum amount of added noise is equivalent to half a photon.\cite{caves1982, caves2012}
For theories of parametric amplification, including JPAs, see Refs.~\onlinecite{roy2018, aumentado2020, clerk2010, eichler2014, carmichael2, haus}.



\subsection{Single-Qubit Gate}
\label{sec:SQG}

\begin{figure}
\centering
\includegraphics{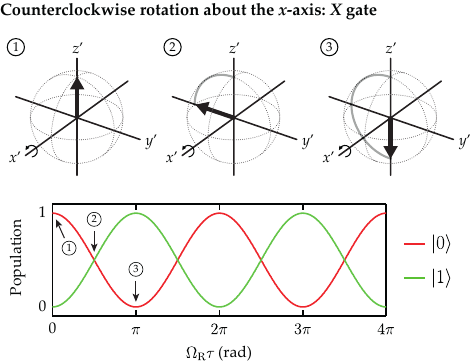}
\caption{Evolution of the quantum state of a single qubit in a Bloch sphere under an external drive.
Gray solid lines in the Bloch spheres indicate the evolution trajectory of the qubit state.
The numbers in circles indicate time steps. The rotation angle of the Bloch vector is determined by $\Omega_\textrm{R} \tau$, where $\Omega_\textrm{R}$ is the Rabi frequency and $\tau$ is the length of the external drive.
The rotation of the Bloch vector results in an oscillation of the qubit population with a period $\Omega_\textrm{R}$; such an oscillation is called the Rabi oscillation.
Note that the population of the qubit is inverted at time \raisebox{.5pt}{\textcircled{\raisebox{-.9pt}{3}}}, thus implementing the $X$ gate.
The frequency of the rotating frame is the drive frequency, which is on-resonance with the qubit.}
\label{fig:Xgate}
\end{figure}

As mentioned in Sec.~\ref{sec:quantumGate}, performing a gate operation is basically engineering the system Hamiltonian such that the resulting unitary evolution of a qubit implements the target gate.
From Eq.~\eqref{eq:uniDef},
\begin{equation}
\hat{U} = \textrm{e}^{-\textrm{i}\hat{\mathcal{H}}t/\hbar} = \hat{U}_\textrm{target}.
\end{equation}
Here, the method we use to engineer the system Hamiltonian is to apply an external drive.

Since the standard qubit readout technique is dispersive readout (Sec.~\ref{sec:readout}), we consider a qubit-resonator system and assume that the qubit is driven via the resonator.
A classical analog of this system is shown in Fig.~\ref{fig:coupledHO}(c).
As we drive oscillator 1 (control oscillator) with $f_\textrm{d} = f_2$, the amplitude of oscillator 2 (target oscillator) increases indefinitely.
However, this does not happen for a qubit because it is intrinsically a nonlinear quantum object.
Instead, the population of the qubit oscillates as a function of time or the amplitude of the drive.
This oscillation in the qubit population is called the Rabi oscillation (Fig.~\ref{fig:Xgate}).

Now we consider the counterclockwise rotation of the Bloch vector about the $x$-axis as an example.
The external drive is usually modeled quantum mechanically\cite{haroche, blais2004}
\begin{equation}\label{eq:drive}
\hat{\mathcal{H}}_\mathrm{d}(t) = 
\hbar \mathcal{E}_\mathrm{r}(t) (\hat{a}^\dagger\mathrm{e}^{-\mathrm{i}\omega_\mathrm{d}t} + \hat{a}\mathrm{e}^{+\mathrm{i}\omega_\mathrm{d}t}),
\end{equation}
where $\mathcal{E}_\mathrm{r}$ and $\omega_\mathrm{d}$ are the amplitude and frequency of the external drive, respectively.
Now, we have to perform multiple transformations.
First, to focus on the dispersive limit, we apply the Schrieffer--Wolff transformation (Sec.~\ref{sec:readout}) to Eq.~\eqref{eq:drive}:
\begin{align}
\hat{\mathcal{H}}_\mathrm{d}^\mathrm{disp} &= 
\hat{U}_\mathrm{disp}\hat{\mathcal{H}}_\mathrm{d}(t)\hat{U}_\mathrm{disp}^\dagger \nonumber\\ 
&\approx 
\hbar \mathcal{E}_\mathrm{r}(t) (\hat{a}^\dagger\mathrm{e}^{-\mathrm{i}\omega_\mathrm{d}t} + \hat{a}\mathrm{e}^{+\mathrm{i}\omega_\mathrm{d}t}) \nonumber\\
&\quad
+ \frac{\hbar \mathcal{E}_\mathrm{r}(t)g}{\Delta_\mathrm{qr}} (\hat{\sigma}_+\mathrm{e}^{-\mathrm{i}\omega_\mathrm{d}t} + \hat{\sigma}_-\mathrm{e}^{+\mathrm{i}\omega_\mathrm{d}t}), \label{eq:driveDisp}
\end{align}
where $\Delta_\mathrm{qr} \equiv \omega_\mathrm{q}-\omega_\mathrm{r}$.
In Eq.~\eqref{eq:driveDisp}, we can see that the first line corresponds to (de)populating the resonator and the second line corresponds to driving the qubit.
Next, we combine Eqs.~\eqref{eq:driveDisp} and \eqref{eq:dispersiveJC2} to obtain the full single-qubit Hamiltonian $\hat{\mathcal{H}}_\mathrm{1q}^\mathrm{full}$:
\begin{align}
\hat{\mathcal{H}}_\mathrm{1q}^\mathrm{full} \equiv
\hat{\mathcal{H}}_\mathrm{JC}^\mathrm{disp} 
+ \hat{\mathcal{H}}_\mathrm{d}^\mathrm{disp}.
\end{align}
Then, we move to the rotating frame whose Hamiltonian is defined by
\begin{equation}
\hat{\mathcal{H}}_0 = \hbar\omega_\mathrm{d} \left( \hat{a}^\dagger \hat{a}
- \frac{\hat{\sigma}_z}{2} \right).
\end{equation}
The final single-qubit Hamiltonian in the rotating frame $\hat{\mathcal{H}}_\mathrm{1q}^\mathrm{rot}$ can be obtained by the unitary transformation:
\begin{align}
\hat{\mathcal{H}}_\mathrm{1q}^\mathrm{rot} &=
\mathrm{e}^{\mathrm{i}\hat{\mathcal{H}}_0 t/\hbar} (\hat{\mathcal{H}}_\mathrm{1q}^\mathrm{full} - \hat{\mathcal{H}}_0) \mathrm{e}^{-\mathrm{i}\hat{\mathcal{H}}_0 t/\hbar} \nonumber\\
&= 
\hbar (\omega_\mathrm{r} - \omega_\mathrm{d}) \hat{a}^\dagger \hat{a}
 + \hbar\mathcal{E}_\mathrm{r}(t)(\hat{a}+\hat{a}^\dagger) \nonumber\\
&\quad
- \hbar[\omega_\mathrm{q} + \chi(1+2\hat{a}^\dagger \hat{a}) - \omega_\mathrm{d}] \frac{\hat{\sigma}_z}{2}
+ \hbar \Omega_\mathrm{R} \frac{\hat{\sigma}_x}{2}, \label{eq:1qCtrl}
\end{align}
where $\Omega_\mathrm{R}$ is the Rabi frequency given by $2\mathcal{E}_\mathrm{r}g/\Delta_\mathrm{qr}$, whose physical meaning will be clarified in Eq.~\eqref{eq:SQG}.
For these transformations, we use the formulas in Table~\ref{tab:formula}.

As mentioned in Sec.~\ref{sec:readout}, the number of photons in the readout resonator has to be close to zero to maintain the phase coherence of the qubit.
Thus, if we choose $\omega_\mathrm{d} = \omega_\mathrm{q} + \chi$, then the term relevant to the qubit dynamics in Eq.~\eqref{eq:1qCtrl} is
\begin{equation}\label{eq:driveRot}
\hat{\mathcal{H}}_\mathrm{d}^\mathrm{rot} =
\hbar\Omega_\mathrm{R} \frac{\hat{\sigma}_x}{2}.
\end{equation}
Using Eq.~\eqref{eq:matrixExp}, we can obtain the evolution driven by Eq.~\eqref{eq:driveRot} for a duration $\tau$:
\begin{align}
\hat{U}(\tau)
&= 
\mathrm{e}^{-\mathrm{i}\hat{\mathcal{H}}_\mathrm{d}^\mathrm{rot}\tau/\hbar} \nonumber\\
&= 
\begin{pmatrix}
\cos(\Omega_\mathrm{R} \tau/2) & -\mathrm{i}\sin(\Omega_\mathrm{R} \tau/2) \\
-\mathrm{i}\sin(\Omega_\mathrm{R} \tau/2) & \cos(\Omega_\mathrm{R} \tau/2) \\
\end{pmatrix}. \label{eq:SQG}
\end{align}
Equation~\eqref{eq:SQG} is a rotation matrix describing the rotation of a Bloch vector around the $x'$-axis, as shown in Fig.~\ref{fig:Xgate}.
This rotation results in an oscillation of the qubit population---the $z'$ component of the Bloch vector---with a period of $\Omega_\mathrm{R}\tau = 2\pi$, if we ignore the global phase factor.
This is known as Rabi oscillation.
Using this, the $X$ gate can be implemented by an external drive satisfying $\Omega_\mathrm{R}\tau = \pi$.
This is the primary mechanism for flipping a qubit.

One might ask how the rotation axis is defined in a real experiment.
The answer is that the reference phase of the instruments, such as the phase of the first pulse in the experiment, defines the rotation axis, which is usually set as the $x$-axis (Ref.~\onlinecite{levitt}).
If we want to change the rotation axis from $x$ to $y$, all we have to do is add a $\pi/2$ phase shift to the subsequent pulses relative to the reference phase of the measurement instruments.
This means that if we set the $x$-axis rotation drive to $\cos(\omega_\mathrm{d} t)$, then the $y$-axis rotation drive is $-\sin(\omega_\mathrm{d} t)$.
(Do not omit the minus sign!)

Note that changing the rotation axis from $x$ to $y$ is actually equivalent to a $z$ rotation.
This suggests that shifting the reference phase of the instruments is functionally equivalent to a rotation about the $z$-axis, which is called a virtual $z$ rotation.\cite{mckay2017}
Since we do not apply a real pulse for this, the virtual $z$ rotation is a nearly perfect and zero-time operation.

\subsection{Two-Qubit Gate}
\label{sec:TQG}

\begin{table*}
\caption{Various two-qubit gates.
From the columns ``Tunable frequency'' to ``Small $\omega_\textrm{q}$ separation'', the name of each column indicates the required condition for implementation.
Here, the condition ``Tunable frequency'' implies that at least one of the qubits has to be out of its sweet spot during the gate operation.
Note that these conditions are minimal conditions; having an additional condition can result in better performance.
For example, it was reported that the CZ gate can be implemented with high fidelity by adiabatic excursion with one tunable qubit and tunable coupling,\cite{chen2014} or by coherent exchange with two tunable qubits and fixed coupling,\cite{barends2019} although the minimal condition for the CZ gate implementation is one tunable qubit as shown in this table.
``Small $\omega_\textrm{q}$ separation'' means how close two qubit frequencies need to be;
the value ``Yes'' means that $|\omega_\textrm{q1}-\omega_\textrm{q2}|$ is comparable to or less than the anharmonicity of the qubit.
``Distance to CNOT'' means the minimum number of two-qubit gates needed to implement the CNOT gate.
Each value in this column is an intrinsic property of the corresponding gate;
values do not depend on the implementation method.
}
\label{tab:TQG}\centering
\begin{ruledtabular}
\begin{tabular}{c c c c c c c c c c c}
\noalign{\smallskip}
Gate	&	Method	&	\begin{tabular}{@{}c@{}}Tunable \\ frequency\end{tabular}	&	\begin{tabular}{@{}c@{}}Tunable \\ coupling\end{tabular}	&	\begin{tabular}{@{}c@{}}Microwave \\ drive\end{tabular}	&	\begin{tabular}{@{}c@{}}Negative \\ anharmonicity\end{tabular}	&	\begin{tabular}{@{}c@{}}Small $\omega_\textrm{q}$ \\ separation\end{tabular}	&	\begin{tabular}{@{}c@{}}Distance \\ to CNOT\end{tabular}	&	Ref. \\
\noalign{\smallskip} \hline \noalign{\smallskip}
iSWAP	&	Coherent exchange		&	Yes	&	No		&	No		&	No		&	No	&	2	& \onlinecite{barends2019} \\
iSWAP	&	Parametric coupling	&	No		&	Yes	&	Yes	&	No		&	No	&	2	&	 \onlinecite{niskanen2007} \\
bSWAP	&	Parametric coupling	&	No		&	Yes	&	Yes	&	No		&	No	&	2	& \onlinecite{niskanen2007} \\
CZ		&	Adiabatic excursion	&	Yes	&	No		&	No		&	Yes	&	No	&	1	& \onlinecite{barends2014},\onlinecite{chen2014} \\
CZ		&	Coherent exchange		&	Yes	&	No		&	No		&	Yes	&	No	&	1	& \onlinecite{barends2019} \\
CZ		&	Parametric coupling	&	No		&	Yes	&	No		&	No		&	No	&	1	& \onlinecite{caldwell2018} \\
CR		&	All microwave control	&	No	&	No		&	Yes	&	No		&	Yes	&	1	& \onlinecite{chow2011},\onlinecite{sheldon2016} \\
\end{tabular}
\end{ruledtabular}
\end{table*}

\begin{figure*}
\centering
\includegraphics{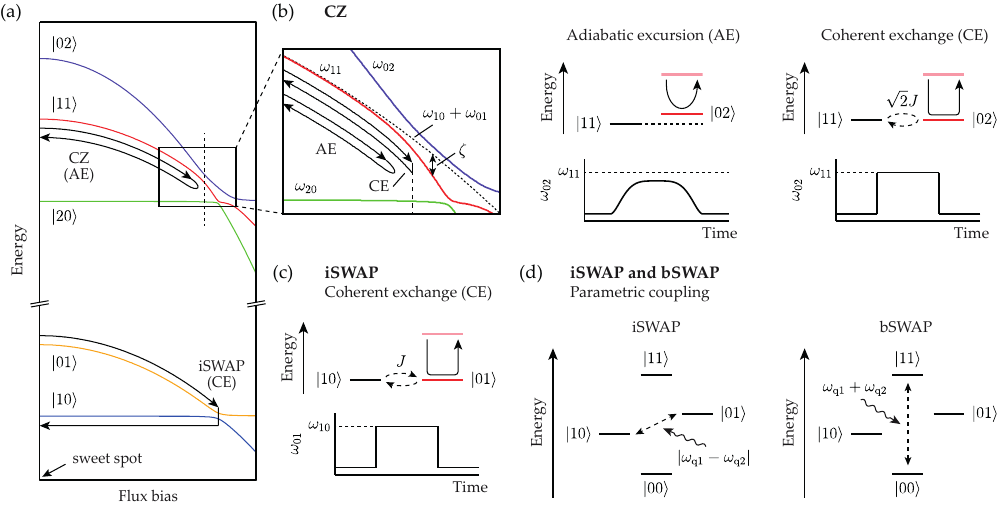}
\caption{(a) Energy levels of a two-transmon system as a function of flux bias.\cite{dicarlo2009}
It is assumed that
(i) qubit 2 is tunable,
(ii) $\omega_\textrm{q2}>\omega_\textrm{q1}$ at the sweet spot, and
(iii) the qubit-qubit coupling is strong as indicated by anticrossings.
The sweet spot of qubit 2 is the leftmost flux bias as indicated by an arrow.
The vertical dashed line in (a) indicates the bias at which the energy levels of $\ket{02}$ and $\ket{11}$ are equal.
The solid lines with arrows show how the energy levels evolve during the gate operation:
the lower one is the trajectory for the iSWAP gate implemented by coherent exchange, and the upper one is for the CZ gate implemented by adiabatic excursion.
Here, CE stands for coherent exchange and AE stands for adiabatic excursion.
(b) Implementation of the CZ gate via adiabatic excursion and coherent exchange.
The left figure shows the energy levels near the anticrossing between $\ket{11}$ and $\ket{02}$ (indicated by the vertical dashed line).
$\zeta$($\equiv \omega_{10}+\omega_{01}-\omega_{11}$) is the effective coupling constant for the $ZZ$ interaction.
In the center and right figures, the upper figure shows the schematic energy level diagram.
The curved solid arrows indicate the relative movement of $\omega_{02}$ with respect to $\omega_{11}$.
The lower figure shows the same movement as a function of time.
The dashed arrow for CE indicates the population swapping to $\ket{02}$ and then back to $\ket{11}$ at the rate $\sqrt{2}J/\pi$ when $\omega_{11}=\omega_{02}$.
Note that, for adiabatic excursion, $\ket{02}$ does not meet $\ket{11}$ to satisfy adiabaticity.
(c) Implementation of the iSWAP gate via coherent exchange.
The upper figure shows the schematic energy level diagram.
The curved solid arrow indicates the relative movement of $\omega_{01}$ with respect to $\omega_{10}$.
The lower figure shows the same movement as a function of time.
The dashed arrows present the two qubits exchanging their energy at the rate $J/\pi$ when $\omega_\textrm{10}=\omega_\textrm{01}$.
(d) Implementation of the iSWAP and bSWAP gates via parametric coupling.
Here, neither qubit needs to be tuned.
The gate operation is performed by applying a microwave to the tunable coupler with the frequency $|\omega_\textrm{q1}-\omega_\textrm{q2}|$ (iSWAP) or $\omega_\textrm{q1}+\omega_\textrm{q2}$ (bSWAP).
}
\label{fig:TQG}
\end{figure*}

Many two-qubit gates have been implemented by various methods.
Among them, four gates and four methods are introduced in this tutorial.
These methods are based on the transverse qubit-qubit interaction whose physics is explored in Sec.~\ref{sec:classicalCoupling}.
The two-qubit gates introduced in this tutorial are summarized in Table~\ref{tab:TQG}.

\subsubsection{iSWAP: Coherent Exchange}
\label{sec:CE_iSWAP}

This method implements a two-qubit gate using changes in the phase and population of the qubit states during the coherent exchange of a photon.
The basic mechanism for this is tuning the transition frequency of one of the qubits so that $\omega_\textrm{q1} = \omega_\textrm{q2}$.
The relevant analogy for this is shown in Fig.~\ref{fig:coupledHO}(a).

Consider two capacitively coupled transmons.
This system is convenient because the dominant qubit-qubit coupling is transverse, as pointed out in Sec.~\ref{sec:qrCoupling}, and it is the most popular qubit system.
For simplicity, we ignore the readout resonators as qubits are usually far detuned from the resonators.
Then, the system Hamiltonian reduces to Eq.~\eqref{eq:2q_H}.
After taking the RWA (Sec.~\ref{sec:qrCoupling}), Eq.~\eqref{eq:2q_H} can be written in a similar manner to Eq.~\eqref{eq:JCH}:
\begin{align}\label{eq:2q_RWA}
\hat{\mathcal{H}}_\textrm{2q} =
- \hbar\omega_\textrm{q1}\frac{\hat{\sigma}_z^{(1)}}{2}
- \hbar\omega_\textrm{q2}\frac{\hat{\sigma}_z^{(2)}}{2}
+ \hbar J \big(
\hat{\sigma}_+^{(1)} \hat{\sigma}_-^{(2)} 
+ \hat{\sigma}_-^{(1)} \hat{\sigma}_+^{(2)}
\big).
\end{align}
The resulting energy levels are shown in Fig.~\ref{fig:TQG}(a).

Now, we assume that $\omega_\textrm{q1} = \omega_\textrm{q2}$.
To focus on the dynamics induced by the qubit-qubit coupling, we move to the rotating frame with the frequency $\omega_\textrm{q1}$.
Then, Eq.~\eqref{eq:2q_RWA} becomes
\begin{align}\label{eq:qq}
\hat{\mathcal{H}}_\textrm{2q}^\textrm{rot} = 
\hbar J \big(
\hat{\sigma}_+^{(1)} \hat{\sigma}_-^{(2)} 
+ \hat{\sigma}_-^{(1)} \hat{\sigma}_+^{(2)}
\big).
\end{align}
This equation is the same as Eq.~\eqref{eq:2qRot} with the RWA.
The time-evolution operator based on $\hat{\mathcal{H}}_\textrm{2q}^\textrm{rot}$ with the time interval $\tau$ is
\begin{align}
\hat{U}(\tau) 
&= \textrm{e}^{-\textrm{i}\hat{\mathcal{H}}_\textrm{2q}^\textrm{rot}\tau/\hbar} \nonumber\\
&=
\begin{pmatrix}
1 & 0 & 0 & 0 \\
0 & \cos(J\tau) & -\textrm{i}\sin(J\tau) & 0 \\
0 & -\textrm{i}\sin(J\tau) & \cos(J\tau) & 0 \\
0 & 0 & 0 & 1 \\
\end{pmatrix}. \label{eq:coherentEx}
\end{align}
Note that Eq.~\eqref{eq:coherentEx} is practically the same as Eq.~\eqref{eq:SQG} because only two states, $\ket{01}$ and $\ket{10}$, contribute to the dynamics.

Equation~\eqref{eq:coherentEx} is easy to derive if we use Eq.~\eqref{eq:matrixExp} with $\alpha = J\tau$ and
\begin{align}
\hat{\sigma}_+^{(1)}\hat{\sigma}_-^{(2)} + \hat{\sigma}_-^{(1)}\hat{\sigma}_+^{(2)}
=
\begin{pmatrix}
0 & 0 & 0 & 0 \\
0 & 0 & 1 & 0 \\
0 & 1 & 0 & 0 \\
0 & 0 & 0 & 0 \\
\end{pmatrix}
\equiv
\begin{pmatrix}
0 & 0 & 0 & 0 \\
0 &  &  & 0 \\
0 & \multicolumn{2}{c}{\smash{\raisebox{.5\normalbaselineskip}{$\hat{A}$}}} & 0 \\
0 &  0 & 0 & 0
\end{pmatrix}. \label{eq:preMatrixExp}
\end{align}

One useful two-qubit gate that can be implemented easily by this method is the iSWAP gate.
The iSWAP gate swaps the populations of $\ket{10}$ and $\ket{01}$ with an additional phase factor $-$i. In the matrix form,
\begin{equation}\label{eq:iSWAP}
\hat{U}_\textrm{iSWAP} =
\begin{pmatrix}
1 & 0 & 0 & 0 \\
0 & 0 & -\textrm{i} & 0 \\
0 & -\textrm{i} & 0 & 0 \\
0 & 0 & 0 & 1 \\
\end{pmatrix}.
\end{equation}
Thus, Eq.~\eqref{eq:coherentEx} with $J\tau = \pi/2$ implements Eq.~\eqref{eq:iSWAP}.

The actual implementation can be done via the following steps [black arrowed path in Fig.~\ref{fig:TQG}(a) and lower figure in Fig.~\ref{fig:TQG}(c)]:
(i) Prepare the initial state.
In Fig.~\ref{fig:TQG}, $\ket{01}$ was chosen as an example.
At this stage, the tunable qubit, qubit 2 in this case, is at its sweet spot.
(ii) Increase the flux bias to the point at which the energy levels of $\ket{10}$ and $\ket{01}$ are equal.
At this bias, the new eigenstates, $(\ket{01} \pm \ket{10})/\sqrt{2}$, exchange their energy at a rate of $J/\pi$.
(iii) Wait for a while to satisfy $J\tau = \pi/2$.
(iv) Decrease the flux bias to the sweet spot.

The bias ramping in steps (ii) and (iv) must be as fast as possible for efficient gate operation. 
Thus, the iSWAP gate implemented using this method is a diabatic gate.
For qubits with negative anharmonicity, such as the two-transmon system shown in Fig.~\ref{fig:TQG}(a), the gate fidelity of a diabatic gate is limited mainly by population leakage out of the computational subspace.
This population leakage is driven by unwanted transitions, such as $\ket{11}$-$\ket{02}$ and $\ket{11}$-$\ket{20}$ transitions, because we pass anticrossings associated with $\ket{11}$ during the flux bias ramping.
Note that this leakage is still a unitary evolution, suggesting that the population leakage oscillates with time.
Thus, the error due to this leakage can be minimized by synchronizing the periods of the leakage and the iSWAP gate time.\cite{barends2019}

\subsubsection{iSWAP and bSWAP: Parametric Coupling}
\label{sec:PC_iSWAP}

The previous implementation of the iSWAP gate relies on the frequency tunability of the qubit.
This means that the qubit must be out of its sweet spot for a while, which potentially degrades the coherence.
To resolve this issue, another scheme that allows qubits to stay at their sweet spots during the gate operation has been developed.\cite{niskanen2006}

In this scheme, the control knob is the qubit-qubit coupling.
Consider the two-qubit Hamiltonian in Eq.~\eqref{eq:2q_H}.
(Here, the qubits do not need to be transmons.)
If $J$ is static and $|\omega_\textrm{q1}-\omega_\textrm{q2}| \gg J$, the qubit-qubit interaction is effectively turned off, i.e., the interaction is very slow compared with the time scale we are interested in, as indicated by in Eq.~\eqref{eq:2qRot}.

Now we modulate the coupling constant as $J(t) = J_0 + J_\textrm{m}\cos(\omega_\textrm{m}t)$.
Here, the modulation frequency $\omega_\textrm{m}$ of the coupler is the control parameter, from which the name ``parametric coupling'' originates.
Such a modulation can be achieved by modulating flux passing through the loop of the coupler shown in Fig.~\ref{fig:couplingQQ}.
If $\omega_\textrm{m} = |\omega_\textrm{q1}-\omega_\textrm{q2}|$, the two qubits exchange their energy as we saw in Fig.~\ref{fig:coupledHO}(b).
This activates the transition between $\ket{01}$ and $\ket{10}$.
In this case, Eq.~\eqref{eq:2qRot} becomes
\begin{align}
\hat{\mathcal{H}}_\textrm{2q}^\textrm{rot} = 
\frac{\hbar J_\textrm{m}}{2}
\left( \hat{\sigma}_+^{(1)} \hat{\sigma}_-^{(2)}
+ \hat{\sigma}_-^{(1)} \hat{\sigma}_+^{(2)} \right).
\end{align}
Using Eq.~\eqref{eq:coherentEx}, we can see that $J_\textrm{m}\tau = \pi$ implements the iSWAP gate.

If $\omega_\textrm{m} = \omega_\textrm{q1} + \omega_\textrm{q2}$, we can activate the transition between $\ket{00}$ and $\ket{11}$.
Using this transition, we can implement the bSWAP gate:\cite{niskanen2007, poletto2012}
\begin{equation}\label{eq:bSWAP}
\hat{U}_\textrm{bSWAP} =
\begin{pmatrix}
0 & 0 & 0 & -\textrm{i} \\
0 & 1 & 0 & 0 \\
0 & 0 & 1 & 0 \\
-\textrm{i} & 0 & 0 & 0 \\
\end{pmatrix}.
\end{equation}
In this case, Eq.~\eqref{eq:2qRot} becomes
\begin{align}
\hat{\mathcal{H}}_\textrm{2q}^\textrm{rot} = 
\frac{\hbar J_\textrm{m}}{2}
\left( \hat{\sigma}_+^{(1)} \hat{\sigma}_+^{(2)}
+ \hat{\sigma}_-^{(1)} \hat{\sigma}_-^{(2)} \right).
\end{align}
The time-evolution operator based on this Hamiltonian is
\begin{align}
\hat{U}(\tau) 
=
\begin{pmatrix}
\cos(J_\textrm{m}\tau/2) & 0 & 0 & -\textrm{i}\sin(J_\textrm{m}\tau/2) \\
0 & 1 & 0 & 0 \\
0 & 0 & 1 & 0 \\
-\textrm{i}\sin(J_\textrm{m}\tau/2) & 0 & 0 & \cos(J_\textrm{m}\tau/2) \\
\end{pmatrix}.
\end{align}
It is clear that $J_\textrm{m}\tau = \pi$ implements the bSWAP gate.

\subsubsection{CZ: Adiabatic Excursion}
\label{sec:AE_CZ}

Any two-qubit gate based on the transverse interaction has to be performed at least twice to implement the CNOT gate.\cite{schuch2003}
Since the CNOT gate is the essential gate for quantum error correction (see Sec.~\ref{sec:QEC}), a more efficient implementation of the CNOT gate is desired.
For this, we need a non-transverse qubit-qubit interaction.
It was soon realized that we have an effective $ZZ$ interaction that comes from the transverse interaction associated with higher excitation levels.\cite{dicarlo2009}
This interaction allows us to implement the controlled-$Z$ (CZ) gate, which is identical to the CNOT gate up to single-qubit rotations:
\begin{align}
\begin{array}{c}
\Qcircuit @C=1em @R=1em {
& \ctrl{1}	& \qw	& \raisebox{-2em}{=}	& & \qw 		& \ctrl{1}  & \qw & \qw   \\
& \targ		& \qw	& 						& & \gate{H}	& \control\qw & \gate{H} & \qw 
} \label{eq:CNOT_CZ}\\
\end{array},
\end{align}
where 
\begin{equation}\label{eq:CZ}
\hat{U}_\textrm{CZ}
= 
\begin{pmatrix}
1 & 0 & 0 & 0 \\
0 & 1 & 0 & 0 \\
0 & 0 & 1 & 0 \\
0 & 0 & 0 & -1 \\
\end{pmatrix}
\equiv
\def\arraystretch{0.01}
\begin{array}{c}
\Qcircuit @C=1em @R=1.3em {
&	\ctrl{1}		& \qw	\\
&	\control\qw	& \qw 
}\\
\end{array}.
\end{equation}
Equation~\eqref{eq:CZ} is read ``apply the $Z$ gate to qubit 2 if the state of qubit 1 is $\ket{1}$.''

As we explained in Sec.~\ref{sec:qqCoupling}, if $J$ is static and $\omega_\textrm{q1}\neq\omega_\textrm{q2}$, the contribution from the transverse qubit-qubit interaction to the dynamics is small [Eq.~\eqref{eq:2qRot}].
In this case, the time-evolution operator has the form 
\begin{equation}\label{eq:CZgen}
\hat{U} = 
\begin{pmatrix}
1 & 0 & 0 & 0 \\
0 & \textrm{e}^{\textrm{i}\theta_{01}} & 0 & 0 \\
0 & 0 & \textrm{e}^{\textrm{i}\theta_{10}} & 0 \\
0 & 0 & 0 & \textrm{e}^{\textrm{i}\theta_{11}} \\
\end{pmatrix},
\end{equation}
where $\theta_{ij}$ is the phase of the state $\ket{ij}$ acquired throughout the evolution for the time interval $\tau$: $\theta_{ij} = \int_0^\tau \omega_{ij}(t) dt$.
Therefore, to implement the CZ gate, we need to have $\theta_{11} - \theta_{10} - \theta_{01} = \pi$.

If we consider the computational subspace only, the difficulty in implementing the CZ gate is that $\omega_{11}$ is always the same as $\omega_{10} + \omega_{01}$; thus, $\theta_{11}=\theta_{10}+\theta_{01}$.
This is why we need a $ZZ$ interaction.
The crucial observation is that $\omega_{11}$ deviates from $\omega_{10} + \omega_{01}$ because of the anticrossing between the $\ket{11}$ and $\ket{02}$ levels [left figure in Fig.~\ref{fig:TQG}(b)].
This gives an effective $ZZ$ interaction whose strength $\zeta$ is $\omega_{10} + \omega_{01} - \omega_{11}$.
As shown in the left figure in Fig.~\ref{fig:TQG}(b), $\zeta$ increases rapidly as the energy levels of $\ket{02}$ and $\ket{11}$ become closer, suggesting that varying the flux bias can be used to tune $\zeta$ by orders of magnitude.\cite{dicarlo2009}

Now we have the $ZZ$ interaction.
Note that, in Eq.~\eqref{eq:CZ}, the population of each state must remain the same.
One way to implement the CZ gate is to tune the energy levels adiabatically.
In addition, another condition to make the CZ gate operation feasible is negative anharmonicity such that the $\ket{11}$-$\ket{02}$ anticrossing appears earlier than the $\ket{10}$-$\ket{01}$ anticrossing.
Otherwise, the transition between $\ket{10}$ and $\ket{01}$ will be activated during the gate operation, resulting in an unwanted population change.
Therefore, the most suitable qubit for the CZ gate operation is a transmon and its variants.

Implementing a high-fidelity and fast CZ gate is then reduced to finding an optimal trajectory satisfying $\int_0^\tau \zeta(t) dt = \pi$.
In general, the flux must be ramped up fast at the beginning to reduce the gate time.
Near the anticrossing, the flux sweep is relatively slow to satisfy adiabaticity and acquire the phase we need [center figure in Fig.~\ref{fig:TQG}(b)].
It was found that the Slepian shape is close to the optimal trajectory.\cite{martinis2014}

\subsubsection{CZ: Coherent Exchange and Parametric Coupling}
\label{sec:CE_CZ}

The CZ gate can also be implemented by coherent exchange between $\ket{11}$ and $\ket{02}$ to acquire the phase factor [right figure in Fig.~\ref{fig:TQG}(b)].\cite{lu2012, barends2019}
Regarding coherent exchange, when $\omega_{11} = \omega_{02}$, we can construct a time-evolution operator similar to Eq.~\eqref{eq:coherentEx}:
\begin{align}
&\hat{U}(\tau) = \nonumber\\
&
\raise 1.6ex \hbox{$
\begin{array}{l c cccccc c}
& & \bra{00} & \bra{01} & \bra{10} & \bra{11} & \bra{02} & \bra{20} & \\[3pt]
\ket{00}\phantom{,}\!\!\!\!\! & \multirow{6}{*}{\Vast(\!\!} & 1 & 0 & 0 & 0 & 0 & 0 & \multirow{6}{*}{\!\!\Vast)} \\
\ket{01}\!\!\!\!\! & & 0 & 1 & 0 & 0 & 0 & 0 & \\
\ket{10}\!\!\!\!\! & & 0 & 0 & 1 & 0 & 0 & 0 & \\
\ket{11}\!\!\!\!\! & & 0 & 0 & 0 & \cos(\sqrt{2}J\tau) & -\textrm{i}\sin(\sqrt{2}J\tau) & 0 & \\
\ket{02}\!\!\!\!\! & & 0 & 0 & 0 & -\textrm{i}\sin(\sqrt{2}J\tau) & \cos(\sqrt{2}J\tau) & 0 & \\
\ket{20}\!\!\!\!\! & & 0 & 0 & 0 & 0 & 0 & 1 & 
\end{array}$}. \label{eq:coherentEx_CZ}
\end{align}
One difference from Eq.~\eqref{eq:coherentEx} is that the coupling constant is $\sqrt{2}J$ instead of $J$.
In general, the coupling constant for the transverse interaction between $\ket{n_1,n_2-1}$ and $\ket{n_1-1,n_2}$ is $J\sqrt{n_1 n_2}$ because $\hat{a}\ket{n} = \sqrt{n}\ket{n-1}$ and $\hat{a}^\dagger\ket{n} = \sqrt{n+1}\ket{n+1}$.
What we want is $\cos(\sqrt{2}J\tau) = -1$; thus, $\sqrt{2}J\tau = \pi$ implements the CZ gate.

Parametric coupling can also be used.\cite{caldwell2018}
Since we do not need to tune the qubit frequencies, negative anharmonicity is not required.
Moreover, we can use the coupling between $\ket{11}$ and either $\ket{02}$ or $\ket{20}$.

\subsubsection{CR: All Microwave Control}
\label{sec:CR}

All implementations of two-qubit gates discussed thus far require some tunability of the qubit frequency or qubit-qubit coupling.
However, the phase coherence is so delicate that any control line potentially degrades $T_2$.
This motivates the development of two-qubit gates purely driven by microwave activation.
Among them, the cross-resonance (CR) gate is the most widely used one.

The CR gate basically excites one qubit (target qubit) through the other qubit (control qubit).
Hence, it is similar to a single-qubit gate, and its classical analogy is Fig.~\ref{fig:coupledHO}(c).
The difference is that, instead of a harmonic oscillator (resonator), a nonlinear oscillator (qubit) is used as a control oscillator.\cite{tripathi2019}
Because of the nonlinearity, the result of the gate operation depends on the state of the control qubit, resulting in a two-qubit gate.

In the following, we repeat the calculations we made in Sec.~\ref{sec:SQG} with slight modifications for a two-qubit system instead of a qubit-resonator system.
We make three assumptions.
First, $J \ll \Delta_\textrm{qq}$, where $\Delta_\textrm{qq} \equiv \omega_\textrm{q1}-\omega_\textrm{q2}$.
This is a kind of dispersive limit.
Secondly, the control qubit, which is assumed to be qubit 1, is an ideal two-level system, i.e., $\alpha_\textrm{1} \gg \Delta_\textrm{qq}$, where $\alpha_\textrm{1}$ is the anharmonicity of qubit 1.
Lastly, there is no spurious cross-talk between the two qubits.

Since the first assumption is also a kind of dispersive limit, we can apply the Schrieffer--Wolff transformation,
\begin{equation}\label{eq:dispersiveTrans_CR}
\hat{U}_\mathrm{disp} = \exp[\frac{J}{\Delta_\mathrm{qq}} (\hat{\sigma}_+^{(1)} \hat{\sigma}_-^{(2)} - \hat{\sigma}_-^{(1)} \hat{\sigma}_+^{(2)} ) ],
\end{equation}
to $\hat{\mathcal{H}}_\mathrm{2q}$ in Eq.~\eqref{eq:2q_RWA} as we did in Sec.~\ref{sec:readout}:
\begin{align}
\hat{\mathcal{H}}_\mathrm{2q}^\mathrm{disp} &= 
\hat{U}_\mathrm{disp}\hat{\mathcal{H}}_\mathrm{2q}\hat{U}_\mathrm{disp}^\dagger \nonumber\\ 
&\approx 
- \hbar \left(\omega_\mathrm{q1} + \frac{J^2}{\Delta_\mathrm{qq}} \right)
\frac{\hat{\sigma}_z^{(1)}}{2}
- \hbar \left(\omega_\mathrm{q2} - \frac{J^2}{\Delta_\mathrm{qq}} \right)
\frac{\hat{\sigma}_z^{(2)}}{2}.
\label{eq:dispersiveQQ}
\end{align}

Regarding the external drive, we use a Hamiltonian similar to Eq.~\eqref{eq:drive}:
\begin{equation}\label{eq:drive_CR}
\hat{\mathcal{H}}_\mathrm{d}(t) = 
\hbar\mathcal{E}_\mathrm{q}(t) (\hat{\sigma}_+^{(1)} \mathrm{e}^{-\mathrm{i}\omega_\mathrm{d}t} 
+ \hat{\sigma}_-^{(1)} \mathrm{e}^{+\mathrm{i}\omega_\mathrm{d}t}),
\end{equation}
where $\mathcal{E}_\mathrm{q}$ and $\omega_\mathrm{d}$ are the amplitude and frequency of the external drive, respectively.
We apply the same Schrieffer--Wolff transformation to Eq.~\eqref{eq:drive_CR}:
\begin{align}
\hat{\mathcal{H}}_\mathrm{d}^\mathrm{disp} &= 
\hat{U}_\mathrm{disp}\hat{\mathcal{H}}_\mathrm{d}(t)\hat{U}_\mathrm{disp}^\dagger \nonumber\\ 
&\approx 
\hbar\mathcal{E}_\mathrm{q}(t) (\hat{\sigma}_+^{(1)} \mathrm{e}^{-\mathrm{i}\omega_\mathrm{d}t} + \hat{\sigma}_-^{(1)} \mathrm{e}^{+\mathrm{i}\omega_\mathrm{d}t}) \nonumber\\
&\quad
- \frac{\hbar\mathcal{E}_\mathrm{q}(t)J}{\Delta_\mathrm{qq}} \hat{\sigma}_z^{(1)}(\hat{\sigma}_+^{(2)}\mathrm{e}^{-\mathrm{i}\omega_\mathrm{d}t} + \hat{\sigma}_-^{(2)}\mathrm{e}^{+\mathrm{i}\omega_\mathrm{d}t}). \label{eq:driveDisp_CR}
\end{align}
In Eq.~\eqref{eq:driveDisp_CR}, we can see that the second line corresponds to driving qubit 2.

Next, we combine Eqs.~\eqref{eq:dispersiveQQ} and \eqref{eq:driveDisp_CR} to obtain the full two-qubit Hamiltonian $\hat{\mathcal{H}}_\mathrm{2q}^\mathrm{full}$:
\begin{align}
\hat{\mathcal{H}}_\mathrm{2q}^\mathrm{full} \equiv
\hat{\mathcal{H}}_\mathrm{2q}^\mathrm{disp} 
+ \hat{\mathcal{H}}_\mathrm{d}^\mathrm{disp}.
\end{align}
Now, we move to the rotating frame whose Hamiltonian is defined by
\begin{equation}
\hat{\mathcal{H}}_0 = 
- \hbar\omega_\mathrm{d} \left( \frac{\hat{\sigma}_z^{(1)}}{2}
+ \frac{\hat{\sigma}_z^{(2)}}{2} \right).
\end{equation}
The final two-qubit Hamiltonian in the rotating frame $\hat{\mathcal{H}}_\mathrm{2q}^\mathrm{rot}$ can be obtained by the unitary transformation:
\begin{align}
\hat{\mathcal{H}}_\mathrm{2q}^\mathrm{rot} &=
\mathrm{e}^{\mathrm{i}\hat{\mathcal{H}}_0 t/\hbar} (\hat{\mathcal{H}}_\mathrm{2q}^\mathrm{full} - \hat{\mathcal{H}}_0) \mathrm{e}^{-\mathrm{i}\hat{\mathcal{H}}_0 t/\hbar} \nonumber\\
&=
- \hbar \left(\omega_\mathrm{q1} + \frac{J^2}{\Delta_\mathrm{qq}} - \omega_\mathrm{d} \right)
\frac{\hat{\sigma}_z^{(1)}}{2} 
+ \hbar\mathcal{E}_\mathrm{q}(t) \hat{\sigma}_x^{(1)} \nonumber\\
&\quad
- \hbar \left(\omega_\mathrm{q2} - \frac{J^2}{\Delta_\mathrm{qq}} - \omega_\mathrm{d} \right)
\frac{\hat{\sigma}_z^{(2)}}{2} 
+ \hbar\Omega_\mathrm{CR}(t) \hat{\sigma}_z^{(1)} \hat{\sigma}_x^{(2)}, \label{eq:2qCtrl}
\end{align}
where $\Omega_\mathrm{CR} \equiv -\mathcal{E}_\mathrm{q}J/\Delta_\mathrm{qq}$.
For these transformations, we use the formulas in Table~\ref{tab:formula}.

If we set $\omega_\mathrm{d} = \omega_\mathrm{q2} - J^2/\Delta_\mathrm{qq}$, the only term contributing to the two-qubit gate operation is the last term in Eq.~\eqref{eq:2qCtrl}, which we denote as $\hat{\mathcal{H}}_\mathrm{d}^\mathrm{rot}$:
\begin{equation}\label{eq:CRdriveRot}
\hat{\mathcal{H}}_\mathrm{d}^\mathrm{rot}
= \hbar\Omega_\mathrm{CR} \hat{\sigma}_z^{(1)} \hat{\sigma}_x^{(2)}.
\end{equation}
Thus, the time-evolution operator is
\begin{align}
&\hat{U}(\tau)
= \mathrm{e}^{-\mathrm{i}\hat{\mathcal{H}}_\mathrm{d}^\mathrm{rot} \tau/\hbar} \nonumber\\ 
&=
\begin{pmatrix} 
\cos(\Omega_\mathrm{CR}\tau) & -\mathrm{i}\sin(\Omega_\mathrm{CR}\tau) & 0 & 0 \\ 
-\mathrm{i}\sin(\Omega_\mathrm{CR}\tau) & \cos(\Omega_\mathrm{CR}\tau) & 0 & 0 \\ 
0 & 0 & \cos(\Omega_\mathrm{CR}\tau) & \mathrm{i}\sin(\Omega_\mathrm{CR}\tau) \\ 
0 & 0 & \mathrm{i}\sin(\Omega_\mathrm{CR}\tau) & \cos(\Omega_\mathrm{CR}\tau)
\end{pmatrix}.
\end{align}
%
The above matrix is read as ``rotate the target qubit state $+\Omega_\mathrm{CR}\tau$ about the $x$-axis if the control qubit state is $\ket{0}$, and rotate the target qubit state $-\Omega_\mathrm{CR}\tau$ about the $x$-axis if the control qubit state is $\ket{1}$.''

Note that the CR gate is related to the CNOT gate by only two single-qubit rotations:\cite{rigetti2010}
\begin{align}
\begin{array}{c}
\Qcircuit @C=1em @R=1em {
& \ctrl{1}	&	\qw	
& \raisebox{-2.9em}{=}	&	
& \gate{R_z(-\frac{\pi}{2})}	& \ctrl{1} 	
& \qw 				& \qw   \\
& \targ		&	\qw	
&								&	
& \qw	& \gate{\textrm{CR}(\frac{\pi}{2})}											
& \gate{R_x(-\frac{\pi}{2})}	& \qw 
} \label{eq:CR_CNOT}\\
\end{array},
\end{align}
where
\begin{align}\label{eq:CR}
\begin{array}{c}
\Qcircuit @C=1em @R=1em {	
& \ctrl{1}							& \qw	\\
& \gate{\textrm{CR}(\frac{\pi}{2})}	& \qw 
}\\
\end{array}
=
\hat{U} \left( \Omega_\textrm{CR}\tau = \frac{\pi}{2} \right)
=
\frac{1}{\sqrt{2}}
\begin{pmatrix}
1 & -\textrm{i} & 0 & 0 \\
-\textrm{i} & 1 & 0 & 0 \\
0 & 0 & 1 & \textrm{i} \\
0 & 0 & \textrm{i} & 1 \\
\end{pmatrix}.
\end{align}

For a qubit such as a transmon, the assumption of an ideal two-level system is not valid.
In this case, we have to consider higher-level contributions.
As the detailed calculation is quite involved, we simply quote the result:\cite{allen2017, tripathi2019}
\begin{align}\label{eq:2qCtrl_higher}
\hat{\mathcal{H}}_\mathrm{d}^\mathrm{rot}
=
\hbar \frac{\mathcal{E}_\mathrm{q}J}{\Delta_\mathrm{qq}+\alpha_1}
\hat{\sigma}_x^{(2)}
+ \hbar \frac{\alpha_1}{\Delta_\mathrm{qq}+\alpha_1}
\Omega_\mathrm{CR} \hat{\sigma}_z^{(1)} \hat{\sigma}_x^{(2)}.
\end{align}
Note that Eq.~\eqref{eq:2qCtrl_higher} becomes Eq.~\eqref{eq:CRdriveRot} in the $\alpha_1 \rightarrow \infty$ limit (ideal two-level system).
In the $\alpha_1 \rightarrow 0$ limit (bosonic system), only the single-qubit drive term (the first term) survives---Eq.~\eqref{eq:2qCtrl_higher} becomes Eq.~\eqref{eq:driveRot}.
Equation~\eqref{eq:2qCtrl_higher} suggests that if $\Delta_\mathrm{qq} \gg \alpha_1$, the CR drive amplitude is greatly reduced.
Therefore, to have a reliable CR drive amplitude, $\Delta_\mathrm{qq}$ must be comparable to $\alpha_1$;
conversely, if $\Delta_\mathrm{qq}$ is too small, the states of the two qubits are hybridized such that the single-qubit gate operation will be nontrivial.
In the actual operation, the unwanted single-qubit drive term in Eq.~\eqref{eq:2qCtrl_higher} can be canceled out by applying an additional pulse.\cite{sheldon2016}

One might wonder about the fate of the effective $ZZ$ interaction, which was a hero for the CZ gate (Sec.~\ref{sec:AE_CZ}).
Here, it turns into a villain that degrades the fidelity of the CR gate.
This unwanted interaction can be canceled out by a refocusing technique,\cite{sheldon2016} which will be explained in Sec.~\ref{sec:refocusing}.

\subsection{Initialization}
\label{sec:reset}

After computation, qubits should be initialized for the next computation.
However, condition~\ref{list:relaxTime} in Sec.~\ref{sec:qubitCriteria} and fast initialization seem to be contradictory.
Here, we introduce two categories of qubit initialization and their working principles.

\subsubsection{Entropy Dumping}
\label{sec:resetDumping}

\begin{figure*}
\centering
\includegraphics{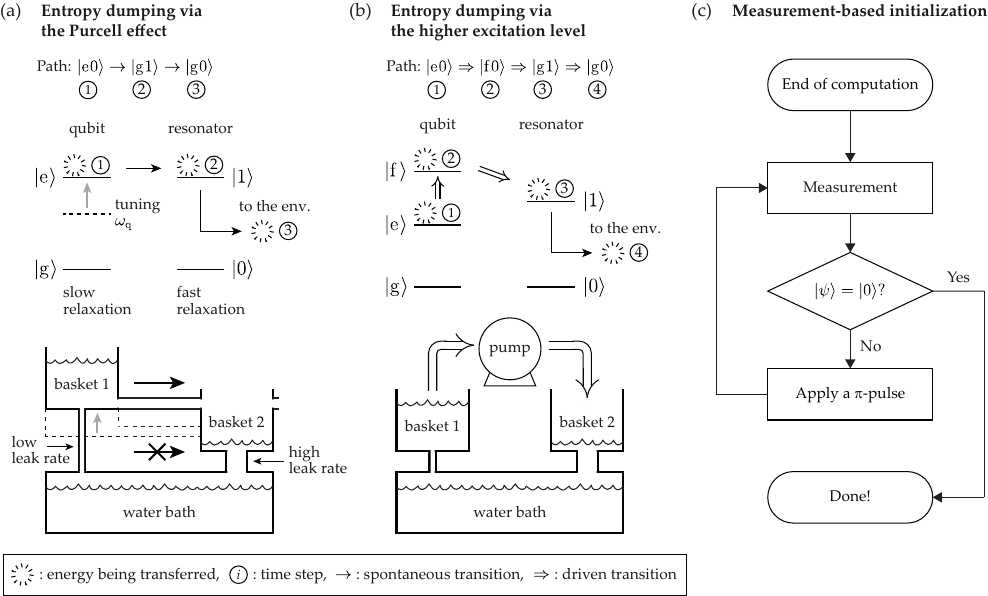}
\caption{Qubit initialization methods.
(a) and (b) Entropy-dumping-based qubit initialization via the Purcell effect (a) and via the higher excitation level $\ket{\textrm{f\,}}$ (b).
The basic strategies for both methods are the same: transfer the energy in the qubit to a resonator whose relaxation process is significantly faster than that of the qubit.
The difference is that, in (a), the qubit state is tuned to satisfy $\omega_\textrm{q} = \omega_\textrm{r}$, such that the energy is transferred directly from qubit state $\ket{\textrm{e}}$ to resonator state $\ket{1}$ via the Purcell effect,
while in (b), all energy levels are fixed and the energy is transferred through $\ket{\textrm{f\,}}$ by our microwave drives.
The diagram named ``Path'' shows the state of the qubit-resonator system during the initialization process.
Below this, the energy levels of the qubit and resonator are shown.
The lower figures show mechanical analogies.
Baskets 1 and 2 correspond to the qubit and resonator, respectively.
Water in each basket corresponds to the entropy of each system.
In (a), the vertical position of the pipe connecting the two baskets must be matched to the position of the holes on the side of basket 2 to yield the water flow.
This represents the condition for the Purcell effect, $\omega_\textrm{q} = \omega_\textrm{r}$.
The dashed boundary represents the moment at which this condition is not satisfied.
In (b), the pump represents our active microwave radiation to the system to activate the transitions $\ket{\textrm{e}0} \leftrightarrow \ket{\textrm{f}0}$ and $\ket{\textrm{f}0} \leftrightarrow \ket{\textrm{g}1}$.
(c) Flow chart describing the measurement-based initialization.}
\label{fig:reset}
\end{figure*}

This method is to pump entropy of the target system to another system, which we call the pumping system, interacting with the target system.
Here, the required condition is that the relaxation of the pumping system has to be much faster than that of the target system.
This idea has been used in magnetic resonance for a long time, with the name ``Dynamic Nuclear Polarization (DNP).''\cite{AG, slichter}
In DNP, the target system is a nuclear spin; the pumping system is an electron spin; and their interaction is mediated by the hyperfine interaction.

In a superconducting circuit, the target system is a superconducting qubit and the pumping system is usually the readout resonator because the qubit must be isolated as much as possible to maintain the coherence, while the resonator needs to be strongly coupled to the microwave feedline for fast readout.
The strategy is to find an efficient and controllable energy transfer path to the environment.

One path is from the qubit state $\ket{\textrm{e}}$ to the resonator state $\ket{1}$ [Fig.~\ref{fig:reset}(a)].\cite{houck2008, reed2010}
For this, the Purcell effect (Sec.~\ref{sec:readout}) is used.
By using a frequency-tunable qubit, we can tune $\omega_\textrm{q}$ to $\omega_\textrm{r}$ to speed up the qubit relaxation process.
Once the energy is emitted from the qubit to the resonator, the energy is quickly dissipated to the environment.

An all-microwave option is also available.\cite{egger2018, magnard2018}
In this case, the higher excitation level $\ket{\textrm{f}\,}$ is inserted into the $\ket{\textrm{e}}$-$\ket{1}$ path [Fig.~\ref{fig:reset}(b)].
The transitions between the steps are induced by microwave drives.
Hence, this method is useful for a transmon with the fixed qubit frequency.
Note that we cannot induce a direct transition between $\ket{\textrm{e0}}$ and $\ket{\textrm{g1}}$ because this transition is forbidden when the qubit-resonator interaction is Jaynes--Cummings-type [Eq.~\eqref{eq:qr}].\cite{blais2007}

These two methods can be understood using mechanical analogues shown in Fig.~\ref{fig:reset}(a) and (b).

\subsubsection{Measurement-Based Initialization}
\label{sec:feedback}

This is a completely different initialization method based on the QND measurement mentioned in Sec.~\ref{sec:readout} and the measurement postulate, which states that a measurement of an observable acting on a quantum state destroys the phase coherence and forces the state to collapse into one of the eigenstates of the observable.\cite{shankar, sakurai}
If a measurement is QND-type and its outcome is $\ket{0}$, then the qubit is initialized.
If the outcome is $\ket{1}$, we simply apply a $\pi$-pulse to flip the qubit state.
The flow chart for this procedure is shown in Fig.~\ref{fig:reset}(c).
This method is often called measurement-based initialization.\cite{riste2012, camibar2013, salanthe2018}

One technical difficulty is that this method heavily relies on high-fidelity projective measurement and fast feedback.
It can suffer from latency due to the classical data processing and the pulse generation and injection.

Note that, in this method, we reduce the entropy of the qubit by extracting information about its state.
This suggests that information processing and thermodynamics are connected intrinsically.
One of the most famous examples showing this connection is Maxwell's demon that is recently implemented in superconducting qubit systems.\cite{cottet2017, masuya2018, naghiloo2018}


\section{Quantum Error Correction}
\label{sec:QEC}

\subsection{Introduction}

Noise in real physical systems, both classical and quantum, cannot be eliminated completely. As quantum algorithms of scientific and/or commercial use consist of many qubits and time-steps, error correction schemes are essential for reliable quantum computation at scale.
The goal of Quantum Error Correction (QEC) is to generate an error-free logical qubit, i.e., two selected quantum states, out of the large Hilbert space of a system composed of multiple quantum systems.
This introduces redundancy that can be used to both detect and correct physical errors.

Classical error correction generally can introduce redundancy by duplicating bits multiple times and then errors can be detected and corrected by comparing copies together.  
However, in QEC, the use of redundancy is different and much more difficult to achieve because of the following reasons.
Firstly, the no-cloning theorem\cite{NC,BCRS} prevents this type of error-correction in quantum information as arbitrary quantum states cannot be copied.  
Secondly, direct measurement of quantum states is not possible as it will collapse the qubit state (measurement postulate in Sec.~\ref{sec:feedback}).  Hence, measurements that are used to detect actual errors have to be performed in an indirect way.
Finally, besides the bit flip error ($\ket{0} \rightarrow \ket{1}$ or $\ket{1} \rightarrow \ket{0}$) which is the standard model for classical information, quantum information can experience a second type of the error, called the phase flip error ($\alpha\ket{0}+\beta\ket{1} \rightarrow \alpha\ket{0}-\beta\ket{1}$).
Consequently, a quantum error correction code must be able to simultaneously correct for both bit flips and phase flips.

As a quantum algorithm creates complex entangled states between the constituent qubits, we require a technique that can detect individual errors on any physical qubits {\em without} extracting any information regarding the computational state of the computing system. 
Parity measurement that measures bit/phase parity of neighboring qubits is such a technique.
For this, additional qubits, which are not used for actual computation, are introduced.
These are commonly referred to as ancilla qubits or syndrome qubits.
Syndrome qubits are entangled with encoded qubits, often called data qubits, within a logical qubit and are used to extract information only related to physical errors that may have occurred on individual data qubits. 
These syndrome qubits are then measured, generating classical information called the error syndrome.
This syndrome extraction procedure is specifically designed to avoid direct measurement of the computational state of any qubit and hence preserves the computation during the error-correction process.
Multiple syndrome measurements of the data qubit are taken and this classical information is decoded.
This decoding procedure determines the most likely physical errors that resulted in the specific set of syndrome measurements that were observed.

Since we have to detect errors without knowing any information about the qubit state, to define a state as an eigenstate of a certain operator in the Heisenberg representation is more convenient than to write the state itself.
A formalism based on this idea is stabilizer formalism.\cite{gottesman1997, gottesman1998}
As it will be shown in the following, the stabilizer formalism describes our action for error detection and logical state construction in a unified manner.

A stabilizer set is a set of commuting multiqubit operators, made up of tensor products of Pauli-$X$, $Y$, and $Z$ operators.
These multiqubit operators are commonly known as stabilizer operators or simply stabilizers.
By using multiqubit projective measurement, we can force an arbitrary quantum state into simultaneous eigenstates of these stabilizers.
One consequence is that, if we repeat these projective measurements in the absence of errors, we will repeatedly measure the same eigenvalue and project the quantum state into the same eigenstate; this is why these operators are called stabilizer operators.
Note that the projective measurement on each stabilizer is actually parity measurement and its outcome is an error syndrome.
Physical errors cause eigenvalues to flip between $+1$ and $-1$, depending on if the physical errors commute or anti-commute with stabilizers.
Bit flip errors will anti-commute with stabilizers made up of Pauli-$Y$ or $Z$ operators and phase flip errors will anti-commute with stabilizers made up of Pauli-$X$ and $Y$ operators, allowing for the correction of both types of errors.

Another consequence is that the size of the stabilizer set determines the size of the restricted subspace of states that satisfy the eigen-conditions, thus defining a logical qubit.
An $N$-qubit state is spanned by $2^N$ possible basis states, where a single qubit is spanned by two ($|0\rangle$ and $|1\rangle$).
Consequently, we say that there are $N$ degrees of freedom for an $N$-qubit state.
Taking $N$ physical qubits and encoding them into a single logical qubit requires us to \emph{fix} $N-1$ degrees of freedom with the one left over to represent the logical qubit.
This is done by requiring the multiqubit state that defines the logical qubit to be in definite eigenstates of stabilizers whose eigenvalue is $+1$.
The logical qubit states, $\ket{0}_\textrm{L}$ and $\ket{1}_\textrm{L}$, both satisfy the eigen-conditions defined by these stabilizers.

In a real system, physical errors do not occur as discrete bit and phase flips, instead either coherent (such as imprecise control errors) or incoherent errors (such as thermalization or dephasing) act to perturb a qubit state in a continuous manner.
However,  measuring the eigenvalues of the stabilizer operators acts to discretize this noise, which translates the continuous nature of the noise into a probability of detecting a discrete error through the measurement of the syndrome qubit.  
An arbitrary error operator can be written as a linear combination of $X$ errors, $Z$ errors, and $Y$ errors.
Therefore, correcting all three is sufficient to correct for all possible errors on a single physical qubit.
Here, as $Y$ error can be decomposed into $Z$ and $X$ errors, $\hat{Y} = \textrm{i}\hat{Z}\hat{X}$, correcting for $Z$ and $X$ errors will together correct for any $Y$ errors.
$Z$ and $X$ errors are conventionally referred to as bit flip and phase flip errors, respectively.

There are a plethora of QEC codes in existence and many of them have been studied extensively. Amongst these, the surface code remains the most suitable scheme for solid-state qubit systems.\cite{kitaev2003, fowler2012}
One of the advantages of the surface code is that it requires each qubit to be coupled to at most, four nearest neighboring qubits.
This allows us to arrange qubits a into two-dimensional lattice  (Fig.~\ref{fig:surfaceCode}).
A single square patch of qubits defines a single, logically encoded qubit.  This patch is generally parameterized by the number of physical qubits along an edge, which is also relate to the distance, $d$, of the underlying quantum code---the distance of a quantum code is the minimum number of physical errors needed to induce a logical error.    
Another advantage is the fault-tolerant threshold of the surface code.
The fault-tolerant threshold is the maximum physical error rate that the code can correct.
For the surface code, the fault-tolerant threshold is approximately one-percent including errors related to state readout and all gate operations.
This error threshold is one of the highest of any error correction scheme to date and remains the highest of a code that is compatible with the architectural constraints of current quantum computing hardware.
However, we emphasize that fault-tolerant thresholds vary significantly, depending on the actual QEC code that is utilized.

In this tutorial, we explain how to construct a logical qubit, how to detect and correct errors, and how to perform gate operations on a logical qubit in the context of the surface code.
Interested readers are referred Refs.~\onlinecite{NC, BCRS, gottesman, devitt} for introductions to QEC and Refs.~\onlinecite{QECbook, terhal, campbell} for comprehensive reviews.

\subsection{Surface Code}
\label{sec:surfaceCode}

\subsubsection{Definition}
\label{sec:surfaceCodeDef}

\begin{figure}
\includegraphics{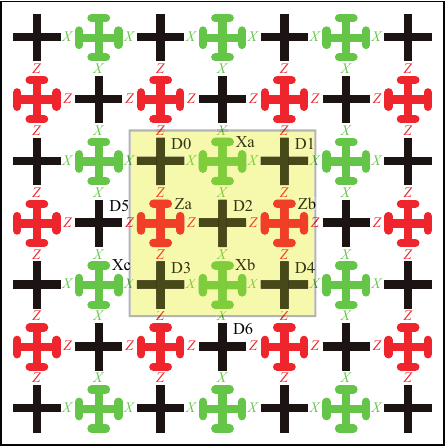}
\caption{Arrangement of physical qubits in the surface code with $d=4$, where $d$ is the number of data qubits along the horizontal or vertical edge at the lattice boundary, which is represented by black solid lines.
The black crosses are data qubits, the red crosses are $Z$ syndrome qubits, and the green crosses are $X$ syndrome qubits.
The data and syndrome qubits are physically the same; only their functions are different.
All qubits in the lattice boundary form a single logical qubit.
In Sec.~\ref{sec:logicalQubit}, we will reduce the lattice boundary to the yellow square to give a simple example of logical qubit construction.
}
\label{fig:surfaceCode}
\end{figure}

The surface code is defined over a two-dimensional lattice of physical qubits that allow for nearest neighbor interactions.
For a distance $d$ code, the size of the lattice is $(2d-1)\times (2d-1)$ physical qubits.
Of the total $4d^2-4d+1$ physical qubits in the lattice, approximately half ($2d^2-2d+1$) are data qubits and the rest of the physical qubits are syndrome qubits.
It is sufficient to detect $\lceil (d-1)/2 \rceil$ single-qubit errors, i.e., errors associated with a single data qubit, or correct $\lfloor (d-1)/2\rfloor$ single-qubit errors.
Thus, the smallest $d$ required to correct any single-qubit error is 3.
Figure~\ref{fig:surfaceCode} illustrates for a $d=4$ surface code, which can detect 2 single-qubit errors and correct 1 single-qubit error.

\subsubsection{Logical Qubit Construction}
\label{sec:logicalQubit}

In this subsection, we give an example for constructing a logical qubit with physical data qubits in the yellow square of Fig.~\ref{fig:surfaceCode}.
Although the yellowed lattice is too small ($d=2$) to specify the position and the type of an arbitrary single-qubit error uniquely (see Sec.~\ref{sec:surfaceCodeDef}), we use this yellowed lattice to give the simplest example for logical qubit construction and logical gate operation.
We first write the logical qubit in the state vector notation.
Then we will re-express the same logical qubit using the stabilizer formalism.
In the state vector notation, the logical qubit states are written as
\begin{equation} \label{eq:codeword} 
\begin{split}  
\ket{0}_\textrm{L} &=
\frac{1}{2} \left( \ket{00000}+\ket{00111}+\ket{11011}+\ket{11100} \right), \\
\ket{1}_\textrm{L} &=
\frac{1}{2} \left( \ket{10010}+\ket{10101}+\ket{01001}+\ket{01110} \right),
\end{split}
\end{equation}
where each ket represents $\ket{\psi_\textrm{D0}\psi_\textrm{D1}\psi_\textrm{D2}\psi_\textrm{D3}\psi_\textrm{D4}}$.
In general, the logical qubit state during computations is $\ket{\psi}_\textrm{L} = \alpha\ket{0}_\textrm{L}+\beta\ket{1}_\textrm{L}$, where $\alpha$ and $\beta$ are complex numbers, and $\abs{\alpha}^2 + \abs{\beta}^2 = 1$.

In the stabilizer formalism, the same logical states are defined by
\begin{equation} \label{eq:stabilizer} 
\ket{\psi} = \hat{S} \ket{\psi}.
\end{equation}
In the literature, Eq.~\eqref{eq:stabilizer} is often read as ``$\ket{\psi}$ is stabilized by the operator set $\hat{S}$.''
Here, $\{\hat{S}\}$ is a stabilizer set in which stabilizers are given by
\begin{equation}\label{eq:stabilizerSet}
\begin{split} 
\hat{S}_0 &= \hat{X}_\textrm{D0} \hat{X}_\textrm{D1} \hat{X}_\textrm{D2} \hat{I}_\textrm{D3} \hat{I}_\textrm{D4}, \\
\hat{S}_1 &= \hat{I}_\textrm{D0} \hat{I}_\textrm{D1} \hat{X}_\textrm{D2} \hat{X}_\textrm{D3} \hat{X}_\textrm{D4}, \\
\hat{S}_2 &= \hat{Z}_\textrm{D0} \hat{I}_\textrm{D1} \hat{Z}_\textrm{D2} \hat{Z}_\textrm{D3} \hat{I}_\textrm{D4}, \\ 
\hat{S}_3 &= \hat{I}_\textrm{D0} \hat{Z}_\textrm{D1} \hat{Z}_\textrm{D2} \hat{I}_\textrm{D3} \hat{Z}_\textrm{D4},
\end{split}
\end{equation}
where $\hat{X}_i$ and $\hat{Z}_i$ are the Pauli matrices applied to data qubit $i$.
Here, $\hat{S}_0$ and $\hat{S}_1$ are $X$ stabilizers, and
$\hat{S}_2$ and $\hat{S}_3$ are $Z$ stabilizers.
All these stabilizers commute with each other and satisfy Eq.~\eqref{eq:stabilizer}.
Since both $\ket{0}_\textrm{L}$ and $\ket{1}_\textrm{L}$ are eigenstates of this stabilizer set with the same eigenvalue, $+1$, repeatedly measuring these operators preserves the logical qubit states without destroying the phase coherence between them.

There are other methods for constructing a logical qubit when logical qubits are defined using defects in a qubit lattice.
But these methods are out of the scope of this tutorial.
The standard introduction to this topic is Ref.~\onlinecite{fowler2012}.

\subsubsection{Error Detection and Correction}
\label{sec:errorDetection}

Error detection requires repeated parity measurements of the stabilizers of the surface code.
The syndrome qubits are used for this.
These syndrome qubits are the only qubits measured during the error detection (syndrome extraction) process to avoid destruction of the logical information stored.

As shown in Fig.~\ref{fig:surfaceCode}, there are two types of syndrome qubits, namely, $Z$ syndrome qubits, which are measured in the $Z$ basis, and $X$ syndrome qubits, which are measured in the $X$ basis.
These syndrome qubits are used to measure the eigenvalue of $Z$ and $X$ stabilizers of the surface code, respectively. 
Having these two types of syndrome qubits allows us to detect both bit and phase flip errors as physical bit flip errors will be detected by the measurement of $Z$ stabilizers and physical phase flip errors will be detected by the measurement of $X$ stabilizers.

The following quantum circuits show one surface code cycle for the Za and the Xb syndrome qubits.
\begin{gather}
\!\!\!\!\!\!\!\!\!\!\!
\begin{array}{c}
\Qcircuit @C=0.9em @R=1.0em {
\lstick{\textrm{D0 }\ket{\psi_\textrm{D0}}} & \ctrl{4} & \qw     & \qw      & \qw      & \qw \\
\lstick{\textrm{D2 }\ket{\psi_\textrm{D2}}} & \qw      & \ctrl{3}& \qw      & \qw      & \qw \\
\lstick{\textrm{D3 }\ket{\psi_\textrm{D3}}} & \qw      & \qw     & \ctrl{2} & \qw      & \qw \\
\lstick{\textrm{D5 }\ket{\psi_\textrm{D5}}} & \qw      & \qw     & \qw      & \ctrl{1} & \qw \\
\lstick{\textrm{Za }\quad\ \ket{0}} & \targ    & \targ   &\targ     & \targ    & \meter
} \\
\end{array}, \label{eq:errorZ} \\
\quad\quad\quad\quad
\begin{array}{c}
\Qcircuit @C=0.8em @R=0.5em {
\lstick{\textrm{D2 }\ket{\psi_\textrm{D2}}}	&\qw	& \targ & \qw		& \qw   & \qw   &\qw	& \qw \\
\lstick{\textrm{D3 }\ket{\psi_\textrm{D3}}}	&\qw	& \qw   & \targ	& \qw   & \qw   &\qw	& \qw \\
\lstick{\textrm{D4 }\ket{\psi_\textrm{D4}}}	&\qw	& \qw   & \qw		& \targ & \qw   &\qw	& \qw \\
\lstick{\textrm{D6 }\ket{\psi_\textrm{D6}}}	&\qw	& \qw   & \qw		& \qw   & \targ &\qw	& \qw \\
\lstick{\textrm{Xb }\quad\,\ket{0}}	& \gate{H} & \ctrl{-4} & \ctrl{-3}	& \ctrl{-2} & \ctrl{-1} & \gate{H} & \meter
} \\
\end{array}. \label{eq:errorX}
\end{gather}
Syndrome qubit measurements in these circuits occur in the computational ($Z$) basis.
Equation~\eqref{eq:errorX} can be rewritten as
\begin{gather}
\quad\quad\quad\ 
\begin{array}{c}
\Qcircuit @C=0.8em @R=0.5em {
\lstick{\textrm{D2 }\ket{\psi_\textrm{D2}}}	& \gate{H}	& \ctrl{4} & \gate{H}		& \qw   & \qw   &\qw	& \qw \\
\lstick{\textrm{D3 }\ket{\psi_\textrm{D3}}}	& \gate{H}	& \qw   & \ctrl{3}	& \gate{H}   & \qw   &\qw	& \qw \\
\lstick{\textrm{D4 }\ket{\psi_\textrm{D4}}}	& \gate{H}	& \qw   & \qw		& \ctrl{2} & \gate{H}   &\qw	& \qw \\
\lstick{\textrm{D6 }\ket{\psi_\textrm{D6}}}	& \gate{H}	& \qw   & \qw		& \qw   & \ctrl{1} & \gate{H}	& \qw \\
\lstick{\textrm{Xb }\quad\,\ket{0}}	& \qw & \targ & \targ	& \targ & \targ & \qw & \meter
} \\
\end{array}, \label{eq:errorX2}
\end{gather}
based on the following identity,
\begin{align}
\begin{array}{c}
\Qcircuit @C=1em @R=1em {
& \ctrl{1}	& \qw	& \raisebox{-2em}{=}	& & \gate{H}& \targ  & \gate{H} & \qw   \\
& \targ		& \qw	& 						& & \gate{H}	& \ctrl{-1} & \gate{H} & \qw 
} \\
\end{array}.
\end{align}
By using the $H$ gates, the CNOTs in Eq.~\eqref{eq:errorX2} can be inverted, ensuring that both circuits have the ancilla qubit as target for all CNOTs.
Of course, for actual implementation, Eq.~\eqref{eq:errorX} is more economical than Eq.~\eqref{eq:errorX2} because of the reduced number of gates.

The quantum circuits [Eqs.~\eqref{eq:errorZ} and \eqref{eq:errorX}], called parity check circuits, are designed to infer the bit and phase parities of neighboring data qubits by measuring the eigenvalue of specific four qubit Pauli operators.
This not only extracts the eigenvalue of the relevant Pauli operator but also projects the four data qubits into the appropriate eigenstate.
In Eq.~\eqref{eq:errorZ}, our measurement on the Za syndrome qubit forces the neighboring data qubits (D0, D2, D3, and D5) into an eigenstate of $\hat{Z}_\textrm{D0} \hat{Z}_\textrm{D2} \hat{Z}_\textrm{D3} \hat{Z}_\textrm{D5}$ ($Z$ stabilizer).
Similarly, in Eq.~\eqref{eq:errorX}, the measurement on the Xb syndrome qubit forces the neighboring data qubits (D2, D3, D4, and D6) into an eigenstate of $\hat{X}_\textrm{D2} \hat{X}_\textrm{D3} \hat{X}_\textrm{D4} \hat{X}_\textrm{D6}$ ($X$ stabilizer).

If the D2 qubit has a phase error, Xa and Xb syndrome qubits [Eq.~\eqref{eq:errorX}] return $-1$ as the error syndromes if there is no error in the syndrome qubits;
if the D2 qubit has a bit flip error, Za and Zb syndrome qubits [Eq.~\eqref{eq:errorZ}] return $-1$.
Note that the existence of both a bit flip error and a phase flip error can be detected simultaneously because $\hat{Z}_\textrm{D0} \hat{Z}_\textrm{D2} \hat{Z}_\textrm{D3} \hat{Z}_\textrm{D5}$ and $\hat{X}_\textrm{D2} \hat{X}_\textrm{D3} \hat{X}_\textrm{D4} \hat{X}_\textrm{D6}$ commute.
Therefore, the parity measurement allows us to know the existence of both the bit flip and the phase flip errors without collapsing the quantum state of data qubits.

These circuits has to be performed across the entire lattice, with measurement of all $X$ syndromes occurring simultaneously followed by simultaneous measurement of all $Z$ syndromes.
The measurement of every $X$ and $Z$ syndrome is referred to as an error correction cycle, which is repeated continuously as the quantum computer is in operation.
Consequently, after each error correction cycle, there will be $d^2-d$ classical bits of information (error syndromes) related to $X$ errors and $d^2-d$ classical bits of information related to $Z$ errors.
After multiple cycles of error-correction, all this information is then processed by a classical error-correction decoder to determine the most likely set of {\em actual} errors that resulted in the syndrome measurements that are observed.\cite{devittdec}

Once the existence of an error is confirmed, the error can be corrected by applying the microwave pulse to flip phase or bit of the D2 qubit.
However, in practice, we can simply record the error in a classical computer and correct measurement outcomes that are affected by the error.
This is known as tracking the Pauli frame.\cite{paler2014}

In the above example, the position and the type (bit or phase flip) of the error were already known. However, determining the position and the type of the error from syndrome measurements are difficult because it is an inverse problem.
Moreover, the error position cannot be determined uniquely for some error patterns.
For example, error syndromes of Xa and Xc resulting from phase errors in D0 and D5 are identical to that from phase errors in D2 and D3.
However, if the number of such an error is reasonably small, the identity of each error can be almost completely inferred by minimum-weight perfect matching algorithm.\cite{fowler2012}
This accuracy in using minimum weight matching to identify the most likely physical errors corresponding to a measured syndrome pattern is what effectively determines the fault-tolerant threshold.
If error rates are too high, or errors spread too much through the quantum circuits used in the syndrome extraction process, then decoding algorithms will not decode physical errors accurately and corrections may induce logical errors.

One interesting consequence is that initialization of a logical qubit can be considered as a kind of error correction starting from a known state, such as $\ket{00000}$.
Initializing the logical qubit from this state only requires projective measurement of the $X$ stabilizers as the state $\ket{00000}$ already satisfies the eigen-conditions of the $Z$ stabilizers.
Additionally, when starting in the $\ket{00000}$ state, we initialize into the logical $\ket{0}_\textrm{L}$, because again our initial state before encoding is already in the $+1$ eigenstate of the logical $Z$ operator.
Further examples can be found in Ref.~\onlinecite{devitt}.


Lastly, we briefly mention how we perform readout of a logical qubit.
The logical readout of an error-corrected qubit ideally requires the direct measurement of all the physical qubits in an encoded block.\cite{fowler2012, horsman2012}
To maintain correct fault-tolerant operation, all of these physical measurements on the encoded block have to take place.
While it is possible to perform readout of an encoded qubit by performing a logical single qubit parity check of the $X$ or $Z$ operator,\cite{horsman2012} this would still require the assistance of a fully encoded ancillary qubit that would still require the physical measurement of all component qubits to perform the readout.

\subsubsection{Logical Gate Operation}
\label{sec:logicalGate}

\begin{figure*}
\includegraphics{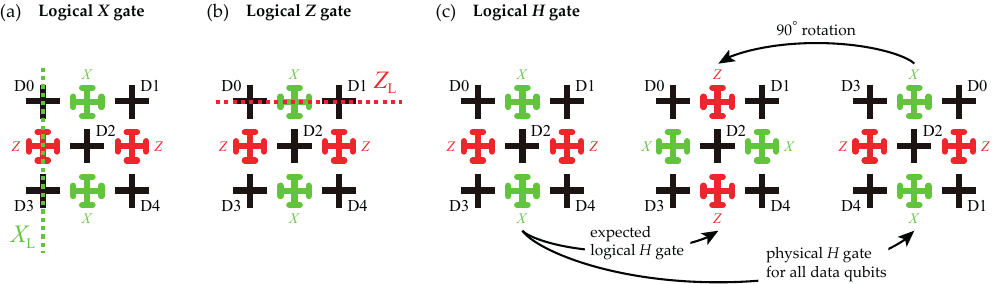}
\caption{Schematic images of (a) logical $X$ gate, (b) logical $Z$ gate, and (c) logical $H$ gate.
As in Fig.~\ref{fig:surfaceCode}, the black crosses are data qubits, the red crosses are $Z$ syndrome qubits, and the green crosses are $X$ syndrome qubits.}
\label{fig:logicalGate}
\end{figure*}

After encoding, computing is performed on the code using logical gates.
Logical gates must preserve the symmetries enforced by the stabilizer operators, but manipulate the operators that define the logical state of the encoded qubit in the same way as operations on physical qubits do.
Consequently, logical gate operators commute with all elements in the stabilizer set $\{\hat{S}\}$, but by definition are not contained in $\{\hat{S}\}$.
For the surface code, for example, the logical $X$ gate, $\hat{X}_\textrm{L}$, corresponds to applying the physical $X$ gate to all data qubits in one of columns---not including the $Z$ syndrome qubits.
The logical $Z$ gate is achieved by applying the physical $Z$ gate to the all data qubits in one of rows---not including the $X$ syndrome qubits.

Figure~\ref{fig:logicalGate} reproduces the yellow area in Fig.~\ref{fig:surfaceCode} with the logical $Z$ and the logical $X$ operations.
Applying $\hat{Z}_\textrm{L} = \hat{Z}_\textrm{D0} \hat{Z}_\textrm{D1} \hat{I}_\textrm{D2} \hat{I}_\textrm{D3} \hat{I}_\textrm{D4}$ and $\hat{X}_\textrm{L} = \hat{X}_\textrm{D0} \hat{I}_\textrm{D1} \hat{I}_\textrm{D2} \hat{X}_\textrm{D3} \hat{I}_\textrm{D4}$ to Eq.~\eqref{eq:codeword} helps to understand these logical operations.
Note that $\hat{Z}_\textrm{L}$ and $\hat{X}_\textrm{L}$ commute with every stabilizers in Eq.~\eqref{eq:stabilizerSet} and anti-commute with each other as they must intersect on an odd number of physical qubits and physical $X$ and $Z$ gates anti-commute.
Consequently, they form a pair of Pauli operators that have the same commutation properties as physical $X$ and $Z$.

Another important single qubit gate is the Hadamard ($H$) gate.
What we expect from the logical $H$ gate is the following operations:
\begin{equation}
\begin{split}
\hat{H}_\textrm{L}\ket{0}_\textrm{L} &= 
\frac{1}{\sqrt{2}} (\ket{0}_\textrm{L} + \ket{1}_\textrm{L}), \\
\hat{H}_\textrm{L}\ket{1}_\textrm{L} &= 
\frac{1}{\sqrt{2}} (\ket{0}_\textrm{L} - \ket{1}_\textrm{L}),
\end{split}
\end{equation}
i.e., the same operations as those for a physical qubit.
In the Heisenberg representation, the role of the physical $H$ gate is to exchange the physical $X$ gate with the physical $Z$ gate, and vice versa.
In other words, a qubit that is in the $+1$ eigenstate of $\hat{Z}$ turns into a state that is in the $+1$ eigenstate of $\hat{X}$ because $\hat{H}^\dagger\hat{X}\hat{H}=\hat{Z}$ and $\hat{H}^\dagger\hat{Z}\hat{H}=\hat{X}$.
As the logical operation must preserve the stabilizer set while interchanging the actual logical operators, the logical $H$ needs to: 
(i) swap the operation of $\hat{Z}_\textrm{L}$ and $\hat{X}_\textrm{L}$ by flipping a vertical chain of $X$ operators into a horizontal chain of $X$ operators; and
(ii) maintain the $X$ and $Z$ stabilizers in Fig.~\ref{fig:logicalGate}(c).
These operations can be achieved by applying physical $H$ gates to all data qubits (this changes all physical $X$ operators to $Z$ operators and visa versa), followed by 90$^\circ$ counterclockwise rotation of the lattice (which interchanges horizontal chains with vertical ones and ensures that the $X$ and $Z$ stabilizer operators stay in the same place in the lattice).

To see how this actually works, we apply the physical $H$ gate to all data qubits composing $\ket{0}_\textrm{L}$ [left qubit lattice in Fig.~\ref{fig:logicalGate}(c)]:
\begin{align}
\hat{H}_\textrm{D0}\hat{H}_\textrm{D1}&\hat{H}_\textrm{D2}\hat{H}_\textrm{D3}\hat{H}_\textrm{D4}\ket{0}_\textrm{L} = \nonumber\\
\frac{1}{2\sqrt{2}} 
( &\ket{00000} + \ket{01101} + \ket{11011} + \ket{10110} \nonumber\\
+ &\ket{00011} + \ket{01110} + \ket{11000} + \ket{01110} ),
\end{align}
which is certainly not $(\ket{0}_\textrm{L}+\ket{1}_\textrm{L})/\sqrt{2}$.
Actually, this is the right qubit lattice in Fig.~\ref{fig:logicalGate}(c), whereas what we want is the center qubit lattice in Fig.~\ref{fig:logicalGate}(c).
Thus, we have to rotate the qubit lattice $90^\circ$ counterclockwise (exchanging the $X$ stabilizers to the $Z$ stabilizers) to obtain the logical $H$ gate we want.
This operation is equivalent to changing indices of the data qubits (i.e. performing SWAP operations):
D0 to D1, D1 to D4, D3 to D0, and D4 to D3.
As a result,
\begin{align}
\hat{H}_\textrm{L}\ket{0}_\textrm{L} &= \nonumber\\
\frac{1}{2\sqrt{2}} 
( &\underbrace{\ket{00000} + \ket{00111} + \ket{11011} + \ket{11100}}_{\ket{0}_\textrm{L}} \nonumber\\
+ &\underbrace{\ket{10010} + \ket{10101} + \ket{01001} + \ket{01110}}_{\ket{1}_\textrm{L}} ).
\end{align}
Now, we have what we expect from the $H$ gate.
The actual operations for rotating the qubit lattice can be found in Ref.~\onlinecite{horsman2012}.

The logical two-qubit gate, such as the logical CNOT gate, is not as simple compared to the logical single-qubit gates.
Implementing a non-Clifford gate, such as the $T$ gate (see Table~\ref{tab:univGate} for its definition), for a logical qubit is even more difficult than implementing the logical CNOT gate.
The reason for this is that non-Clifford gates cannot be operated directly on the stabilizer codes in a fault-tolerant manner, i.e., ensuring errors do not cascade out of control;
whereas gates in the Clifford group (gates that map Pauli operators to Pauli operators) can generally be applied directly to encoded data.  
It should be noted that not all stabilizer codes can enact the full Clifford group of logical operations directly.
In fact, the surface code cannot enact the $S$ gate (Table~\ref{tab:univGate}) directly and must use other constructions.\cite{fowler2012}
Thus, these gates will not be covered in this tutorial.
Interested readers are referred Ref.~\onlinecite{horsman2012} for the logical CNOT gate and Refs.~\onlinecite{bravyi2005, meier2013} for non-Clifford logical gates.

\subsection{Proposed Device Architectures}


\begin{figure*}
\includegraphics{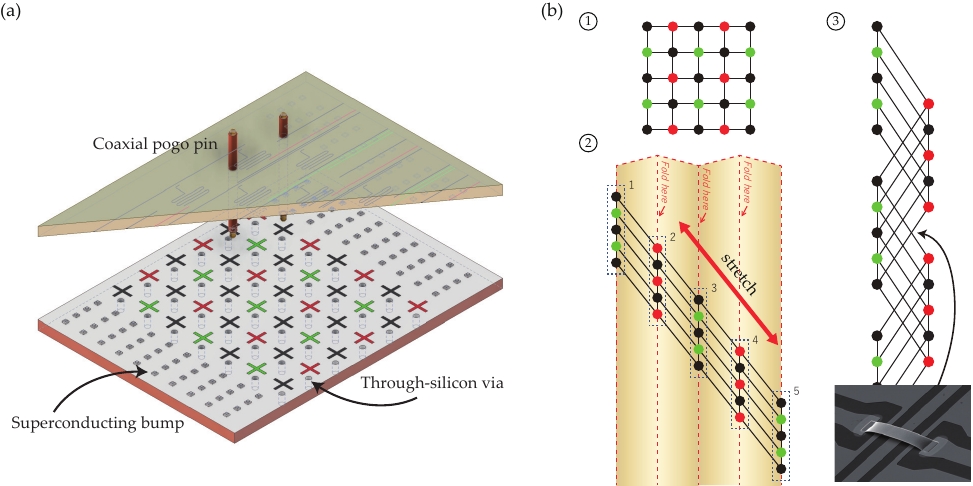}
\caption{(a) Schematic image of 3D wiring techniques for quantum circuit: flip-chip bonding using superconducting bumps, pogo pins, and through-silicon vias.
(b) Pseudo-2D architecture.
All three figures have the same lattice structure and circles represent qubits.
The numbers in circles indicate how the original qubit lattice evolves to the pseudo-2D architecture. 
The inset shows a scanning electron microscopy image of an air bridge, which can be used at resonator intersections.}
\label{fig:architecture}
\end{figure*}

One of the biggest problems in scaling up a superconducting qubit system is thet so-called wiring problem.\cite{vandersypen2017}
The wiring problem is that the number of wires required to operate qubits increases too fast with the number of qubits because each qubit requires multiple channels such as control lines and measurement devices.
In such a situation, it will be difficult to access a qubit inside a chip.

One natural idea that avoids this problem and fits well with Figs.~\ref{fig:surfaceCode} and \ref{fig:logicalGate} is the use of the third dimension [Fig.~\ref{fig:architecture}(a)].\cite{barends2014}
For a classical solid-state circuit, a three-dimensional (3D) structure is made by employing a silicon oxide film as a interlayer insulator.
However, this method cannot be used for quantum systems because such layers are so lossy that they can be severe decoherence channels.
To get around this problem, flip-chip bonding is often used.
On one chip, qubits are arranged into a square lattice; 
on another chip, other circuit components such as control lines and readout resonators are fabricated.
Then, the two chips are combined face-to-face using superconducting bumps.\cite{rosenberg2017, foxen2018}
To introduce microwaves to qubits from the top or back side of the wafer, pogo pins\cite{bronn2018} and through-silicon vias\cite{vahidpour2017, yost2020} can be used, respectively.
Here, a through-silicon via is a coaxial structure that passes through a silicon wafer.


Recently, it has been reported that the 2D lattice for the surface code can be folded like origami as shown in Fig.~\ref{fig:architecture}(b).\cite{mukai2020}
In this architecture, qubits and control lines can be fabricated on the same plane if the coupling resonators are allowed to intersect by air bridges [inset in Fig.~\ref{fig:architecture}(b)].\cite{bronn2018}
Because of these cross-connections, this method was named a pseudo-2D architecture.
The appealing point of this architecture is that all qubits and their associated lines can exist on the same chip.
Moreover, it can be made by utilizing the standard 2D microwave technology so that we can avoid the complex techniques required for a 3D architecture.

Possible concerns are cross-talk between intersecting resonators and the degradation of the quality factor caused by air bridges.
In Ref.~\onlinecite{mukai2020}, it was shown that the cross-talk is at most about $-50$ dB;
in addition, resonators with 15--20 air bridges showed an internal quality factor in a range where the infidelity is lower than the threshold value of the surface code.


\section{Characterizing a Quantum System}
\label{sec:characterize}


\begin{table*}
\caption{Minimal procedure for extracting system parameters and related experimental methods.
}
\label{tab:char}\centering
\begin{ruledtabular}
\begin{tabular}{l l l}
\noalign{\smallskip}
Measured quantities	&	Method	& Comment \\
\noalign{\smallskip} \hline \noalign{\smallskip}
Resonator: resonance frequency, quality factor	&	Single-tone spectroscopy \\
Anticrossing in the flux bias sweep (if existing):	&	 \\
$\ \ $ qubit-resonator coupling constant	&	 \\
\noalign{\smallskip} \hline \noalign{\smallskip}
Qubit: transition frequency, sweet spot	&	Two-tone spectroscopy \\
Anticrossing in the flux bias sweep (if existing):	&	 \\
$\ \ $ qubit-qubit coupling constant	&	 \\
\noalign{\smallskip} \hline \noalign{\smallskip}
Rabi oscillation: pulse strength calibration	&	Time domain	&	ready for time-domain measurements \\
Dispersive shift: qubit-resonator coupling constant	&	Pulsed spectroscopy	&	ready for setting up the Hamiltonian \\
$T_1$ and $T_2$	&	Time domain	&	show the quality of qubits \\
\end{tabular}
\end{ruledtabular}
\end{table*}

\begin{figure*}
\centering
\includegraphics{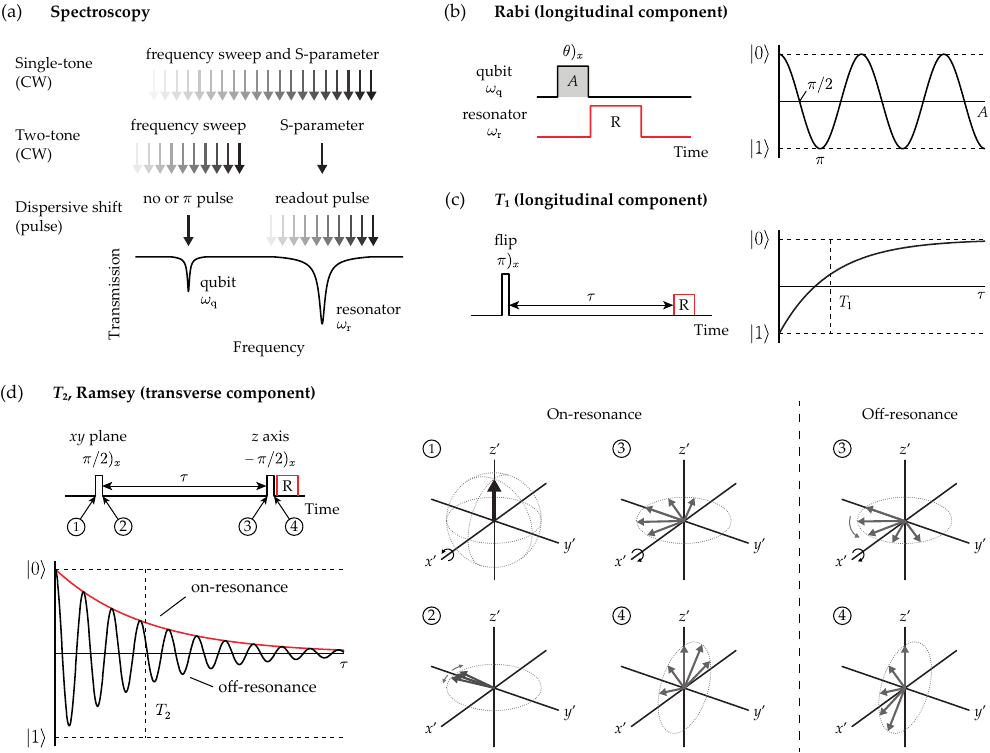}
\caption{(a) Spectroscopy (frequency domain).
The control parameter of this type of measurement is the frequency of the applied microwave.
Arrows represent microwave excitation, and the gradient of the arrows represents the frequency sweep direction from low to high frequency.
The resonator and qubit signals are detected by the changes in the S-parameters of the system.
For the circuit in Fig.~\ref{fig:couplingQR}(a), the qubit and the resonator signals can be detected through the transmission.
``CW'' stands for continuous wave.
(b--d) Time domain measurements.
Each figure represents the pulse sequence and typical measurement result for (b) Rabi oscillation, (c) $T_1$, and (d) $T_2$ measured via the Ramsey fringes.
Note that each graph represents the averaged result of a set of identical measurements; the result of a single-shot readout is digital.
The readout pulses are in red and labeled ``R''.
$\tau$ indicates the time interval, which is the independent variable.
``$A$'' in (b) indicates the area of the drive pulse.
$\theta)_x$ is shorthand notation of $\hat{R}_x(\theta)$.
The circled numbers in (d) describe the evolution of the qubit state in the Bloch sphere. (The longitudinal relaxation process is ignored for clarity.)
Each arrow represents the qubit state for each measurement.
The average $z'$-component in \raisebox{.5pt}{\textcircled{\raisebox{-.9pt}{4}}} is recorded as a result.
Note that, in (d), $\pm\pi/2)_x$ pulses are required at the beginning and end of the sequence to transfer the $z'$-component of the Bloch vector to the $y'$-component, and vice versa.
If the external drive and qubit transition frequency are off-resonance, the Bloch vector rotates about the $z'$-axis.
This results in an oscillation called Ramsey fringes.}
\label{fig:pulseSequence}
\end{figure*}

To control a quantum system precisely, we have to know the Hamiltonian of the system.
In other words, we must determine a set of parameters, called system parameters, that characterize the system.
Since the system parameters are based on our model describing the system, appropriate modeling is essential.
For example, if we treat our qubit as an ideal two-level system as we did in Sec.~\ref{sec:couplingConcept}, knowing $\omega_\mathrm{q}$ might be sufficient to set up the Hamiltonian.
However, for high-fidelity gate operations, we must consider higher excitation levels;
hence, we have to extract more information, such as the transition frequency between $\ket{1}$ and $\ket{2}$.
If our target operation requires a strong drive, we may also have to consider the nonlinearity of the readout resonator.

In this section, we explain the minimal procedure for extracting system parameters and the related experimental methods.
These are summarized in Table~\ref{tab:char}.
One useful reference regarding this topic is Ref.~\onlinecite{gao_tutorial}.

\subsection{Spectroscopy}
\label{sec:spectroscopy}

\subsubsection{Single-Tone Spectroscopy}
\label{sec:singleTone}

The first step in characterizing a superconducting qubit system is finding the resonance frequency of the readout resonator, $\omega_\textrm{r}$.
For this, we inject a microwave continuously (Continuous Wave, CW) and measure the S-parameters of the system as a function of microwave frequency [Fig.~\ref{fig:pulseSequence}(a)].
This task is usually done using a vector network analyzer (VNA).
Since a single microwave source, which is a part of the VNA, is used in this step, this type of measurement is called single-tone spectroscopy.

Since the transition frequency of the qubit, $\omega_\textrm{q}$, is usually designed to be far detuned from $\omega_\textrm{r}$ for dispersive readout (Sec.~\ref{sec:readout}), most of the microwave power whose frequency is close to $\omega_\textrm{q}$ is filtered out by the resonator.
As a result, the qubit signal is not visible in single-tone spectroscopy.
This is why we need another type of spectroscopy, called two-tone spectroscopy.

If the qubit and the readout resonator are on-resonance at a certain external bias, we can estimate the qubit-resonator coupling constant $g$ by observing the peak splitting as explained in Sec.~\ref{sec:strongCoupling}.
If the qubit is not tunable, we can still estimate $g$ using the dispersive shift (Sec.~\ref{sec:dispersiveShift}).

\subsubsection{Two-Tone Spectroscopy}
\label{sec:twoTone}

Once $\omega_\textrm{r}$ is known, we fix the excitation frequency of the VNA near $\omega_\textrm{r}$ for readout.
Subsequently, we inject another microwave to the circuit to drive the qubit.
$\omega_\textrm{q}$ is found by sweeping the frequency of the second microwave, called the drive frequency $\omega_\textrm{d}$, while monitoring the changes in the S-parameters of the readout resonator.
If $\omega_\textrm{d}$ becomes close to $\omega_\textrm{q}$, the S-parameters of the readout resonator will vary because of the dispersive shift in resonance frequency [Fig.~\ref{fig:readout}(a)].
This type of measurement is called two-tone spectroscopy.

When we characterize $\omega_\textrm{q}$, we have to minimize the excitation power of the VNA; if the excitation power is too high, then the readout resonator is populated by multiple photons, resulting in the shift or splitting of the qubit spectrum as mentioned in Sec.~\ref{sec:readout}.

We can repeat this procedure for different external biases to obtain the full bias dependence of $\omega_\textrm{q}$, which informs us of the position of the sweet spot.
If the transition frequencies of two qubits coincide at a certain bias, we can see an anticrossing.
From this, we can estimate the qubit-qubit coupling constant.

\subsubsection{Dispersive Shift}
\label{sec:dispersiveShift}

As explained in Sec.~\ref{sec:readout}, the dispersive shift $\chi$ is the qubit-state-dependent frequency shift of the readout resonator.
From this, we can estimate the qubit-resonator coupling $g$ using $\chi = g^2/\Delta_\textrm{qr}$, where $\Delta_\textrm{qr} = \omega_\textrm{r}-\omega_\textrm{q}$.

To measure $\chi$, we first prepare the qubit state of either $\ket{0}$ or $\ket{1}$.
In this step, a $\pi$-pulse is required to prepare $\ket{1}$.
Because of this, the dispersive shift measurement has to be preceded by the Rabi oscillation measurement (Sec.~\ref{sec:Rabi}).
Once the qubit state is prepared, we apply the readout pulse and measure the S-parameters.
The sweep parameter is the frequency of the readout pulse [Fig.~\ref{fig:pulseSequence}(a)].
Hence, the measurement of the dispersive shift is like pulsed single-tone spectroscopy with the qubit state preparation.
By comparing the spectrum of the readout resonator with two different qubit states, we can obtain $\chi$ as shown in Fig.~\ref{fig:readout}(a).

\subsection{Time-Domain Measurement}
\label{sec:timeDomain}

According to quantum theory, the measurement forces the qubit state to collapse to either the $\ket{0}$ or $\ket{1}$ state, as emphasized in Secs.~\ref{sec:feedback} and \ref{sec:QEC}.
Hence, to extract the parameters mentioned in this section, we have to make a set of measurements and take the average.

\subsubsection{Rabi Oscillation}
\label{sec:Rabi}

In time-domain measurements, we calibrate the qubit drive by observing the Rabi oscillation [Fig.~\ref{fig:pulseSequence}(b)].
First, the qubit drive pulse with $\omega_\textrm{q}$ is applied to excite the qubit.
After the drive pulse, the readout pulse near $\omega_\textrm{r}$ is applied, and the S-parameters of this readout pulse are measured via quadrature detection.\cite{mit1, levitt}
The sweep parameter is the area of the drive pulse; in actual experiments, it can be either the length or amplitude of the drive pulse.

Since the drive pulse rotates the qubit state in the Bloch sphere (Fig.~\ref{fig:Xgate}), and the dispersive measurement detects the longitudinal component of the Bloch vector (Sec.~\ref{sec:readout}), the amplitude of the signal oscillates with the drive pulse area: 
this oscillation is the Rabi oscillation.
The Rabi oscillation provides a correspondence between the nutation angle of the Bloch vector and the drive pulse area.
The names of frequently used pulses, such as $\pi$- and $\pi/2$-pulses, indicates the nutation angles induced by such pulses.

If the sweep parameter is the length of the drive pulse $\tau_\textrm{p}$, the following function is used for fitting:
\begin{align}
f(\tau_\textrm{p}) = 
A_0 + A_1 \cos(\Omega_\textrm{R} \tau_\textrm{p} + A_2) \exp(-\frac{\tau_\textrm{p}}{T_\textrm{R}}),
\end{align}
where the fitting parameters are $A_0$, $A_1$, $A_2$, $\Omega_\textrm{R}$, and $T_\textrm{R}$.
Among them, only $\Omega_\textrm{R}$ (Rabi frequency) and $T_\textrm{R}$ (characteristic decay time of the Rabi oscillation) are physically meaningful quantities.
Note that the exponential decay is also required to fit the data properly.
The main reason for this decay is finite $T_1$ and fluctuations in $\Omega_\textrm{R}$.\cite{ithier2005, yoshihara2014}

With the extracted system parameters, we can establish the correspondence between the control parameters we set and the actual response of the system.
This process is called calibration.
The Rabi oscillation measurement is the simplest calibration procedure;
however, the Rabi frequency estimated from this method is usually not accurate enough for high-fidelity control.
Probably the next simplest and more accurate one is to apply a train of pulses;
interested readers should see Refs.~\onlinecite{burum1981, vlastakisThesis}.
Optimal control theory can also be adopted; see Refs.~\onlinecite{kelly2014, werninghaus2020}.
More advanced techniques for Google's devices can be found in Refs.~\onlinecite{kelly2016, kelly2018} and the supplementary materials for Ref.~\onlinecite{neill2018}.

\subsubsection{Relaxation Time: $T_1$}
\label{sec:T1}

The measurement procedure for the longitudinal relaxation time ($T_1$) is shown in Fig.~\ref{fig:pulseSequence}(c).
Since the dispersive readout measures the longitudinal component of the Bloch vector, all we have to do is apply a pulse (often but not necessarily a $\pi$-pulse) and detect the population of $\ket{0}$ as a function of the time interval.
During this time interval, the qubit relaxes back to $\ket{0}$.
By fitting the population of $\ket{0}$ with an exponential function, we can extract $T_1$.
Hence, the fitting function has the following form:
\begin{align}
f(\tau) = A_0 + A_1 \exp(-\frac{\tau}{T_1}),
\end{align}
where $\tau$ is the time interval between the drive and readout pulses, and the fitting parameters are $A_0$, $A_1$, and $T_1$.

\subsubsection{Relaxation Time: $T_2$}
\label{sec:Ramsey}

The standard measurement procedure for the transverse relaxation time ($T_2$) is to observe the Ramsey fringes [Fig.~\ref{fig:pulseSequence}(d)].
Since we need to detect the transverse component of the Bloch sphere, we apply a $\pi/2$ pulse at the beginning of the time interval and then a $-\pi/2$ pulse to transfer the transverse component back to the longitudinal component.
Subsequently, we make a detection.
The time constant for the decay of the $\ket{0}$ population is $T_2$.

If $\omega_\textrm{d} \! = \! \omega_\textrm{q}$ (on-resonance), we observe single exponential decay shown on the left side of Fig.~\ref{fig:pulseSequence}(d).
However, if $\omega_\textrm{d} \! \neq \! \omega_\textrm{q}$ (off-resonance), the Bloch vector will acquire an additional rotation in the $x'y'$ plane during the time interval, resulting in an oscillation with the frequency $\omega_\textrm{d}-\omega_\textrm{q}$.
The reason for the oscillation is that only the transverse component perpendicular to the rotating axis is transferred to the longitudinal component.
This oscillation is called Ramsey fringes.

The fitting function is
\begin{align}
f(\tau) = A_0 + A_1 \cos(\omega_\textrm{qd} \tau + A_2) \exp(-\frac{\tau}{T_2}),
\end{align}
where $\tau$ is the time interval between the $\pm\pi/2$ pulses, and the fitting parameters are $A_0$, $A_1$, $A_2$, $\omega_\textrm{qd}$, and $T_2$.
Here, $\omega_\textrm{qd}$ is the frequency detuning ($\equiv \omega_\textrm{d}-\omega_\textrm{q}$) from which we can obtain a precise value of $\omega_\textrm{q}$.

If the dephasing process is dominated by low-frequency noise, the line shape of the qubit spectrum is close to Gaussian rather than Lorentzian.\cite{yoshihara2006, bylander2011}
This happens easily when the flux bias is out of the sweet spot.
In this case, the fitting function ($\omega_\textrm{qd}$ is assumed to be zero for simplicity)
\begin{align}\label{eq:T2fit2}
f(\tau) = A_0 + A_1 \exp(-\frac{\tau^2}{T_\textrm{G}^2}) \exp(-\frac{\tau}{2T_1})
\end{align}
fits the data better than the simple exponential decay function because the Fourier transformation of a Gaussian is also a Gaussian.
Here, $T_1$ is not a fitting parameter;
it has to be measured independently (Sec.~\ref{sec:T1}).
The fitting parameter $T_\textrm{G}$ can be practically considered as $1/\Gamma_\varphi$, although it is not mathematically identical to $1/\Gamma_\varphi$ because $\Gamma_\varphi$ is defined in an exponential decay function (see Sec.~\ref{sec:relaxConcept}).


\section{Controlling a Quantum System}
\label{sec:control}


In this section, we discuss how to control a superconducting qubit system.
As a quantum computing system becomes larger, its precise control becomes as important as making the system itself.
Without efficient control, we cannot reduce errors enough to perform quantum error correction.

What does ``controlling a quantum system'' mean?
Consider the Hamiltonian
\begin{equation}\label{eq:qControl}
\hat{\mathcal{H}}(t) = 
\hat{\mathcal{H}}_0 + \sum_{k=1}^K u_k(t) \hat{\mathcal{H}}_k,
\end{equation}
where $\hat{\mathcal{H}}_0$ is the system Hamiltonian, $\hat{\mathcal{H}}_k$ is the control Hamiltonian, $u_k$ are the control parameters, and $K$ is the number of the control parameters.
Controlling a quantum system means finding $u_k(t)$ that drive a quantum state of the system to the desired state.\cite{d'alessandro}
Single qubit gates in Sec.~\ref{sec:SQG} are good examples.
In this case, the control Hamiltonian is $\hat{\mathcal{H}}_\textrm{d}$ [Eq.~\eqref{eq:drive}], and the control parameters are $\mathcal{E}_\textrm{r}(t)$ and the phase of the drive pulse, which selects the rotation axis.

To achieve a high-fidelity gate operation, a control pulse must satisfy the following conditions:
\begin{enumerate}
\itemsep-0.1em 

\item \emph{Fast qubit manipulation}: 
The control pulse must be as short as possible to avoid loss of coherence.

\item \emph{Narrow excitation bandwidth}: 
The excitation bandwidth has to be sufficiently narrow to minimize the information leakage through an unwanted transition.
For a multiqubit system, the excitation bandwidth of the readout pulse is also important because the readout of a certain qubit might induce the dephasing of other qubits by populating readout resonators for these qubits (see Sec.~\ref{sec:readout}).

\item \emph{Decoupling from unwanted interactions}: 
Couplings to uncharted or unaccountable external degrees of freedom, which induce unwanted interactions and information leakage, should be minimized.

\item \emph{Self-compensating}: 
The gate operation must compensate for the complex response of classical control electronics: bandwidth and long-time transients or nonlinearity such as kinetic inductance in a superconducting resonator, amplifier nonlinearity, or mixer imbalance.
Although these nonlinearity is controllable in principle,\cite{hincks2015} it is a difficult task.
Thus, the best option is to minimize any unnecessary nonlinearity by careful device design and operating the amplifier in the linear regime, which will set the power limit.

\item \emph{Robustness against experimental imperfections}: 
The control must be robust against uncertainties and stochastic variations in the system's internal and control Hamiltonians, such as amplitude and phase noises in the control pulse.

\end{enumerate}

This section is presented to answer the question of how to optimize the control pulse to satisfy the above criteria.
A good introduction to this topic can be found in Refs.~\onlinecite{wilhelm,rembold2020}.

\subsection{Elementary Pulse Shaping}
\label{sec:easyPulse}

\subsubsection{Excitation Bandwidth}
\label{sec:bandWidth}

In this subsection, we estimate the excitation bandwidth by Fourier transforming the pulse applied to the qubit.
We stress that this method is valid only if the system's response is linear, while the evolution on the Bloch sphere is intrinsically nonlinear.\cite{ernst, freeman}
To control the excitation bandwidth precisely, we must solve the full equation of motion of the system.
Here, we use the concept of Fourier transformation because it simplifies our discussion and helps develop intuition.

\begin{figure}
\centering
\includegraphics{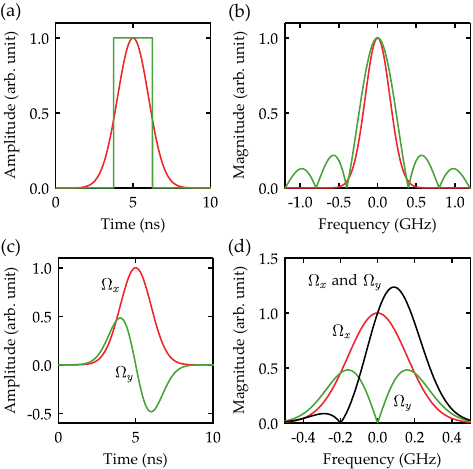}
\caption{(a) Square (green) and Gaussian (red) pulses with the same area in the time domain.
(b) In the frequency domain, it is evident that the excitation bandwidth of a Gaussian pulse (red) is narrower than that of a square pulse (green). (The Fourier transformation of a Gaussian is also Gaussian.)
(c) and (d) DRAG pulse implemented in Gaussian pulse with anharmonicity $\alpha = 2\pi(-200)$ MHz. $\Omega_x$ and $\Omega_y$ indicate the quadratures of the pulse [Eq.~\eqref{eq:ctrl_DRAG}].
In (d), note that the Fourier component of the DRAG pulse (black curve) is suppressed at the position of anharmonicity, resulting in the suppression of the information leakage from $\ket{1}$ to $\ket{2}$.
}
\label{fig:DRAG}
\end{figure}

To grasp the concept of controlling the excitation bandwidth, we compare two widely used pulses: square and Gaussian pulses.
As shown in Fig.~\ref{fig:DRAG}(a) and (b), the excitation bandwidth of a Gaussian pulse is significantly narrower than that of a square pulse.\cite{bauer1984}
One drawback of a Gaussian pulse is that it does not have well-defined starting and end points; hence, the pulse envelope must be truncated somewhere.\cite{gambetta2011}
Because of this, a cosine pulse is also widely used.

For qubits like transmons, however, a Gaussian pulse is still not enough to perform nanosecond qubit control because of the bandwidth of 100 MHz order, which is comparable to the typical anharmonicity of a transmon [Fig.~\ref{fig:DRAG}(b)].
A pulse with this bandwidth is likely to induce transitions not only between $\ket{0}$ and $\ket{1}$ but also between $\ket{1}$ and $\ket{2}$, resulting in information leakage out of the computational subspace.
A more advanced pulse resolving this issue is the Derivative Removal by Adiabatic Gate (DRAG) pulse.\cite{motzoi2009}
The DRAG scheme is very effective for weakly nonlinear qubits, such as transmons;
all high-fidelity controls achieved in superconducting qubit systems are based on the DRAG pulse and its variants.

Here, we explain the DRAG pulse in an intuitive way using the Fourier transformation following Ref.~\onlinecite{motzoiThesis}.
Consider a single qubit driven by the following Hamiltonian (in the rotating frame):
\begin{equation} \label{eq:ctrl_DRAG}
\hat{\mathcal{H}}_\textrm{d}^\textrm{rot} =
\hbar\Omega_x(t)\frac{\hat{\sigma}_x}{2}
+ \hbar\Omega_y(t)\frac{\hat{\sigma}_y}{2},
\end{equation}
where $\Omega_{x(y)}$ is the amplitude for the rotation about the $x(y)$-axis.
This is just an extension of Eq.~\eqref{eq:driveRot}.
For a qubit system whose anharmonicity is $\alpha$, the DRAG condition is given by 
\begin{equation}\label{eq:DRAG}
\Omega_y(t) = -\frac{\dot{\Omega}_x(t)}{\alpha}.
\end{equation}
That is, the imaginary part of the pulse is the derivative of the real part as shown in Fig.~\ref{fig:DRAG}(c).

In the frequency domain, taking the derivative gives sharp suppression at a certain frequency [black curve in Fig.~\ref{fig:DRAG}(d)].
The coefficient $-1/\alpha$ in Eq.~\eqref{eq:DRAG} matches this suppressed frequency and the transition frequency between $\ket{1}$ and $\ket{2}$.
This idea is very intuitive and easy to apply; for example, if we want to suppress two frequencies, one higher and the other lower than our working frequency, we can simply take a second derivative of the pulse as the imaginary part.\cite{motzoiThesis}
Although we explained the DRAG scheme for a Gaussian pulse, the concept of the DRAG pulse can also be applied to other pulse shapes.

One interesting point of view is to consider the DRAG scheme as an extension of ``shortcuts to adiabaticity'', which are fast routes to the final
results of slow, adiabatic changes of the controlling parameters of a system.\cite{STAreview}
From this point of view, the imaginary part of the pulse makes the qubit couple to the non-computational subspace only adiabatically, resulting in the removal of population leakage.\cite{DRAGreview}
Interested readers should see Refs.~\onlinecite{STAreview, DRAGreview}.

\subsubsection{Pulse Distortion}
\label{sec:pulseDist}

\begin{figure}
\centering
\includegraphics{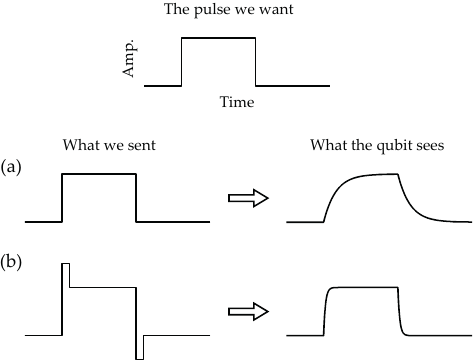}
\caption{Pulse distortion due to the finite response time of various electronics.
}
\label{fig:pulseDist}
\end{figure}

Although we carefully design a microwave or a DC pulse for qubit control, the shape of the pulse is distorted as it travels from the pulse generator to the qubit.
This is mainly due to the finite bandwidth of various electrical components, such as cables, filters, and resonators.
The problem is even worse for the readout pulse because, in this case, the pulse is close to the resonance of the high-quality superconducting resonator.
A typical example of pulse distortion is shown in Fig.~\ref{fig:pulseDist}(a).
If the time scale of the transient is not negligible compared with $T_1$, which is often the case, the readout fidelity will be limited.
The simplest solution is to add overshoot and negative pulses at the beginning and end of the readout pulse as shown in Fig.~\ref{fig:pulseDist}(b).\cite{hoult1979}

A more advanced way to solve this problem is to model the transient behavior using various filter functions\cite{butscherThesis, rol2020} or an RLC circuit.\cite{borneman2012, hincks2015}

\subsection{Numerical Optimization}
\label{sec:numOpt}

The goal of optimal control is to find a set of control parameters that minimize the cost function.
Here, the cost function, also called the performance measure, is a measure of how far the performance of a system is from the target one.\cite{kirk}
If our interest is unitary gate optimization, gate infidelity $J_\textrm{gate}$ is a natural choice as the cost function.
Mathematically, $J_\textrm{gate}$ is defined by
\begin{align}\label{eq:gateInfid}
J_\textrm{gate} \equiv 1-\Big|\textrm{tr}\Big\{ \hat{U}_\textrm{target}^\dagger \hat{U}(T) \Big\} \Big|^2,
\end{align}
where $\hat{U}_\textrm{target}$ and $\hat{U}(T)$ are the target and the final unitary gates, respectively.
For ideal gate operation, $J_\textrm{gate}=0$.
If we want to let the system evolve to the target state (state-to-state transfer), state infidelity $J_\textrm{state} (\equiv 1-\left|\braket{\psi_\textrm{target}}{\psi(T)} \right|^2)$ is used as the cost function.

There are two types of optimization algorithm:
gradient-based and gradient-free.
In gradient-based algorithms, the performance of the pulse is evaluated by the cost function.
Then, the control parameters are updated for the next iteration based on the derivative of the cost function with respect to the control parameters.
The advantage of gradient-based algorithms is that they are much faster than gradient-free algorithms.
However, gradient-based algorithms do not work well if the cost function is discontinuous or noisy because calculating the gradient is difficult in such a case.
Thus, they are difficult to use for direct optimization on an experimental system, i.e., closed-loop optimal control.

Despite their slowness, gradient-free algorithms are simple to implement and work well with a noisy measurement outcome, suggesting that we can perform closed-loop optimal control using these algorithms.
Thus, gradient-free algorithms are useful for the calibration or optimization of pulses defined by a limited number of parameters.\cite{kelly2014, werninghaus2020}

There are many numerical optimization algorithms that have been used widely in quantum control.
One helpful ``decision tree'' for the choice of an optimization algorithm is available in Ref.~\onlinecite{optTree}.

\subsubsection{GRAPE Algorithm}

We introduce GRadient Ascent Pulse Engineering (GRAPE)\cite{khaneja2005} because it has been found to be one of the most reliable algorithms for controlling quantum systems.
As the name implies, GRAPE is a gradient-based algorithm.
Hence, it is fast but the control parameters are determined in simulations.
This suggests that the GRAPE algorithm requires complete system modeling.

\begin{figure}
\centering
\includegraphics{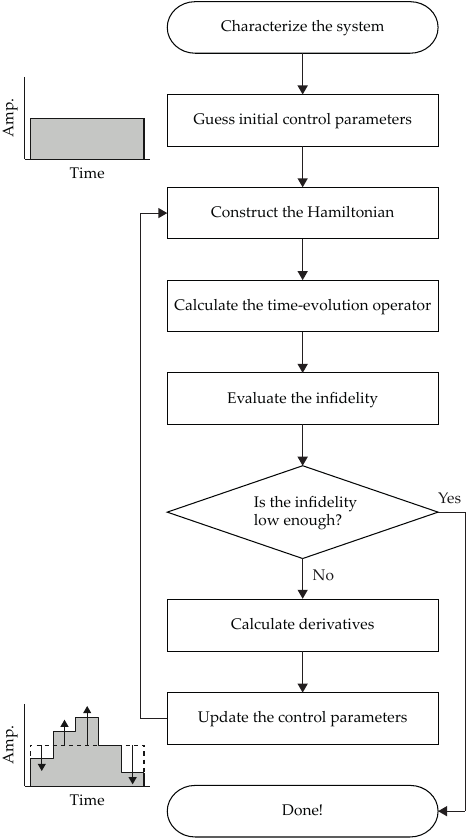}
\caption{Procedure for typical gradient-based numerical pulse optimization.
The upper inset shows an example of the initial pulse shape.
In the lower inset, the dashed line indicates the initial pulse, and the vertical arrows represent gradients, indicating how each amplitude should be modified in the next iteration in order to reduce the cost function.
The solid line represents the resulting updated pulse shape.
}
\label{fig:pulseOpt}
\end{figure}

In the GRAPE algorithm, the control pulses are defined as a collection of pulses with a piecewise constant amplitude and phase over a number of intervals $N$, each of length $\Delta t$, which yields an overall pulse length of $T=N\Delta t$ [see Figs.~\ref{fig:pulseOpt}(insets) and \ref{fig:GRAPE}].
Hence, the control parameters are a set of amplitudes and phases (or amplitudes of two quadratures) of these piecewise pulses.

A brief description of the GRAPE algorithm for unitary gate optimization is as follows (Fig.~\ref{fig:pulseOpt}):
\begin{enumerate}
\itemsep-0.1em 

\item \emph{Characterize the system}:
In this step, the system parameters are extracted as described in Sec.~\ref{sec:characterize}.
From these parameters, we set up the model Hamiltonian that describes the dynamics of the system properly [$\hat{\mathcal{H}}_0$ in Eq.~\eqref{eq:qControl}].

\item \emph{Guess initial control parameters and construct the Hamiltonian}: 
With a set of initial control parameters, we can construct the full Hamiltonian [$\hat{\mathcal{H}}$ in Eq.~\eqref{eq:qControl}].
An example of the initial pulse shape is shown in the upper inset of Fig.~\ref{fig:pulseOpt}.

\item \emph{Calculate the time-evolution operator}: 
The time evolution of the system during a time step $j$ is given by the operator [Eq.~\eqref{eq:neumannSol}]
\begin{align}
\hat{U}_j = \exp[-\textrm{i} \, \frac{\Delta t}{\hbar} \left\{ \hat{\mathcal{H}}_0 + \sum_k^K u_k(j) \hat{\mathcal{H}}_k \right\} ].
\end{align}
Thus, $\hat{U}(T) = \hat{U}_N \cdots \hat{U}_1$.

\item \emph{Evaluate the infidelity}: 
In this step, it is determined how close the resulting state is to the target one via $J_\textrm{gate}$.

\item \emph{Calculate derivatives and update the control parameters}:
The update of the control parameters can be written as
\begin{equation}
u_k(j) \rightarrow u_k(j) + \epsilon\frac{\delta J_\textrm{gate}}{\delta u_k(j)},
\end{equation}
where $\epsilon$ is an adjustable small step size.
An example is shown in the lower inset of Fig.~\ref{fig:pulseOpt}.
For small enough $\Delta t$, the gradients are calculated using the following formula:\cite{khaneja2005}
\begin{align}
\frac{\delta J_\textrm{gate}}{\delta u_k(j)} 
\approx
-2\Re \left\{
\textrm{tr} \left( \textrm{i} \frac{\Delta t}{\hbar} \hat{\mathcal{H}}_k \hat{V}_j \hat{\Lambda}_j^\dagger \right)
\textrm{tr}\left( \hat{V}_j^\dagger \hat{\Lambda}_j \right)
\right\},
\end{align}
where $\hat{V}_j$ and $\hat{\Lambda}_j$ are defined by
\begin{align}
J_\textrm{gate} 
&= 1 - 
\Big|\textrm{tr}\Big( \underbrace{\hat{U}_\textrm{target}^\dagger \hat{U}_{N} \cdots \hat{U}_{j+1}}_{\equiv \hat{\Lambda}_j^\dagger} \, \underbrace{\hat{U}_j \cdots \hat{U}_1}_{\equiv \hat{V}_j} \Big) \Big|^2 \nonumber\\
&= 1 - 
\textrm{tr}\Big( \hat{V}_j \hat{\Lambda}_j^\dagger \Big) \,
\textrm{tr}\Big( \hat{V}_j^\dagger \hat{\Lambda}_j \Big).
\end{align}
Here, the property that the trace is invariant under cyclic permutations was used.

\end{enumerate}

Note that the piecewise constant amplitude and phase assumption is not valid in reality because of the pulse distortion mentioned in Sec.~\ref{sec:pulseDist}.
However, this does not make the pulse optimization particularly harder.
Once the transient behavior is suitably modeled, it can be integrated into the GRAPE algorithm easily by introducing an additional level of discretization.
One discretization is for integrating the state evolution and the other is for the control parameters.\cite{borneman2012, quantumUtils, motzoi2011}

\subsubsection{Example: Controlling Excitation Bandwidth}

\begin{figure}
\centering
\includegraphics{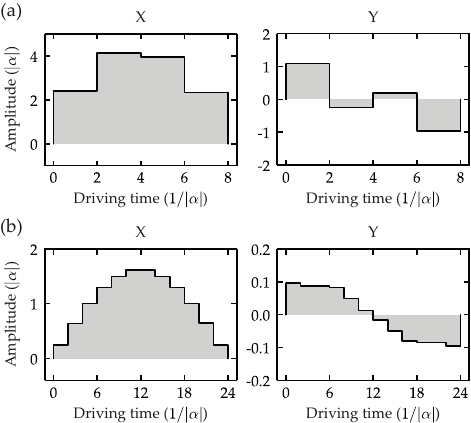}
\caption{Examples of optimal solutions (smallest length and maximum amplitude) for $\pi$-rotation obtained from the GRAPE algorithm.
The fidelity is higher than $0.99999$.
The left figures show the $\Omega_x$ (X) quadrature and the right figures show the $\Omega_y$ (Y) quadrature.
(a) When the width of the time slice is $2/|\alpha|$, the minimal number of slices required to reach $>\!0.99999$ fidelity is 4.
(b) If there are amplitude constraints caused by hardware limitations, we can find an optimal solution with a longer gating time in a limited power range.
Note that the amplitude of the Y quadrature is only 1/10 that in (a).
This is due to the longer pulse having narrower bandwidth; thus, less phase correction is needed.
Also note that the shape of the Y quadrature is close to the derivative of the X quadrature, implying that GRAPE takes the idea of the DRAG scheme.
For a system with $\alpha = 2\pi(-200)$ MHz, the unit step $1/|\alpha|$ corresponds to 0.8 ns. In this case, the gating times for (a) and (b) are 6.4 ns and 19.2 ns, respectively.
The initial control parameters for the X quadrature are a Gaussian-like shape whose area is roughly $\pi$; for the Y quadrature, they are all zeros.
For these figures, the QuantumUtils package was used.\cite{quantumUtils}
}
\label{fig:GRAPE}
\end{figure}

Here, we use GRAPE to control the excitation bandwidth.
For simplicity, we consider a single transmon system (Sec.~\ref{sec:transmon}) controlled by an on-resonance external drive, $\omega_\textrm{d}=\omega_\textrm{q}$.

\paragraph{Goal.}

The quantum state of a transmon can evolve easily outside of the computational subspace because of the small anharmonicity.
This causes the qubit state to acquire an unwanted phase that degrades the fidelity.
Hence, our goal is to find the shortest pulse shape that suppresses the qubit evolution outside of the computational subspace.
If our target gate $U_\textrm{target}$ is the $X$ gate, it can be written in matrix form as
\begin{align}
U_\textrm{target}\, = 
\raise 1.6ex \hbox{$
\begin{array}{l c ccc c}
& & \bra{0} & \bra{1} & \bra{2} & \\[3pt]
\ket{0}\phantom{,}\!\!\!\!\! & \multirow{3}{*}{\midvast(\!\!} & 0 & 1 & 0 & \multirow{3}{*}{\!\!\midvast)} \\
\ket{1}\!\!\!\!\! & & 1 & 0 & 0 & \\
\ket{2}\!\!\!\!\! & & 0 & 0 & \textrm{e}^{\textrm{i}\phi_2} &
\end{array}$}.
\end{align}
where $\phi_2$ is the relative phase of $\ket{2}$ acquired during the evolution.
Although we do not need $\phi_2$ for the gate operation, the fidelity must be calculated as a function of $\phi_2$ because only a certain range of $\phi_2$ gives a high-fidelity gate solution.

\paragraph{Hamiltonian.}

Our system Hamiltonian $\hat{\mathcal{H}}_0$ in the rotating frame, whose angular velocity is the same as $\omega_\textrm{q}$, is given by
\begin{equation}
\hat{\mathcal{H}}_0 = 
\begin{pmatrix}
0 	&	0	&	0 \\ 
0	&	0	&	0 \\
0	&	0	&	\alpha
\end{pmatrix},
\end{equation}
where $\alpha$($\equiv\! \omega_{12}-\omega_\textrm{q}$) is the anharmonicity.

The control parameter is the amplitude of the external drive.
Thus, the control Hamiltonian $\hat{\mathcal{H}}_\textrm{ctrl}$ is given by
\begin{equation}
\hat{\mathcal{H}}_\textrm{ctrl}(t) = 
\frac{\hbar\Omega_x(t)}{2}
\begin{pmatrix}
0 	&	1	&	0 \\ 
1	&	0	&	\lambda \\
0	&	\lambda	&	0
\end{pmatrix}
+
\frac{\hbar\Omega_y(t)}{2}
\begin{pmatrix}
0 	&	-\textrm{i}	&	0 \\ 
\textrm{i}	&	0	&	-\textrm{i}\lambda \\
0	&	\textrm{i}\lambda	&	0
\end{pmatrix}.
\end{equation}
Here, $\Omega_{x(y)}$ is the amplitude of the rotation about the $x(y)$-axis, and the matrix next to $\Omega_{x(y)}$ is the three-level-system version of the $\hat{\sigma}_{x(y)}$ operator with the relative strength of the $\ket{1}$-$\ket{2}$ transition with respect to the $\ket{0}$-$\ket{1}$ transition denoted as $\lambda$. 
For a transmon, $\lambda = \sqrt{2}$ (Ref.~\onlinecite{koch2007}).
We also use this number.
We emphasize that the $y$-rotation is implemented by $-\sin(\omega_\textrm{d}t)$, not by $+\sin(\omega_\textrm{d}t)$, as mentioned in Sec.~\ref{sec:SQG}.

\paragraph{Constraint.}

The constraint is usually set by our instruments.
In this case, it is the maximal power of the drive.

\paragraph{Result.}

The optimal pulse shape found by GRAPE is shown in Fig.~\ref{fig:GRAPE}.
If the system is an ideal two-level system, i.e., $\lambda = 0$, we do not need one of the quadratures, say $\Omega_y$, and any shape satisfying $\int_0^{\tau} \mathcal{E}_x(t)dt = \pi$ achieves a perfect $\pi$-rotation, where $\tau$ is the total gate time.
However, because of the existence of $\ket{2}$, the numerical solutions found by GRAPE have $\Omega_y(t)$ whose shape is similar to that of the DRAG pulse as shown in Fig.~\ref{fig:GRAPE}.

\subsection{Refocusing Technique}
\label{sec:refocusing}

\begin{figure*}
\centering
\includegraphics{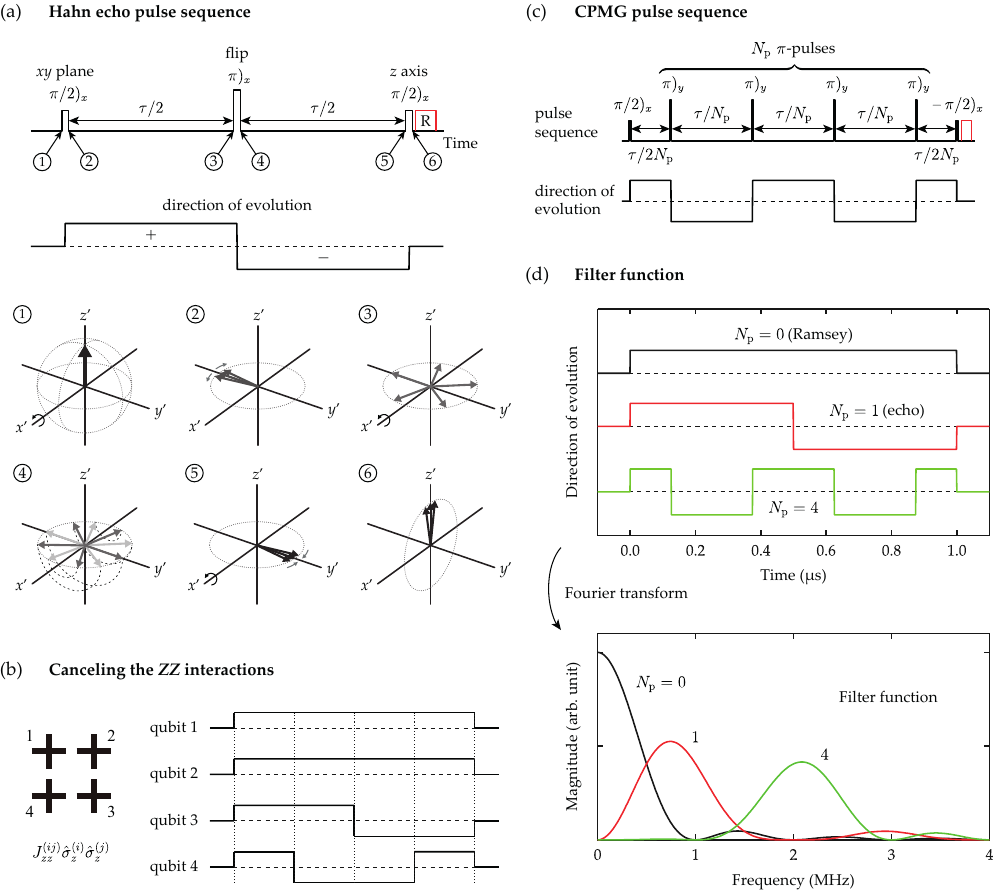}
\caption{
(a) Hahn echo pulse sequence.
$\tau$ is the time interval of the pulse sequence.
The red pulse labeled ``R'' is the readout pulse.
The circled numbers describe the evolution of the qubit state in the Bloch sphere. (The thermalization process is ignored for clarity.)
(b) Refocusing scheme for a four-qubit system (four crosses) designed to control the $ZZ$ interactions between qubit $i$ and qubit $j$, $J_{zz}^{(ij)}$ (Ref.~\onlinecite{VC}).
In this sequence, all possible $J_{zz}^{(ij)}$ will be canceled out, except $J_{zz}^{(12)}$.
The time interval is divided into four slices of equal duration by dotted lines.
(c) Carr--Purcell--Meiboom--Gill (CPMG) pulse sequence.
The red pulse is the readout pulse.
Note that $\pi$-rotations are about the $y$-axis, while $\pi/2$-rotations are about the $x$-axis.
This change of the rotation axis prevents the accumulation of the error caused by imperfect $\pi$-pulses.\cite{slichter}
(d) Concept of filter function.
The modulation of the direction of evolution acts as a filter function in the frequency domain.
}
\label{fig:CPMG}
\end{figure*}

Although a qubit must interact with external systems to be controlled or read, the interaction must be turned on only when we need it.
Any unwanted interaction makes the qubit lose its coherence and degrades the performance of gate operation.
The problem is that interactions associated with qubits are not always controllable;
even if we can control some of the interactions using a tunable coupler or flux bias, such control knobs always introduce additional noises.
A refocusing technique\cite{sodickson1998} allows us to cancel the evolution caused by unwanted interactions by applying appropriate pulses that can effectively reverse the direction of evolution.
In this context, such a pulsing technique is also called dynamical decoupling.\cite{souza, lidar}

The simplest and most well-known refocusing scheme is the Hahn echo sequence [Fig.~\ref{fig:CPMG}(a)].\footnote{Although the majority of the literature, including this tutorial, calls the pulse sequence in Fig.~\ref{fig:CPMG}(a) the Hahn echo sequence, it was proposed by Carr and Purcell.\cite{carr1954} In Hahn's original paper, he applied pulses with the same rotation angle.\cite{hahn1950}}
The Bloch spheres with circled numbers show how the qubit state evolves in repeated identical measurements.
If the transition frequency of a qubit fluctuates for each measurement because of noise or unwanted interactions, then the qubit will lose its phase coherence [\raisebox{.5pt}{\textcircled{\raisebox{-.9pt}{3}}} in Fig.~\ref{fig:CPMG}(a)] (Sec.~\ref{sec:dephasing}).
Here, the role of the $\pi$ pulse is to flip the population of the qubit, thus reversing the direction of evolution [\raisebox{.5pt}{\textcircled{\raisebox{-.9pt}{4}}} in Fig.~\ref{fig:CPMG}(a)].
This makes the net area of the ``direction of evolution'' in Fig.~\ref{fig:CPMG}(a) zero, indicating that the unwanted evolution is canceled out.
Thus, $T_2$ measured using this pulse sequence is usually much longer than that measured via the Ramsey fringes (Sec.~\ref{sec:Ramsey}).
To distinguish the relaxation time constants obtained from the Ramsey fringes and Hahn echo sequence, notations such as $T_2^*$ and $T_2^\textrm{echo}$ are often used.

Now, we explain more precisely how the Hahn echo sequence works.
The first thing we have to do is distinguish the state preparation from the actual refocusing operation.
For this, we rewrite the Hahn echo sequence as
\begin{equation}\label{eq:hahnEcho}
\pi/2)_x \textrm{ - } \tau/2 \textrm{ - } \pi)_x 
\textrm{ - } \tau/2 \textrm{ - } \pi)_x \textrm{ - } \!-\!\pi/2)_x.
\end{equation}
Combining the last two pulses gives the sequence in Fig.~\ref{fig:CPMG}(a).
Note that, in the pulse sequence for the Ramsey fringes [Fig.~\ref{fig:pulseSequence}(d)], the $\pm \pi/2$-pulse transfers the longitudinal component to the transverse component and vice versa for the state preparation and measurement.
The role of the $\pm \pi/2$-pulse in Eq.~\eqref{eq:hahnEcho} is the same---it is not an essential part of the refocusing process.
The refocusing for an arbitrary qubit state is done by two $\pi$-pulses.

Consider an interaction between a qubit of interest and the coupled system, such as other qubits or the environment, with the following form:
\begin{align}\label{eq:qeInter}
\hat{\mathcal{H}}_\textrm{qe} = 
\hbar J_{z} \hat{\sigma}_z^{(\textrm{q})} \hat{A}^{(\textrm{e})},
\end{align}
where $J_{z}$ is the coupling constant,
$\hat{\sigma}_z^{(\textrm{q})}$ is the Pauli operator for the qubit, and
$\hat{A}^{(\textrm{e})}$($\in \{\hat{I}^{(\textrm{e})}, \hat{\sigma}_x^{(\textrm{e})}, \hat{\sigma}_y^{(\textrm{e})}, \hat{\sigma}_z^{(\textrm{e})} \}$, where $\hat{I}$ is the identity operator) is an operator associated with the coupled system.
If we ignore this interaction during the time when the pulse is turned on, i.e., the pulse length is very small compared with $\tau$, the time-evolution operator for the Hahn echo sequence can be written as [using Eq.~\eqref{eq:neumannSol}]
\begin{align}
\hat{U}_\textrm{echo} =
\hat{X}\,\hat{U}_\textrm{qe}(\tau/2)\,\hat{X}\,\hat{U}_\textrm{qe}(\tau/2),
\end{align}
where
\begin{align}
\hat{U}_\textrm{qe}(t) =
\textrm{e}^{-\textrm{i} \hat{\mathcal{H}}_\textrm{qe} t / \hbar},
\end{align}
and $\hat{X}$ is the $X$ gate acting on the qubit.
Using the two identities
\begin{equation}
\hat{U}\textrm{e}^{\hat{B}}\hat{U}^\dagger = \textrm{e}^{\hat{U}\hat{B}\hat{U}^\dagger}, \quad
\hat{\sigma}_x \hat{\sigma}_z \hat{\sigma}_x = - \hat{\sigma}_z,
\end{equation}
where $\hat{U}$ is a unitary operator and $\hat{B}$ is an arbitrary operator,
it is easy to show that
\begin{align}
\hat{X}\,\hat{U}_\textrm{qe}(\tau/2)\,\hat{X}
&=
\hat{\sigma}_x^{(\textrm{q})}\,
\textrm{e}^{-\textrm{i} \hat{\mathcal{H}}_\textrm{qe} \tau /2 \hbar}\,
\hat{\sigma}_x^{(\textrm{q})}
\nonumber\\
&=
\textrm{e}^{-\textrm{i} \hat{\sigma}_x^{(\textrm{q})}\, \hat{\mathcal{H}}_\textrm{qe}\,\hat{\sigma}_x^{(\textrm{q})}\, \tau /2 \hbar}
\nonumber\\
&=
\textrm{e}^{-\textrm{i} J_{zz}
\left( \hat{\sigma}_x^{(\textrm{q})}\hat{\sigma}_z^{(\textrm{q})}\hat{\sigma}_x^{(\textrm{q})} \right)
\hat{A}^{(\textrm{e})} \tau /2}
\nonumber\\
&=
\textrm{e}^{-\textrm{i} J_{zz}
\left( -\hat{\sigma}_z^{(\textrm{q})} \right)
\hat{A}^{(\textrm{e})} \tau /2}
\nonumber\\
&= \hat{U}_\textrm{qe}(-\tau/2), \label{eq:timeRev}
\end{align}
where the global phase was ignored.
This suggests that two $\pi$-pulses act as a time-reversal operator.
Thus, the Hahn echo sequence cancels any static interaction in the form of $\hat{\sigma}_z^{(\textrm{q})} \hat{A}^{(\textrm{e})}$.

Note that the Hahn echo sequence also cancels the $\hat{\sigma}_y^{(\textrm{q})} \hat{A}^{(\textrm{e})}$ interaction because $\hat{\sigma}_x \hat{\sigma}_y \hat{\sigma}_x = - \hat{\sigma}_y$.
If we use an additional axis for $\pi$-pulses, either the $y$- or $z$-axis, we can also remove the $\hat{\sigma}_x^{(\textrm{q})} \hat{A}^{(\textrm{e})}$ interaction.
Here, we choose the $z$-axis as an example: 
\begin{align}
\hat{U}_\textrm{XY4} =&
\hat{Z}\,\hat{U}_\textrm{echo}\,\hat{Z}\,\hat{U}_\textrm{echo} \nonumber\\
=&
\hat{Z}\hat{X}\,\hat{U}_\textrm{qe}\,\hat{X}\,\hat{U}_\textrm{qe}\,
\hat{Z}\hat{X}\,\hat{U}_\textrm{qe}\,\hat{X}\,\hat{U}_\textrm{qe} \nonumber\\
=&
\hat{Y}\,\hat{U}_\textrm{qe}\,\hat{X}\,\hat{U}_\textrm{qe}\,
\hat{Y}\,\hat{U}_\textrm{qe}\,\hat{X}\,\hat{U}_\textrm{qe},
\label{eq:XY4}
\end{align}
where $\hat{Z}$($\hat{Y}$) is the $Z$($Y$) gate acting on the qubit, and we omit the time interval in $\hat{U}_\textrm{qe}$ to simplify the notation.
This pulse sequence is called XY4 (Refs.~\onlinecite{maudsley1986, gullion1990}).

The Hahn echo sequence is also useful for a multiqubit system.
Figure~\ref{fig:CPMG}(b) shows an example of engineering $ZZ$ interactions in a four-qubit system.\cite{VC}
We can intuitively see which interaction will be eliminated by multiplying the two directions of evolution curves and checking if the net area is zero.

One might notice that, if we use the Hahn echo sequence to remove the $ZZ$ interaction for the CR gate operation as mentioned in Sec.~\ref{sec:CR},
our gate operation based on the $ZX$ interaction will also be removed.
To prevent such a situation, we apply the $\pi$-pulses to the control qubit and change the sign of the CR drive after the first $\pi$-pulse.\cite{sheldon2016}
As a result, the $ZX$ interaction survives, while the $ZZ$ interaction is canceled out.

The Hahn echo sequence works well only when the unwanted interaction is static in the time scale of $\tau$.
This means that only low frequency noise can be canceled out by the Hahn echo sequence.
The Carr--Purcell--Meiboom--Gill (CPMG) pulse sequence [Fig.~\ref{fig:CPMG}(c)], an extension of the Hahn echo sequence, can remove a wider frequency range of noise by applying multiple $\pi$-pulses.\cite{carr1954,meiboom1958}
This can be understood using the concept of the filter function, which is basically the Fourier transformation of the direction of evolution.\cite{bylander2011, cywinski2008}
Figure~\ref{fig:CPMG}(d) clearly shows that increasing the number of $\pi$-pulses filters out a wider range of low-frequency noise.

As the number of qubits increases, the dimension of the Hilbert space associated with the qubits increase exponentially and cannot be efficiently simulated on a classical computer, which is a requirement for optimization as suggested in Sec.~\ref{sec:numOpt}.
Furthermore, the mere task of perfectly characterizing all multiqubit Hamiltonian terms becomes impossible.
Nevertheless, the increase in $T_2$ by refocusing techniques shows us that we can engineer the system dynamics without full knowledge of the system-environment couplings and the state of the environment in some large Hilbert space that cannot be characterized.
This success of the refocusing techniques arises from treating the problem perturbatively for a class of possible perturbations.
These same ideas can be employed when controlling multiqubit systems whose Hilbert space can be divided into small manageable sectors (e.g., either one or a few qubits) with unwanted/uncharacterized Hamiltonian terms treated as perturbations.
An approach based on this philosophy is effective Hamiltonian engineering.
The most popular theoretical tool for this is average Hamiltonian theory.
Interested readers should see Refs.~\onlinecite{VC, ernst, haeberlen, brinkmann, haas2019}.

\subsection{Evaluation of Gate Operation}
\label{sec:RB}

\begin{figure*}
\centering
\includegraphics{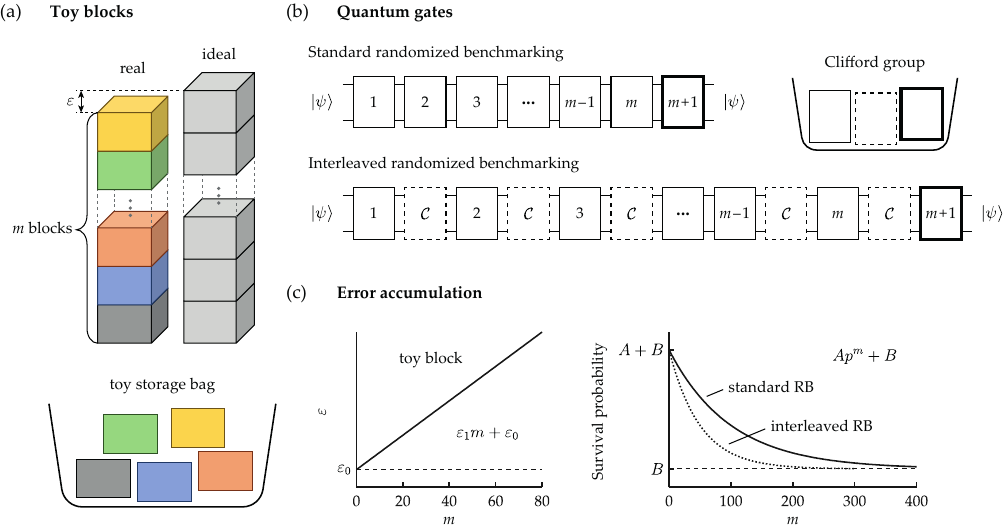}
\caption{(a) Toy block analogy of the Clifford group randomized benchmarking (standard RB).
The toy storage bag corresponds to the Clifford group,
and each toy block corresponds to a single gate operation.
$\varepsilon$ is the cumulative error in the total height.
(b) In the standard RB, $m$ gates randomly selected from the Clifford group are applied sequentially.
In the interleaved RB, a random gate and the gate of interest $\mathcal{C}$ (dashed rectangle) appear alternately.
Here, the $(m+1)$th gate (thick rectangle) is chosen so that the net sequence is the identity operation.
(c) In the toy block analogy, the cumulative height error $\varepsilon$ is fitted with the function $\epsilon_1 m + \epsilon_0$.
This separates the average height error $\epsilon_1$ (what we want) from an unintended position offset $\epsilon_0$ (what we do not want).
In RB, the survival probability, which is the probability that the initial state is not changed by the gate sequence, is fitted with the function $Ap^m+B$.
Here, the average gate fidelity is estimated from $p$ only; 
the effects of state preparation and measurement errors are reflected in $A$ and $B$.
Therefore, RB separates errors due to gate operations from that due to state preparation and measurement.
In the graph, the survival probability for the interleaved RB decays faster than that for the standard RB.
This is because there are twice as many applied gates in the interleaved RB.
}
\label{fig:RB}
\end{figure*}

For precise quantum control, a quantitative measure that shows how close the actual gate operation is to the target operation is required.
The quantity showing this is the error rate or gate infidelity $J_\textrm{gate}$ [see Eq.~\eqref{eq:gateInfid} for the definition].
For the surface code, $J_\textrm{gate} \lesssim 0.01$ is required.

The standard method of estimating $J_\textrm{gate}$ in the early days of quantum computation (before the 2010s) was quantum process tomography (QPT).\cite{NC}
Although it gives complete information about the dynamics occurring in the system during a given time, QPT has two drawbacks.
First, it is not scalable---the time for fidelity estimation increases exponentially with the number of qubits.
Second, QPT is susceptible not only to errors associated with gate operation but also to errors associated with state preparation and measurement (SPAM), resulting in inaccurate gate fidelity estimation.

Nowadays, the Clifford group randomized benchmarking (standard RB) is the standard measure of error rates associated with a set of gate operations, i.e., average gate fidelity, because of the following advantages.
First, the Clifford group RB is scalable: the time for fidelity estimation increases polynomially with the number of qubits.
Secondly, the estimated error is insensitive to the type of gate.
Lastly, this protocol is robust against SPAM errors.

The idea of RB is to observe how the gate error accumulates with the number of gate operations, while SPAM errors remain similar with increasing number of gate operations.
We can understand the idea of RB with a toy block analogy in the following way [Fig.~\ref{fig:RB}(a)].
Imagine that we have lots of toy blocks in a toy storage bag with different height errors.
Our job is to measure the average height error of these blocks.
For this, we pick a set of $m$ blocks randomly from the toy storage bag.
Then, we stack these blocks and measure their total height.
By comparing the measured height and our expected height from the design, we can estimate the error $\varepsilon$ as shown in Fig.~\ref{fig:RB}(a).
We can repeat this process by varying $m$ to obtain $\varepsilon$ as a function of $m$ [Fig.~\ref{fig:RB}(c)].
The slope from the fitting indicates the average height error of the toy blocks [$\varepsilon_1$ in Fig.~\ref{fig:RB}(c)].
The advantage of measuring the slope is that it is free from errors originating from an unintended position offset of the stack [$\varepsilon_0$ in Fig.~\ref{fig:RB}(c)].

The procedure of RB is essentially the same as that of the toy blocks explained above.
A toy block corresponds to a quantum gate, and the toy storage bag corresponds to the Clifford group.
The main difference is that the experimentally measured quantity for the toy blocks is the error in height, which increases linearly with $m$,
whereas the measured quantity for quantum gates is the survival probability, which decays exponentially with $m$.
Here, the survival probability means the probability that the initial state is not changed by the gate sequence.
After the entire process, the survival probability is fitted with the function $Ap^m+B$.
The crucial point is that the average gate fidelity is estimated from $p$ only; 
the effects of SPAM errors are reflected in $A$ and $B$.

The detailed protocol of the standard RB is as follows:\cite{magesan2011}
\begin{enumerate}
\itemsep-0.1em 

\item Generate a sequence of $m+1$ gates picked uniformly at random from the Clifford group. Here, the last gate is chosen so that the net sequence is the identity operation.

\item Prepare a state, such as $|0\rangle$ for a single-qubit system or $|00\rangle$ for a two-qubit system.

\item Perform the gate operation.

\item Repeat steps 1--3 to measure the survival probability.

\item Repeat steps 1--4 for various $m$ to obtain the survival probability as a function of $m$.

\item Fit the survival probability with the decay function $A p^m + B$.

\item The average gate infidelity is given by $J_\textrm{gate} = (d-1)(1-p)/d$, where $d \equiv 2^n$ is the dimension of the Hilbert space and $n$ is the number of qubits.
\end{enumerate}

The standard RB gives a single value, the average gate (in)fidelity.
Although such convenience is an advantage of RB, we often need to know the fidelity of a specific gate.
One example is the fidelity evaluation for the closed-loop optimization of a certain gate operation.\cite{kelly2014, werninghaus2020}
In this case, we use the interleaved RB.\cite{magesan2012}

\begin{table*}
\caption{Useful formulas. Note that, because the vector representation of the qubit basis is $\ket{0} \doteq (1, 0)^\top$ and $\ket{1} \doteq (0, 1)^\top$, the relations in the last line are different from those in standard quantum mechanics textbooks.}
\label{tab:formula}\centering
\begin{ruledtabular}
\begin{tabular}{c}
\noalign{\smallskip}
$\mathrm{e}^{\hat{A}}\hat{B}\mathrm{e}^{-\hat{A}} =
\hat{B} + [\hat{A},\hat{B}] 
+ \frac{1}{2!}[\hat{A},[\hat{A},\hat{B}]]
+ \frac{1}{3!}[\hat{A},[\hat{A},[\hat{A},\hat{B}]]]
+ \frac{1}{4!}[\hat{A},[\hat{A},[\hat{A},[\hat{A},\hat{B}]]]]
+ \cdots,$ \\
\noalign{\smallskip}
$[\hat{A} \hat{B},\hat{C}] = \hat{A}[\hat{B},\hat{C}] + [\hat{A},\hat{C}]\hat{B}, \quad
[\hat{a},\hat{a}^\dagger] = 1, \quad
[\hat{a}^\dagger\hat{a},\hat{a}^\dagger] = \hat{a}^\dagger, \quad
[\hat{a}^\dagger\hat{a},\hat{a}] = -\hat{a},$ \\
\noalign{\smallskip}
$[\hat{\sigma}_x,\hat{\sigma}_y] = 2\mathrm{i}\hat{\sigma}_z, \quad
[\hat{\sigma}_y,\hat{\sigma}_z] = 2\mathrm{i}\hat{\sigma}_x, \quad
[\hat{\sigma}_z,\hat{\sigma}_x] = 2\mathrm{i}\hat{\sigma}_y,$ \\
\noalign{\smallskip}
$\hat{\sigma}_+ \equiv \ketbra{1}{0} = \frac{1}{2}(\hat{\sigma}_x - \mathrm{i}\hat{\sigma}_y), \quad
\hat{\sigma}_- \equiv \ketbra{0}{1} = \frac{1}{2}(\hat{\sigma}_x + \mathrm{i}\hat{\sigma}_y), \quad
[\hat{\sigma}_+,\hat{\sigma}_-] = -\hat{\sigma}_z, \quad
\hat{\sigma}_+ \hat{\sigma}_- = \frac{1}{2}(\hat{I}-\hat{\sigma}_z), \quad
[\hat{\sigma}_\pm,\hat{\sigma}_z] = \pm 2\hat{\sigma}_\pm.$ \\
\noalign{\smallskip}
\end{tabular}
\end{ruledtabular}
\end{table*}

The procedure for the interleaved RB is similar to that for the standard RB.
The differences are
(i) a random gate and the gate of interest $\mathcal{C}$ appear alternately;
(ii) the infidelity associated with the gate of interest, $r_\mathcal{C}$, is given by $r_\mathcal{C} = (d-1)(1-p_\mathcal{C}/p)/d$, where $p_\mathcal{C}$ is the fitting parameter in the decay function and $p$ is the fitting parameter obtained from the standard RB.
Note that this formula works only when the gate error is due to a stochastic process, such as thermalization or dephasing.
If the gate error is due to coherent errors, such as imprecise control errors, we might have interferences between these errors.
In this case, we have to estimate the upper and lower bounds of $p_\mathcal{C}$ assuming the best and worst cases of interference.\cite{dugas2019}

Lastly, we point out that, although RB is scalable in principle, it is not clear in practice.
The reason for this is that the implementation of the $N$-qubit Clifford operation requires $\mathcal{O}(N^2/\log N)$ primitive two-qubit gate operations,\cite{aaronson2004}
which implies that, even if the gate fidelity of a primitive two-qubit gate looks reasonably good, the quality of the gate degrades rapidly with increasing number of qubits.
This increases the number of measurements required to estimate the fidelity.
One of the scalable alternatives is cycle benchmarking:
in this protocol, the uncertainty of the fidelity estimate is independent of the number of qubits.
Interested readers should see Ref.~\onlinecite{erhard2019}.

\newpage

\begin{acknowledgments}
S.K. thanks Sota Ino, Paul Magnard, Hiroto Mukai, Shotaro Shirai, Teruaki Yoshioka, and Dengke Zhang for helpful discussions and Giuseppe Falci, Holger Haas, Ian Hincks, Anita Fadavi Roudsari, Rui Wang, Alex Wozniakowski, and anonymous reviewers for valuable comments on the manuscript.
This work was supported by CREST, JST (Grant No. JPMJCR1676) and the New Energy and Industrial Technology Development Organization (NEDO).
\end{acknowledgments}

\end{document}